\documentclass[useAMS, useAMSmath,usenatbib, usegraphicx, twocolumn]{mn2e}

\usepackage{amsbsy}

\title[Galactic rotation curves in brane world models]{Galactic rotation curves in brane world models}
\author[Gergely, Harko, Dwornik, Kupi, Keresztes]{L. \'{A}. Gergely$^{1,}$$^2$\thanks{E-mail:
gergely@physx.u-szeged.hu}, T. Harko$^{3}$\thanks{E-mail:
harko@hkucc.hku.hk}, M. Dwornik$^{1,}$$^2$\thanks{E-mail:
marek@titan.physx.u-szeged.hu}, G. Kupi$^{4}$ \thanks{E-mail:
gabor.kupi@weizmann.ac.il}, Z. Keresztes$^{1,}$$^2$ \thanks{E-mail:
zkeresztes@titan.physx.u-szeged.hu}\\
$^{1}$Department of Theoretical Physics, University of Szeged, Tisza Lajos krt 84-86, Szeged 6720, Hungary \\
$^{2}$ Department of Experimental Physics, University of Szeged, D\'{o}m T\'{e}r 9, Szeged 6720, Hungary \\
$^{3}$Department of Physics and Center for Theoretical and Computational Physics, The University of Hong Kong, \\
Pok Fu Lam Road, Hong Kong, Hong Kong SAR, P. R. China \\
$^{4}$Weizmann Institute of Science, Rehovot 76100, Israel \\
}

\begin{document}



\maketitle


\def\apj{Astrophys. J.}
\def\mnras{Month. Not. Roy. Astr. Soc.}
\def\prd{Phys. Rev. D}
\def\prl{Phys. Rev. Lett.}

\date{Accepted 1988 December 15. Received 1988 December 14; in original form 1988 October 11}

\pubyear{2011}

\maketitle

\label{firstpage}

\begin{abstract}
In the braneworld scenario the four dimensional effective
Einstein equation has extra source terms, which arise from the embedding
of the 3-brane in the bulk. These non-local effects, generated by
the free gravitational field of the bulk, may provide an
explanation for the dynamics of the neutral hydrogen clouds at
large distances from the galactic center, which is usually
explained by postulating the existence of the dark matter.
In the present paper we consider the asymptotic behavior
of the galactic rotation curves in the brane world models, and we compare the theoretical results with observations
of both High Surface Brightness and Low Surface Brightness galaxies.
 For the chosen sample of galaxies we determine first
the baryonic parameters by fitting the photometric data to the adopted galaxy model;
then we test the hypothesis of the Weyl fluid acting as dark matter on the chosen sample
of spiral galaxies by fitting the tangential velocity equation of the  combined baryonic-Weyl
model to the rotation curves.
We give an analytical expression for the rotational velocity of a test
particle on a stable circular orbit in the exterior region to a galaxy,
with Weyl fluid contributions included. The model parameter
ranges for which the $\chi^2$ test provides agreement (within 1$\sigma$ confidence level)
with observations on the velocity fields of the chosen galaxy sample are then determined. There is a good agreement between the theoretical predictions and observations, showing that extra-dimensional models can be effectively used as a viable alternative to the standard dark matter paradigm.

\end{abstract}









%

\section{Introduction}

Gravitational effects which require more matter than what is visible can be
explained in terms of a mysterious dark matter, the nature of which remains
a long-standing problem in modern astrophysics. Two important observational
issues, the behavior of the galactic rotation curves and the mass
discrepancy in clusters of galaxies, led to the necessity of considering the
existence of dark matter both at galactic and extra-galactic scales.

The rotation curves of spiral galaxies~\citep{Bi87, Pe96, Bo01} are among
the best evidences showing the problems Newtonian gravity and/or standard
general relativity have to face on the galactic/intergalactic scale. In
these galaxies neutral hydrogen clouds are observed at large distances from
the center, much beyond the extent of the luminous matter. Since these
clouds move in circular orbits with velocity $v_{tg}(r)$, the orbits are
maintained by the balance between the centrifugal acceleration $v_{tg}^2/r$
and the gravitational attraction $GM(r)/r^2$ of the total mass $M(r)$
contained within the orbit. This allows the expression of the mass profile
of the galaxy in the form $M(r)=rv_{tg}^2/G$.

Observations show that the rotational velocities increase near the center of
the galaxy, in agreement with the theory, but then remain nearly constant at
a value of $v_{tg\infty }\sim 200-300$ km/s~\citep{Bi87}, which leads to a
mass profile $M(r)=rv_{tg\infty }^2/G$. Consequently, the mass within a
distance $r$ from the center of the galaxy increases linearly with $r$, even
at large distances where very little luminous matter has been detected.

The second astrophysical evidence for dark matter comes from the study of
the clusters of galaxies. The total mass of a cluster can be estimated in
two ways. Knowing the motions of its member galaxies~\cite{giov94}, the virial theorem
gives one estimate, $M_{VT}$, say. The second is obtained by separately
estimating the mass of the individual members, and summing up these masses,
to give a total baryonic mass $M_{B}$. Almost without exception it is found
that $M_{VT}$ is considerably larger than $M_{B}$, $M_{VT}>M_{B}$, typical
values of $M_{VT}/M_{B}$ being about 20-30~\citep{Bi87}.

This behavior of the galactic rotation curves and of the virial mass of
galaxy clusters is usually explained by postulating the existence of some
dark (invisible) matter, distributed in a spherical halo around the
galaxies. The dark matter is assumed to be a cold, pressure-less medium.
There are many possible candidates for dark matter, the most popular ones
being the weakly interacting massive particles (WIMP) (for a review of the
particle physics aspects of the dark matter see~\citet{OvWe04}. Their
interaction cross sections with normal baryonic matter, while extremely
small, are expected to be non-zero, and we may expect to detect them
directly. Models based on right-handed (sterile) neutrinos \citep
{sterile,sterileDM} as a form of warm dark matter were also advanced.

It has also been suggested that the dark matter in the Universe might be
composed of super-heavy particles, with mass $\geq 10^{10}$ GeV. But
observational results show that the dark matter can be composed of
super-heavy particles only if these interact weakly with normal matter, or
if their mass is above $10^{15}$ GeV~\citep{AlBa03}. Scalar fields or other
long range coherent fields coupled to gravity have also intensively been
used to model galactic dark matter~\citep{Fu04, He04, Gi05, gulo, scal7, Bri, Ha}.

However, up to now no non-gravitational evidence for dark matter has been
found, and no direct evidence or annihilation radiation from it has been
observed yet.

Therefore, it seems that the possibility that Einstein's (and the Newtonian)
theory of gravity breaks down at the scale of galaxies cannot be excluded 
\textit{a priori}. Several theoretical models, based on a modification of
Newton's law or of general relativity, have been proposed to explain the
behavior of the galactic rotation curves~\citep{Mi83, Sa84, Mo96, Ma97,
 Ro04, Bo07a, Bo07b, Ber, Bo08, alt3, brownstein06}.

The idea of embedding our Universe in a higher dimensional space has
attracted a considerable interest recently, due to the proposal by Randall
and Sundrum~\citep{RS99a, RS99b} that our four-dimensional (4D) space-time
is a three-brane, embedded in a 5D space-time (the bulk). According to the
brane world scenario, the physical fields (electromagnetic, Yang-Mills etc.)
in our 4D Universe are confined to the three brane. Only gravity can freely
propagate in both the brane and bulk space-time, with the gravitational
self-couplings not significantly modified. Even if the fifth dimension is
uncompactified, standard 4D gravity is reproduced on the brane in the
appropriate limit. Hence this model allows the presence of large, or even
infinite non-compact extra dimensions. Our brane is identified to a domain
wall in a 5D anti-de Sitter space-time. For a review of the dynamics and
geometry of brane universes, see e.g.~\citet{Mar04}. In the brane world
scenario, the fundamental scale of gravity is not the Planck scale, but
another scale which may be at the TeV level. The gravitons propagating
through the bulk space give rise to a Kaluza-Klein tower of massive
gravitons on the brane. These gravitons couple to the energy-momentum term
of the standard model fields, and could be produced under the appropriate
circumstances as real or virtual particles.

Due to the correction terms coming from the extra dimensions, significant
deviations from the standard Einstein theory occur in brane world models at
very high energies~\citep{SMS00, Sh00, Ma03, Decomp, VarBraneTensionPRD, Eotvos}.
Gravity is largely modified at the electro-weak scale of about 1~TeV. The
cosmological and astrophysical implications of the brane world theories have
been extensively investigated in the physical literature.

The static vacuum gravitational field equations on the brane depend on the
generally unknown Weyl stresses, which can be expressed in terms of two
functions, called the dark radiation $U$ and the dark pressure $P$ terms
(the projections of the Weyl curvature of the bulk, generating non-local
brane stresses)~\citep{Da00, GeMa01, BlackString, Bo07c, Mar04}. Generally, the vacuum
field equations on the brane can be reduced to a system of two ordinary
differential equations, which describe all the geometric properties of the
vacuum as functions of the dark pressure and dark radiation terms~\citep
{Ha03}. In order to close the system of vacuum field equations on the brane
a functional relation between these two quantities is necessary.

The full 5-dimensional Einstein equations have been solved numerically for static, 
spherically symmetric matter localized on the brane in \citet{Wise}, yielding regular 
geometries in the bulk with axial symmetry. The same data that specifies stars in 4-dimensional 
gravity, uniquely constructs a 5-dimensional solution. An upper mass limit is observed for these 
small stars, and the geometry shows no global pathologies. The intrinsic geometry of large stars, 
with radius several times the AdS length, is described by four-dimensional General Relativity. 
The results obtained in \citet{Wise} show that Randall-Sundrum gravity, with localized brane matter, 
reproduces relativistic astrophysical solutions, such as neutron stars and massive black holes, 
in a way which is consistent with observations.

Several classes of spherically symmetric solutions of the static
gravitational field equations in the vacuum on the brane have been obtained
in~\citet{Ha03, Ma04, Ha05, Bo07c, Collapse, TidalLens}. As a possible physical application of
these solutions the behavior of the angular velocity $v_{tg}$ of the test
particles in stable circular orbits has been considered~\citep{Ma04, Ha06, 
Bo07c, Ra08}. The general form of the solution, together with two constants
of integration, uniquely determines the rotational velocity of the particle.
In the limit of large radial distances, and for a particular set of values
of the integration constants the angular velocity tends to a constant value.
This behavior is typical for massive particles (hydrogen clouds) outside
galaxies~\citep{Bi87}, and is usually explained by postulating the 
existence of the dark matter.

Thus, the rotational galactic curves can be naturally explained in brane
world models, without introducing any additional 
hypothesis \citep{Ma04, Ha06, Bo07c}. The galaxy is embedded in a modified, spherically symmetric
geometry, generated by the non-zero contribution of the Weyl tensor from the
bulk. The extra terms, which can be described in terms of the dark radiation
term $U$ and the dark pressure term $P$, act as a ``matter'' distribution
outside the galaxy. The particles moving in this geometry feel the
gravitational effects of $U$, which can be expressed in terms of an
equivalent mass (the dark mass) $M_U$. The dark mass is linearly increasing
with the distance, and proportional to the baryonic mass of the galaxy, 
$M_{U}(r)\approx M_{B}(r/r_{0})$~\citep{Ma04,Ha06, Bo07c}. The exact 
galactic metric, the dark radiation and the dark pressure in the flat 
rotation curves region in the brane world scenario has been obtained 
in~\citet{Ha06}.

Similar interpretations of the dark matter as bulk effects have been also
considered in~\citet{Pal} and \citet{Pa08}, where it was also shown that it is possible 
to model the X-ray profiles of clusters of galaxies, without the need for 
dark matter.

Hence, a first possible approach to the study of the vacuum brane consists
in adopting an explicit equation of state for the dark pressure as a
function of the dark radiation. A second method consists in closing the
field equations on the brane by imposing the condition of the constancy of
the rotational velocity curves for particles in stable orbits.

It is the purpose of the present paper to consider the general behavior of
the vacuum gravitational field equations in the brane world model in the
region of constant tangential rotational velocity of test particles in
stable circular orbits, and to compare the predictions of our model with the
existing observational data. Physically, this situation is characteristic
for particles gravitating in circular orbits around the galactic center 
\citep{Bi87}. As a first step in our study we derive, under the assumption
of spherical symmetry, the basic equations describing the structure of the
vacuum on the brane, and the equation giving the behavior of the tangential
velocity of the test particles as functions of the dark radiation and of the
dark pressure.

To obtain the tangential velocity of test particles we analyze the geodesics
for a 4D motion by reducing the problem to the motion of a test particle in
an effective potential, also containing the effects of the bulk. By assuming
that the brane is a fixed point of the bulk, we derive the equation giving
the tangential velocity as a function of the $g_{tt}$ component of the metric
tensor only. In the region of constant tangential velocities, the general
solutions of the gravitational equations can be obtained in an exact
analytic form.

To test the viability of the model, and in order to apply it to realistic
systems, we compare the theoretical rotation curves with a sample of 18
galaxies that includes both low and high surface brightness galaxies, with
measured rotation curves extending in the dark matter dominated region. From
this comparison we obtain the numerical values of the parameters
characterizing the equation of state of the Weyl fluid, as well as the
parameters giving the behavior of the rotation curves in the brane world
models.

\citet{Pe96} found that spiral rotation curves show that the 
rotation velocity at any radius depends only on the luminosity. This implies 
that the galactic rotation curves can be modeled by a Universal Rotation 
Curve. These curves show a slightly declining velocity in the galactic halo.

The present paper is organized as follows. The field equations in the brane
world models, as well as the tangential velocity of test particles in stable
circular orbits are presented in Section 2. The structure of the vacuum in
the brane world models is analyzed in Section 3. The explicit form of the
tangential velocity is obtained, for a specific form of the equation of
state of the Weyl fluid, in Section 4. A general theoretical comparison of
the tangential velocity relation with the observational constraints is
performed in Section 5. In Section 6, by fitting the theoretical predictions
of the model with the observational data for a sample of 9 HSB galaxies, the
numerical values of the parameters describing the Weyl fluid are determined.
We use the same method for the LSB galaxies in Section 7. We discuss 
and conclude our results in Section 8.

\section{Field equations and tangential velocity of test particles}

In the present Section we present the field equations for static, spherically symmetric vacuum branes, 
and obtain the velocity of the test particles in stable circular
orbits around the galactic center.

\subsection{The field equations in the brane world models}

We start by considering a five dimensional (5D) space-time (the bulk), with
a single four-dimensional (4d) brane, on which ordinary matter is confined,
only gravity can probe the extra dimensions. The 4D brane world $
({}^{(4)}M,g_{ab})$ is located at a hypersurface $\left( B\left(X^{A}\right) =0\right) $ in the 5d bulk space-time $
({}^{(5)}M,{}^{(5)}g_{AB})$, of which coordinates are described by $
X^{A},A=0,1,...,4$. The 4D coordinates on the brane are $
x^{a},a=0,1,2,3$. All tensors $T_{AB}$ and vectors $V_{A}$ satisfy 
$T\,_{IJ}=\,T\,_{ab}\delta _{I}^{a}\delta _{J}^{b}$
 and $V_{A}=V_{a}\delta _{A}^{a}$ at the hypersurface.

The action of the system is given by~\citep{SMS00} 
\begin{equation}
S=S_{bulk}+S_{brane},  \label{bulk}
\end{equation}
where 
\begin{equation}
S_{bulk}=\int_{{}^{(5)}M}\sqrt{-{}^{(5)}g}\left[ \frac{1}{2k_{5}^{2}}{}%
^{(5)}R+{}^{(5)}L_{m}+\Lambda _{5}\right] d^{5}X,
\end{equation}
and 
\begin{eqnarray}
S_{brane} &=&\int_{{}^{(4)}M}\sqrt{-{}^{(4)}g}\times  \nonumber \\
&&\left[ \frac{1}{k_{5}^{2}}K^{\pm }+L_{brane}\left( g_{ac},\psi \right)
+\lambda _{b}\right] d^{4}x,
\end{eqnarray}
where $k_{5}^{2}=8\pi G_{5}$ is the 5D gravitational constant, ${}^{(5)}R$
and ${}^{(5)}L_{m}$ are the 5D scalar curvature and the matter Lagrangian in
the bulk, representing non-standard model fields, $L_{brane}\left(g_{ab},\psi \right) $ is the 4D Lagrangian, which is given by a generic
functional of the brane metric $g_{ab}$ and of the matter fields $\psi $, $
K^{\pm }$ is the trace of the extrinsic curvature on either side of the
brane, and $\Lambda _{5}$ and $\lambda _{b}$ (the constant brane tension)
are the negative vacuum energy densities in the bulk and on the brane,
respectively.

The Einstein field equations in the bulk can be obtained as~\citep{SMS00} 
\begin{equation}
{}^{(5)}G_{IJ}=k_{5}^{2}{}^{(5)}T_{IJ},
\end{equation}
where 
\begin{equation}
{}^{(5)}T_{IJ}\equiv -2\frac{\delta {}^{(5)}L_{m}}{\delta {}^{(5)}g^{IJ}}
+{}^{(5)}g_{IJ}{}^{(5)}L_{m},
\end{equation}
is the energy-momentum tensor of bulk matter fields. The energy-momentum
tensor of the matter localized on the brane, $T_{\mu \nu }$, is defined by 
\begin{equation}
T_{ab}\equiv -2\frac{\delta L_{brane}}{\delta g^{ab}}+g_{ab}L_{brane}.
\end{equation}
(The minus sign in the above definitions appears because the variation is
done with respect to $g^{ab}$ rather then $g_{ab}$, and there are two terms
because the Lagrangian is varied, rather then the Lagrangian density.)

The delta function $\delta \left( B\right) $ denotes the localization of
brane contribution. In the 5D space-time a brane is a fixed point of the $
Z_{2}$ symmetry. The basic equations on the brane are obtained by
projections onto the brane world. The induced 4D metric is $
g_{IJ}={}^{(5)}g_{IJ}-n_{I}n_{J}$, where $n_{I}$ is the space-like unit
vector field normal to the brane hypersurface ${}^{(4)}M$. In the following
we assume ${}^{(5)}L_{m}=0$.

Assuming a metric of the form $ds^{2}=(n_{I}n_{J}+g_{IJ})dx^{I}dx^{J}$, with 
$n_{I}dx^{I}=d\chi $ the unit normal to the $\chi =\mathrm{constant}$
hypersurfaces and $g_{IJ}$ the induced metric on $\chi =\mathrm{constant}$
hypersurfaces, the effective 4d gravitational equation on the brane (the
effective Einstein equation), takes the form~\citep{SMS00}: 
\begin{equation}
G_{ab}=k_{4}^{2}T_{ab}+k_{5}^{4}S_{ab}-E\,_{ab},  \label{Ein}
\end{equation}
where $S_{ab}$ is the local quadratic energy-momentum correction 
\begin{equation}
S_{ab}=\frac{1}{12}TT_{ab}-\frac{1}{4}T_{a}{}^{c}T_{bc}+\frac{1}{24}%
g_{ab}\left( 3T^{cd}T_{cd}-T^{2}\right) ,
\end{equation}
and $E\,_{ab}$ is the non-local effect from the free bulk gravitational
field, the transmitted projection of the bulk Weyl tensor $C_{IAJB}$, $
E\,_{IJ}=C_{IAJB}n^{A}n^{B}$, with the property $E\,_{IJ}\rightarrow
\,E\,_{ab}\delta _{I}^{a}\delta _{J}^{b}$ as $\chi \rightarrow 0$. We have
also denoted $k_{4}^{2}=8\pi G$, with $G$ the usual 4D gravitational
constant. In the limit $\lambda _{b}^{-1}\rightarrow 0$ we recover standard
general relativity~\citep{SMS00}.

The Einstein equation in the bulk and the Codazzi equation also imply the
conservation of the energy-momentum tensor of the matter on the brane, $
\nabla _{c}T_{a}{}^{c}=0$, where $\nabla _{a}$ denotes the brane covariant
derivative. Moreover, from the contracted Bianchi identities on the brane it
follows that the projected Weyl tensor obeys the constraint $\,\nabla
_{c}E_{a}{}^{c}=k_{5}^{4}\nabla _{c}S_{a}{}^{c}$. Thus the
4-divergence of $E_{ab}$ is constrained by the matter on the
brane \citep{Sh00}.

The generic traceless $E\,_{ab}$ can be decomposed irreducibly with respect
to a chosen $4$-velocity field $u^{a}$ as~\citep{Mar04} 
\begin{equation}
E\,_{ab}=-k_{4}^{4}\left[ U\left( u_{a}u_{b}+\frac{1}{3}h_{ab}\right)
+P_{ab}+2Q_{(a}u_{b)}\right] ,  \label{WT}
\end{equation}
where the induced metric $h_{ab}=g_{ab}+u_{a}u_{b}$ projects orthogonal to $
u^{a}$, the "dark radiation" term $U=-k_{4}^{-4}\,E\,_{ab}u^{a}u^{b}$ is a
scalar, $Q_{a}=k_{4}^{-4}h_{a}^{c}\,E\,_{cd}u^{d}$ is a spatial vector,
finally $P_{ab}=-k_{4}^{-4}\left[ h_{(a}^{c}h_{b)}^{d}-\frac{1}{3}
h_{ab}h^{cd}\right] \,E\,_{cd}$ is a spatial, symmetric and trace-free
tensor.

In the following we neglect the effect of the cosmological constant on the
geometry and dynamics of the galactic particles. In the case of the vacuum
state we have $\rho =p=0$, $T_{\mu \nu }\equiv 0$, and consequently $
S_{ab}=0 $. Therefore the field equation describing a static brane takes the
form 
\begin{equation}
R_{ab}=-\,E\,_{ab},
\end{equation}
with the trace $R$ of the Ricci tensor $R_{ab}$ satisfying the condition $
R=R_{a}^{a}=0$.

In the vacuum case $E\,_{ab}$ also satisfies the constraint $\nabla
_{c}E_{a}{}^{c}=0$. In an inertial frame at any point on the brane we have $
u^{a}=\delta _{0}^{a}$ and $h_{ab}=\mathrm{diag}(0,1,1,1)$ . In a static
vacuum $Q_{a}=0$ and the constraint for $E\,_{ab}$ takes the form~%
\citep{GeMa01} 
\begin{equation}
\frac{1}{3}D_{a}U+\frac{4}{3}UA_{a}+D^{b}P_{ab}+A^{b}P_{ab}=0,  \label{DU}
\end{equation}
where $A_{a}=u^{b}\nabla _{b}u_{a}$ is the 4-acceleration and $D_{a}$
denotes the covariant derivative associated to the metric $h_{ab}$. In the
static spherically symmetric case we may chose $A_{a}=A(r)r_{a}$ and $
P_{ab}=P(r)\left( r_{a}r_{b}-\frac{1}{3}h_{ab}\right) $, where $A(r)$ and $
P(r)$ (the "dark pressure" although the name dark anisotropic stress might
be more appropriate) are some scalar functions of the radial distance~$r$,
and~$r_{a}$ is a unit radial vector~\citep{Da00}. Thus the expression (\ref
{WT}) simplifies to 
\begin{equation}
E\,_{ab}=-k_{4}^{4}\left[ Uu_{a}u_{b}+\frac{U-P}{3}h_{ab}+Pr_{a}r_{b}\right]
.
\end{equation}

\subsection{The motion of particles in stable circular orbits on the brane}

In brane world models test particles are confined to the brane.
Mathematically, this means that the equations governing the motion are the
standard 4d geodesic equations \citep{Mar04}. However, the bulk has an
effect on the motion of the test particles on the brane via the metric.
Since the projected Weyl tensor effectively serves as an additional matter
source, the metric is affected by these bulk effects, and so are the
geodesic equations. This has to be contrasted with Kaluza-Klein theories,
where matter travels on 5d geodesics.

In order to obtain results which are relevant to the galactic dynamics, in
the following we will restrict our study to the static and spherically
symmetric metric given by 
\begin{equation}
ds^{2}=-e^{\nu(r)}dt^{2}+e^{\lambda(r)}dr^{2}+ r^{2}d\Omega^2,  \label{metr1}
\end{equation}
where $d\Omega^2 = d\theta^{2}+\sin^{2}\theta d\phi^{2}$.

The Lagrangian $\mathcal{L}$ for a massive test particle traveling on the
brane reads 
\begin{equation}
\mathcal{L}=\frac{1}{2}\left(-e^{\nu}\dot{t}^{2}+e^{\lambda}\dot{r}^{2}+
r^{2}\dot{\Omega}^2\right),  \label{lag}
\end{equation}
where the dot means differentiation with respect to the affine parameter.

Since the metric tensor coefficients do not explicitly depend on $t$ and $
\Omega $, the Lagrangian~(\ref{lag}) yields the following conserved
quantities (generalized momenta): 
\begin{equation}
-e^{\nu (r)}\dot{t}=E,\qquad r^{2}\dot{\Omega}=L,  \label{cons}
\end{equation}
where $E$ is related to the total energy of the particle and $L$ to the
total angular momentum. With the use of conserved quantities we obtain from
Eq.~(\ref{lag}) the geodesic equation for massive particles (for which $2 
\mathcal{L}=-1$ holds) in the form 
\begin{equation}
e^{\nu +\lambda }\dot{r}^{2}+e^{\nu }\left( 1+\frac{L^{2}}{r^{2}}\right)
=E^{2},  \label{geod1}
\end{equation}

The second term of the left-hand side can, in some cases, be interpreted as
an effective potential. For instance, for the Schwarzschild space-time,
where $e^{\nu+\lambda}=1$, the kinetic term is position independent. In that
case the notion of an effective potential is appropriate. In other cases,
even one can still compute the turning points of the kinetic term, however,
the effective potential interpretation is lost.

For particles in circular and stable orbits the following conditions must be
satisfied: a) $\dot{r}=0$ (circular motion) b) $\partial V_{eff}/\partial r$ 
$=0$ (extreme motion) and c) $\partial ^{2}V_{eff}/\partial r$ $^{2}\left.
{}\right\vert _{{\rm extr}}>0$ (stable orbit), respectively. Conditions a) and b) 
immediately give the conserved quantities as 
\begin{equation}
E^{2}=e^{\nu}\left(1+\frac{L^{2}}{r^{2}}\right),  \label{cons1}
\end{equation}
and 
\begin{equation}
\frac{L^{2}}{r^{2}}=\frac{r\nu^{\prime}}{2}e^{-\nu}E^{2},  \label{cons2}
\end{equation}
respectively. Equivalently, these two equations can be rewritten as 
\begin{equation}
E^{2}=\frac{e^{\nu}}{1-r\nu^{\prime}/2},\qquad L^{2}=\frac{r^{3}\nu
^{\prime}/2}{1-r\nu ^{\prime}/2}.
\end{equation}

We define the tangential velocity $v_{tg}$ of a test particle on the brane,
as measured in terms of the proper time, that is, by an observer located at
the given point, as~\citep{LaLi}) 
\begin{equation}
v_{tg}^{2}=e^{-\nu}r^{2}\dot{\Omega}^{2}\dot{t}^{-2}=e^{-\nu }\frac{L^{2}}{
r^{2}}\dot{t}^{-2}.  \label{vtgbr}
\end{equation}
In the second equality we have employed the second Eq.~(\ref{cons}). By
eliminating $L$ with Eq.~(\ref{cons2}) and subsequently $E$ with the first
Eq.~(\ref{cons}), we obtain the expression of the tangential velocity of a
test particle in a stable circular orbit on the brane as ~\citep{Matos, Nu01}
\begin{equation}
v_{tg}^{2}=\frac{r\nu^{\prime}}{2}.  \label{vtg}
\end{equation}

Let us emphasize again that the function $\nu^{\prime}$ is obtained by
solving the field equations containing the bulk effects as additional matter
terms; we consider this in Section~\ref{sec:3}.

\subsection{The gravitational field equations for a static spherically
symmetric brane}

For the metric given by Eq.~(\ref{metr1}) the gravitational field equations and the effective energy-momentum tensor conservation equation in the vacuum
take the form~\citep{Ha03,Ma04} 
\begin{equation}
-e^{-\lambda}\left(\frac{1}{r^{2}}-\frac{\lambda ^{\prime }}{r}\right) + 
\frac{1}{r^{2}}=3\alpha _bU,  \label{f1}
\end{equation}
\begin{equation}
e^{-\lambda}\left(\frac{\nu^{\prime}}{r}+\frac{1}{r^{2}}\right) -\frac{1 
}{r^{2}}=\alpha _b\left( U+2P\right) ,  \label{f2}
\end{equation}
\begin{equation}
e^{-\lambda }\frac{1}{2}\left(\nu ^{\prime \prime}+\frac{\nu^{\prime 2}}{2 
}+\frac{\nu ^{\prime }-\lambda ^{\prime }}{r}-\frac{\nu ^{\prime }\lambda
^{\prime }}{2}\right) =\alpha _b\left( U-P\right)  ,  \label{f3}
\end{equation}
\begin{equation}
\nu ^{\prime }=-\frac{U^{\prime }+2P^{\prime }}{2U+P}-\frac{6P}{r\left(
2U+P\right) },  \label{f4}
\end{equation}
where $^{\prime }=d/dr$, and we have denoted 
$\alpha _b=k_{4}^{4}/3$. 
Note that Eq.~(\ref{f4}) is a consequence of Eqs.~( 
\ref{f1}), (\ref{f2}) and (\ref{f3}), respectively.

As for the motion of the test particle on the brane we assume that they
follow stable circular orbits, with tangential velocities given by Eq. (\ref%
{vtg}). Thus, the rotational velocity of the test body is determined by the
metric coefficient $\exp \left( \nu \right) $ only.

The field equations~(\ref{f1})--(\ref{f2}) yield the following effective
energy density $\rho ^{\mathrm{eff}}$, radial pressure $P^{\mathrm{eff}}$
and orthogonal pressure $P_{\perp }^{\mathrm{eff}}$, respectively, 
\begin{eqnarray}
\rho ^{\mathrm{eff}} &=&3\alpha _bU,  \label{eq1} \\
P^{\mathrm{eff}} &=&\alpha _b \left(U+2P\right) ,  \label{eq2} \\
P_{\perp }^{\mathrm{eff}} &=&\alpha _b\left(U-P\right) ,  \label{eq3}
\end{eqnarray}
which obey 
$\rho ^{\mathrm{eff}}-P^{\mathrm{eff}}-2P_{\perp }^{\mathrm{eff}}=0$. 
This is expected for the `radiation' like source, the projection of the bulk 
Weyl tensor, which is trace-less, $E_{a}^{a}=0$.

\section{Structure equations of the vacuum in the brane world models}\label{sec:3}

Eq.~(\ref{f1}) can 
immediately be integrated to give 
\begin{equation}
e^{-\lambda }=1-\frac{C_{b}}{r}-\frac{GM_{U}\left( r\right) }{r},  \label{m1}
\end{equation}
where $C_{b}$ is an arbitrary constant of integration, and we denoted 
\begin{equation}
GM_{U}\left( r\right) =3\alpha _b\int_{0}^{r}U(r)r^{2}dr.
\end{equation}

The function $M_{U}$ is the gravitational mass corresponding to the dark
radiation term (the dark mass). For $U=0$ the metric coefficient given by
Eq.~(\ref{m1}) must tend to the standard general relativistic Schwarzschild
metric coefficient, which gives $C_{b}=2GM$, where $M=\mathrm{constant}$ is
the baryonic (usual) mass of the gravitating system.

By substituting $\nu ^{\prime }$ given by Eq.~(\ref{f4}) into Eq.~(\ref{f2})
and with the use of Eq.~(\ref{m1}) we obtain the following system of
differential equations satisfied by the dark radiation term $U$, the dark
pressure $P$ and the dark mass $M_{U}$, describing the vacuum gravitational
field, exterior to a massive body, in the brane world model~\citep{Ha03}: 
\begin{equation}
\frac{dU}{dr}=-\frac{2v_{tg}^{2}\left(2U+P\right) }{r}-2\frac{dP}{dr}-\frac{
6P}{r},  \label{e1}
\end{equation}
\begin{equation}
\frac{dM_{U}}{dr}=\frac{3\alpha _b}{G}r^{2}U,  \label{e2}
\end{equation}
with the tangential velocity given as 
\begin{equation}
v_{tg}^{2}=\frac{1}{2}\frac{2GM+GM_{U}+\alpha _b\left(U+2P\right) r^{3}}{
r\left( 1-\frac{2GM}{r}-\frac{GM_{U}}{r}\right)}.  \label{vtg2}
\end{equation}

In order to close the system a supplementary functional relation between one
of the unknowns $U$, $P$, $M_{U}$ and $v_{tg}$ is needed. Once this relation
is known, Eqs.~(\ref{e1})--(\ref{vtg2}) give a full description of the
geometrical properties and of the motion of the particles on the brane.

The system of equations~(\ref{e1}) and~(\ref{e2}) can be transformed to an
autonomous system of differential equations by means of the transformations 
\begin{eqnarray}\label{trans}
\theta &=&\ln r, \;q =1-e^{-\lambda}=\frac{2GM}{r}+\frac{GM_{U}}{r},  \\
\mu &=&3\alpha _br^{2}U, \;p=3\alpha _b r^{2}P,
\end{eqnarray}

We shall call $\mu $ and $p$ the "reduced" dark radiation and pressure, respectively.

With the use of the new variables given by Eqs.~(\ref{trans}), Eqs.~(\ref{e2}) and (\ref{e1}) become 
\begin{equation}
\frac{dq}{d\theta}=\mu -q,  \label{aut1}
\end{equation}
\begin{equation}
\frac{d\mu}{d\theta }=-\frac{\left(2\mu +p\right) \left[q+\frac{1}{3}
\left(\mu +2p\right)\right]}{1-q}-2\frac{dp}{d\theta }+2\mu -2p.
\label{aut2}
\end{equation}

Eqs.~(\ref{e1}) and~(\ref{e2}), or, equivalently,~(\ref{aut1}) and~(\ref
{aut2}) may be called the structure equations of the vacuum on the brane. In
order to close this system an ``equation of state'', relating the reduced
dark radiation and the dark pressure terms is needed. Generally, this
equation of state is given in the form $P=P(U)$.

In the new variables the tangential velocity of a particle in a stable
circular orbit on the brane is given by 
\begin{equation}
v_{tg}^{2}=\frac{1}{2}\frac{q+\frac{1}{3}\left( \mu +2p\right) }{1-q}.
\label{vtgx}
\end{equation}
By using the expression of the tangential velocity, Eq.~(\ref{aut2}) can be
rewritten as 
\begin{equation}
\frac{d}{d\theta }\left(\mu +2p\right) =-2\left(2\mu +p\right)
v_{tg}^{2}+2\mu -2p.  \label{muvtg}
\end{equation}

Eqs.~(\ref{vtgx}) and (\ref{muvtg}) allow the easy check of the physical consistency of some 
simple equations of state for the dark pressure. The equation of state $\mu 
+2p=0$ immediately gives $v_{tg}^2=1$ and $q=2/3$, respectively, implying that all test particles in 
stable circular motion on the brane move at the speed of light. This result contradicts 
the assumption that the test particles are time-like, as well as the observations on 
galactic scale. Therefore the equation of state $\mu
+2p=0$ is not consistent with the rotation curves. The equation of 
state $ 2\mu +p=0$ gives $\mu =\mu _0/r^2$, where $\mu _0$ = constant is an
arbitrary integration constant, $U=\mu _0/3\alpha _br^4$ and $
GM_U=-\mu _0/3r^3$, respectively. In the limit of large $r$, the
tangential velocity $v_{tg}^2$ tends to zero, $v_{tg}\rightarrow 0$.
Therefore, this model seems also to be ruled out by observations. The case $
\mu =p$ gives $\mu \left( \theta \right) =\mu _{0}\exp \left[ -2\int {
v_{tg}^2\left( \theta \right) }d\theta \right] $, $\mu _{0}=\mathrm{
constant}$, and $q(\theta )=\left( 2v_{tg}^2-\mu \right) /\left(
1+2v_{tg}^2\right) $.

\section{Tangential velocity for galaxies for a linear equation of state of
the Weyl fluid}\label{IV}


\subsection{Linear equation of state for the Weyl fluid}

In order to close the system of equations Eqs.~(\ref{aut1}) and (\ref{aut2}),
we need to specify the equation of state relating the dark
pressure to the dark radiation. In the full 5-dimensional approach of %
\citet{Wise} the Einstein field equations were solved numerically for
static, spherically symmetric matter localized on the brane, yielding
regular geometries in a bulk with axial symmetry. For this a density
profile, taken as a deformed top hat function, was imposed.

An alternative approach for closing the system of field equations
can be obtained from the 3+1+1 covariant approach in brane worlds, developed
in \citet{3+1+1,3+1+1b}. First, we impose: i) cosmological vacuum in
5-dimensional spacetime, ii) the brane embedding is symmetrical, iii)
fine-tuning on the brane (the brane cosmological constant vanishes), iv) no
matter on the brane, v) the brane is static and spherical symmetric, with
metric given by Eq.~(\ref{metr1}). The set of gravito-electro-magnetic
quantities in the 3+1+1 covariant approach is presented in Appendix I. Due
to assumptions ii) and iv) the Lanczos equations impose that the tensorial
and vectorial projections of the extrinsic brane curvature along the brane
normal $n^{A}$ vanish: $\widehat{\sigma}_{ab}=0=\widehat{K}_{a}$. 
Therefore $\mathcal{E}_{ab}=\widetilde{E}_{ab}-k_{4}^{4}P_{ab}/2$ 
 and $\mathcal{H}_{ab}=\widetilde{H}_{ab}$, respectively.

From the above assumptions we also find that the only nonzero
kinematical quantity related to the normal $u^{a}$ of the $t=$
const. hypersurfaces is its acceleration $A_{a}=u^{b}D_{b}u_{a}$ 
(here $D_{a}$ is the covariant derivative on the brane).
The kinematical quantities of the vector congruences $u^{a}$ and $
r^{a}$ are enlisted in Tables \ref{ukin} and \ref{rkin}.

\begin{table}
\begin{center}
{
\begin{tabular}{|c|c|c|c|}
acceleration & expansion & shear & vorticity \\ \hline\hline
$A$ & $0$ & $0$ & $0$ \\ \hline
\end{tabular}
}
\caption{Kinematical scalars of the congruence $u^{b}$ in the 3+1 
decomposition of $\nabla _{a}u_{b}$. }
\label{ukin}
\end{center}
\end{table}
\begin{table}
\begin{center}
{
\begin{tabular}{|c|c|c|c|}
acceleration & expansion & shear & vorticity \\ \hline\hline
$0$ & $\widetilde{\Theta }$ & $0$ & $0$ \\ \hline
\end{tabular}
}
\caption{Kinematical scalars of the congruence $r^{b}$ in the 2+1 
decomposition of $D_{a}r_{b}$.}
\label{rkin}
\end{center}
\end{table}

Due to the spherical symmetry, assumption v), it is convenient to
decompose the brane spacetime into a 2+1+1 form. Then the
gravito-electro-magnetic quantities appearing in the brane equations further
reduce to $H_{ab}=0$, $\widetilde{E}_{ab}=\widetilde{E}\left(r\right)
\left(r_{a}r_{b}-h_{ab}/3\right) $, $k_{4}^{4}U\left(r\right) $, $Q_{a}=0$ 
and $k_{4}^{4}P_{ab}=k_{4}^{4}P\left( r\right) \left(r_{a}r_{b}-h_{ab}/3\right)$, 
respectively, while $A_{a}=A\left(r\right) r_{a}$. Therefore under the assumption of spherical
symmetry the Weyl fluid is characterized by the set $U\left(r\right)
,~P\left(r\right)$; the electric part of 4-dimensional Weyl tensor
by $\widetilde{E}\left( r\right)$; and the acceleration of the
time-like normal to the 3-space by $A\left(r\right)$.

We also introduce the covariant derivative $^{\left(3\right)
}D_{a}$\ associated with the 3-metric $h_{ab}$ and the
expansion $\widetilde{\Theta }\left(r\right) $ of the radial
geodesics in the local three-dimensional space (obeying $r^{a}~^{\left(3\right) }D_{a}r_{b}=0$) by the relation $^{\left(3\right)
}D^{a}r_{a}=\widetilde{\Theta }\left( r\right)$. The 4 independent
field equations for the variables $U,~P,~\widetilde{E},~A,$ and $
\widetilde{\Theta }$, arising in the 2+1+1 brane formalism
are given in Appendix II. There are two algebraic equations [Eqs. (\ref{beq5})
and (\ref{rcoord})] determining $U$ and $P$ as
function of $\widetilde{E},A,$ and $\widetilde{\Theta }$: 
\begin{equation}
k_{4}^{4}U=\widetilde{\Theta }\left(A-\frac{\widetilde{\Theta }}{4}\right) +%
\frac{4\widetilde{E}}{3}\,+\frac{1}{r^{2}},  \label{Ueq}
\end{equation}
\begin{equation}
k_{4}^{4}P=\widetilde{\Theta }\left(A+\frac{\widetilde{\Theta }}{2}\right) -%
\frac{2\widetilde{E}}{3}\,-\frac{2}{r^{2}}\,\,,  \label{Peq}
\end{equation}
and two first order non-linear ordinary differential equations [Eqs.
(\ref{beqr1}) and (\ref{beqr2})] for the variables $A$, $
\widetilde{\Theta }$ (also containing $\widetilde{E}$): 
\begin{equation}
\frac{r\widetilde{\Theta }}{2}\widetilde{\Theta }^{\prime }+\frac{\widetilde{
\Theta}^{2}}{2}+\frac{4\widetilde{E}}{3}+A\widetilde{\Theta }=0\,,
\label{Thetaprime}
\end{equation}
\begin{equation}
\frac{r\widetilde{\Theta}}{2}A^{\prime }+A^{2}+\frac{\widetilde{\Theta }^{2}
}{4}-\frac{4\widetilde{E}}{3}\,-\frac{1}{r^{2}}=0\ .  \label{Aprime}
\end{equation}
The last two equations contain only quantities defined and
determined by the brane dynamics. In order to obtain a special solution of
the system (\ref{Thetaprime})-(\ref{Aprime}), we need a third relation
between the brane kinematic quantities $A$, $\widetilde{\Theta }$ 
and electric Weyl brane curvature $\widetilde{E}$. In order
to establish this, first we remark, that in the Schwarzschild case these
quantities obey 
\begin{equation}
\frac{2\widetilde{E}}{3}\,+\widetilde{\Theta }A=\widetilde{\Theta }\left( 
\frac{\widetilde{\Theta }}{4}\,\,+A\right) -\frac{1}{r^{2}}\,=0.
\label{idsSch}
\end{equation}
By virtue of these the two Eqs. (\ref{Thetaprime})-(\ref{Aprime})
are found to coincide, thus $A$, $\widetilde{\Theta }$ and
 $\widetilde{E}$ are still defined by a set of three equations.
For a spherically symmetic solution on the brane we allow for a slightly
modified identity as compared to the first Eq. (\ref{idsSch}): 
\begin{equation}
\frac{2\widetilde{E}}{3}\,+\widetilde{\Theta }A=\mathcal{A}\widetilde{\Theta 
}\left( \frac{\widetilde{\Theta }}{4}\,\,+A\right) -\frac{\mathcal{B}}{r^{2}}%
\,,  \label{newEq}
\end{equation}%
with $\mathcal{A}$ and $\mathcal{B}$ two
constants, both reducing to $1$ in Schwarzschild case. By employing
Eqs. (\ref{Ueq}), (\ref{Peq}), this equation can be rewritten as a simple
equation of state\ for the Weyl fluid:
\begin{equation}
P=\left(a-2\right) U-\frac{B}{k_{4}^{4}r^{2}}\ .  \label{EOS}
\end{equation}
Here we have introduced the new constants $a,B$ by
redefining $\mathcal{A}=a/\left(2a-3\right) $, $\mathcal{B}=\left(
a-B\right) /\left( 2a-3\right) $. We also remark that in the
variables given by Eq.~(\ref{trans}) the equation (\ref{EOS}) takes the
simple linear form 
\begin{equation}
p\left(\mu \right) =\left( a-2\right) \mu -B\ .  \label{eqstate}
\end{equation}
The system (\ref{Thetaprime})-(\ref{Aprime}) and (\ref{newEq}),
being a first order ordinary system of differential equations, determines
the variables $A$, $\widetilde{\Theta }$ and $\widetilde{E
}$. Their existence is assured by the Cauchy-Peano theorem. When
rewritten in metric variables, the solution valid in the region where
rotation curve data is available will be obtained in the following
subsection.

One can be rightously worried about the compatibility of Eq. (\ref
{EOS}) with the full 5-dimensional gravitational dynamics, however this is
but a boundary condition imposed on the brane at some arbitrary time. Then
the static character of the problem assures that it is preserved by the
dynamics throughout the temporal evolution.

Looking to this problem from a mathematical point of view, for the
system of first order differential equations describing the the extra
dimensional evolution \citet{3+1+1}, the choice of the Weyl fluid variables
represents an "initial condition" in any suitable off-brane parameter (for
example, the one associated with the integral curves of the vector field $
n^{A}$). Such "initial values" could be chosen arbitrarily except
when there is a constraint to be obeyed on the "initial surface". This
surface being the brane, the condition to be obeyed is the constraint
equation for the Weyl fluid, Eq. (\ref{DU}). As shown in Appendix II, on a
static and spherically symmetric brane Eq. (\ref{DU}) simplifies to Eq. (\ref
{beqr2}). It is not very difficult to see, that this equation follows from
the system (\ref{Ueq})-(\ref{Aprime}). Therefore the constraint for the Weyl
fluid is trivially satisfied in the static and spherically symetric case\
and the particular choice of "initial conditions" (\ref{newEq}) is allowed,
as would be any other choice. The advantage of chosing Eq. (\ref{newEq}) is,
however, that it leads to the simple linear equation of state in the reduced
Weyl variables $p$ and $\mu $. We will see that the
confrontation of the model with galactic rotation curve data supports the
choice (\ref{newEq}) from a physical point of view either.

\subsection{The metric and tangential velocity on the brane}

The dark radiation and the dark pressure can be obtained as a functions of
the tangential velocity in a closed analytical form for the equation of
state given by Eq.~(\ref{eqstate}). The reduced dark radiation can be
obtained as 
\begin{eqnarray}
\mu \left(\theta \right) &=&\theta ^{2\left( 3-a\right) /\left( 2a-3\right)
}\exp \left[ -\frac{2a}{2a-3}\int v_{tg}^{2}\left( \theta \right) d\theta 
\right] \times  \nonumber \\
&&\left\{ C_{0}-\frac{3B}{2a-3}\int \left[ 1+v_{tg}^{2}\left( \theta \right) 
\right] \theta ^{-2\left( 3-a\right) /\left( 2a-3\right) }\right. \times 
\nonumber \\
&&\left. \exp \left[ \frac{2a}{2a-3}\int v_{tg}^{2}\left( \theta \right)
d\theta \right] \right\} ,
\end{eqnarray}%
where $C_{0}$ is an arbitrary integration constant. Hence, if the velocity
profile of a test particle in stable circular motion is known, one can
obtain all the relevant physical parameters for a static spherically
symmetric system on the brane.

By using the linear equation of state of the dark pressure Eq.~(\ref{aut2})
takes the form 
\begin{eqnarray}
\left( 2a-3\right) \frac{d\mu }{d\theta }&=&-\frac{\left( a\mu -B\right) 
\left[ q+\left( 2a-3\right) \mu /3-2B/3\right] }{1-q}+  \nonumber \\
&&2\left( 3-a\right) \mu +2B,  \label{mueq}
\end{eqnarray}
where we have neglected the possible effect of the cosmological constant on
the structure of the cluster. Due to the mathematical structure of Eq.~(\ref
{mueq}) there are two cases that can be considered separately, $a=3/2,$ and $
a\neq 3/2$, respectively.

For a galactic dark matter halo with mass of the order of $M=10^{12}M_{\odot
}$ and radius $R=100$ kpc, the quantity $2GM/R$ is of the order of $
9.6\times 10^{-7}$, which is much smaller than one. Since observations show
that inside the galaxy the mass is a linearly increasing function of the
radius $r$, the value of this ratio is roughly the same at all points in the
galaxy. Therefore from its definition it follows that generally $q<<1$, and $
1-q\approx 1$. Moreover, the quantities $q^{2}$ and $qdq/d\theta $ are also
very small as compared to $q$. Eq.~(\ref{aut1}) gives $\mu =q+dq/d\theta $, $
d\mu /d\theta =dq/d\theta +d^{2}q/d\theta ^{2}$.

\subsubsection{The case $a\neq 3/2$}

Hence, by neglecting the second order terms and assuming $a\neq 3/2$, we
obtain for $q$ the following differential equation: 
\begin{equation}
\frac{d^{2}q}{d\theta ^{2}}+m\frac{dq}{d\theta }-nq=b,  \label{lineq}
\end{equation}
where we have denoted 
\begin{equation}  \label{m}
m=1-\frac{B}{3}-\frac{2}{3}\frac{a\left( B-3\right) +9}{2a-3},a\neq \frac{3}{
2},
\end{equation}
\begin{equation}  \label{n}
n=\frac{2}{3}\frac{a\left( 2B-3\right) +9}{2a-3},a\neq \frac{3}{2},
\end{equation}
and 
\begin{equation}
b=\frac{2}{3}\frac{B(B-3)}{3-2a},a\neq \frac{3}{2},
\end{equation}
respectively. The general solution of Eq. (\ref{lineq}) is given by 
\begin{equation}
q\left( \theta \right) =v_{0}+C_{1}e^{l_{1}\theta }+C_{2}e^{l_{2}\theta
},a\neq \frac{3}{2},
\end{equation}
where $C_{1}$ and $C_{2}$ are arbitrary constants of integration, and we
denoted 
\begin{equation}  \label{v0}
v_{0}=-\frac{b}{n}=\frac{B(B-3)}{a\left( 2B-3\right) +9},
\end{equation}
and 
\begin{equation}  \label{l12}
l_{1,2}=\frac{-m\pm \sqrt{m^{2}+4n}}{2},  \label{l}
\end{equation}
respectively. The reduced dark radiation term is given by 
\begin{equation}
\mu \left( \theta \right) =v_{0}+C_{1}\left( 1+l_{1}\right) e^{l_{1}\theta
}+C_{2}\left( 1+l_{2}\right) e^{l_{2}\theta }.
\end{equation}

The tangential velocity of a test particle in the ''dark matter'' dominated
region is given by 
\begin{equation}
v_{tg}^{2}\approx \frac{1}{2}\left[ q+\frac{1}{3}\left( \mu +2p\right) 
\right] \approx \frac{1}{2}\left[ q+\frac{\left( 2a-3\right) \mu -2B}{3} 
\right] ,
\end{equation}
or 
\begin{equation}
v_{tg}^{2}\left( \theta \right) \approx v_{tg\infty }^{2}+\gamma
e^{l_{1}\theta }+\eta e^{l_{2}\theta },
\end{equation}
where we have denoted 
\begin{equation}  \label{vtgin}
v_{tg\infty }^{2}=\frac{1}{3}\left( av_{0}-B\right) ,
\end{equation}
\begin{equation}  \label{gamma}
\gamma =\frac{1}{2}\left[ \frac{2a-3}{3}\left( 1+l_{1}\right) +1\right]
C_{1},
\end{equation}
\begin{equation}  \label{eta}
\eta =\frac{1}{2}\left[ \frac{2a-3}{3}\left( 1+l_{2}\right) +1\right] C_{2}.
\end{equation}

In the initial radial coordinate $r$ the tangential velocity can be
expressed as 
\begin{equation}
v_{tg}^{2}\left( r\right) \approx v_{tg\infty }^{2}+\gamma r^{l_{1}}+\eta
r^{l_{2}}.  \label{v2core}
\end{equation}
In order to have an asymptotically constant $v_{tg}^{2}\left( r\right) ,$
both $l_{1}$ and $l_{2}$ should be negative numbers, $l_{1}<0$ and $l_{2}<0$
, respectively. This can be achieved if $m>0$, $n<0$ hold simultaneously.

In the original radial variable $r$ we obtain for the dark radiation and the
mass distribution inside the cluster the expressions 
\begin{equation}
3\alpha _bU(r)=\frac{v_{0}}{r^{2}}+C_{1}\left( 1+l_{1}\right)
r^{l_{1}-2}+C_{2}\left( 1+l_{2}\right) r^{l_{2}-2},  \label{UDARK}
\end{equation}
and 
\begin{equation}
GM_{U}(r)=r\left( v_{0}+C_{1}r^{l_{1}}+C_{2}r^{l_{2}}\right) -2GM,
\label{UMASS}
\end{equation}
respectively, where we have neglected the possible effect of the
cosmological constant. The first term in the mass profile of the dark mass
given by Eq.~(\ref{UMASS}) is linearly increasing with $r$, thus having a
similar behavior to the dark matter outside galaxies. The second and third
terms in Eq.~(\ref{UMASS}) are either constants (for $l_{1,2}=-1$) or they
decrease with increasing $r$, as $l_{1,2}<0$.

Finally, the metric coefficient $\nu $ can be calculated from the equation 
\begin{equation}
\frac{d\nu \left( \theta \right) }{d\theta }=q\left( \theta \right) +\frac{
2a-3}{3}\mu \left( \theta \right) -\frac{2B}{3},
\end{equation}
giving 
\begin{eqnarray}
e^{\nu \left( r\right) }&=&C_{\nu }r^{2v_{tg\infty }^{2}}\times  \nonumber \\
&&\exp \left[ C_{1}\frac{3+\left( 2a-3\right) \left( 1+l_{1}\right) }{3l_{1}}
r^{l_{1}}\right.+  \nonumber \\
&&\left.C_{2} \frac{3+\left( 2a-3\right) \left( 1+l_{2}\right) }{3l_{2}}
r^{l_{2}}\right] ,  \label{NU}
\end{eqnarray}
where $C_{\nu }$ is an arbitrary integration constant. In the limit of large
distances $e^{\nu \left(r\right) }$ behaves like $e^{\nu \left( r\right)
}\approx C_{\nu }r^{2v_{tg\infty }^{2}}$. Inside the galaxy we can
approximate $e^{-\lambda }\approx 1-C/r-GM_{U}(r)/r$.

A few comments are in order. First, in the particular case $B=0$, we have $
v_{0}\equiv 0$, and in consequence the obtained mass profile is not
consistent in general with observations, as for such an equation of state
the dark radiation usually cannot play the role of the dark matter (it has
no $r^{-1}$ term). For example in the case $a=4$, corresponding to an
equation of state $P=2U$, the dark mass is given by $
GM_{U}(r)=C_{1}/r^{1.97}+C_{2}/r^{0.82}-C_{b}$.

Secondly, the effective, geometry induced mass must satisfy the condition $
M_{U}\geq 0$. Therefore the solution obtained in the present Section is
valid only for values of the coordinate radius $r$ so that $
v_{0}r+C_{1}r^{l_{1}+1}+C_{2}r^{l_{2}+1}-C\geq 0$. In the limit of small $r$
, taking into account that $l_{1}<0$ and $l_{2}<0$, and by assuming that the
coefficient of the dominant term (either $v_{0},~C_{1}$ or $C_{2}$) is
positive, the dark mass diverges at the center of the cluster, $
\lim_{r\rightarrow 0}M_{U}(r)=\infty $. Otherwise there is a point $r_{0}$,
where $M_{U}(r_{0})\approx 0$ and $M_{U}(r)<0$, for $r<r_{0}$. In this case
the solution has physically acceptable properties only in the region $r\geq
r_{0}$.

\section{Further simplification from observational constraints}

In Section~\ref{IV} the assumption $q\ll 1$ was made in order to derive the
analytical solution. As it can be seen from its definition, he quantity $q$
is basically a post-Newtonian parameter and as such the assumption is
justified when studying galactic rotation curves. Further, as we are
studying bounded motions in the galaxy, due to the virial theorem $
v_{tg}^{2} $ should be of the same order. Eq.~(\ref{vtgx}), and the remark
preceding Subsection IV.A , stating $\mu \ll 1$, together imply $p\ll 1$.
Then, from the definition of $p$, Eq.~(\ref{eqstate}) the condition $B\ll
a-2 $ stems out. In other words, $B/a$ should be of the same order of
magnitude as $\mu $ and $q$, respectively.

In what follows, we will use the above remark for simplifying the generic
results established for $a\neq 3/2$ in the previous Section. For the
parameters $m,~n$ we obtain 
\begin{equation}
m\approx \frac{4a-9}{2a-3}~,\qquad n\approx -2\frac{a-3}{2a-3},
\end{equation}
which give%
\begin{equation}
l_{1}\approx -1,\qquad l_{2}\approx -1+\frac{3}{2a-3}.
\end{equation}
In order $l_{2}<0$ to hold (or equivalently $m>0$ and $n<0$), the constant $
a $ must take values in the range $a\in \left( -\infty ,3/2\right) \cup
\left( 3,\infty \right) $.

With these values of $l_{1,2}$ the constants $C_{1.2}$ simplify to $\gamma
\approx C_{1}/2$ and $\eta \approx C_{2}$. We also get 
\begin{equation}
v_{0}=\frac{B\left(B-3\right)/a}{2B-3+9/a}\approx \frac{B}{a-3},
\end{equation}
for any $\ a$ in the allowed domain. The rotational velocity at infinity
then becomes 
\begin{equation}
v_{tg\infty }^{2}=\frac{a}{3}\left( v_{0}-\frac{B}{a}\right) \approx \frac{B 
}{a-3},  \label{v2tginfty}
\end{equation}
Obviously $v_{tg\infty }^{2}\geq 0$, therefore {\it either } $a<3/2$ 
{\it and } $B\leq 0$ {\it or } $a>3$ {\it and } $B>0$.

The metric functions approximate as 
\begin{equation}  \label{U}
3\alpha _bU(r)\approx \frac{v_{0}}{r^{2}}+\frac{3C_{2}}{2a-3}r^{-3\left( 1- 
\frac{1}{2a-3}\right) },
\end{equation}
\begin{equation}
GM_{U}(r)\approx v_{0}r+C_{1}+C_{2}r^{\frac{3}{2a-3}}-2GM,
\end{equation}
\begin{equation}
e^{\nu \left( r\right) }\approx C_{\nu }r^{2v_{tg\infty }^{2}}\exp \left[
-C_{1}r^{-1}-C_{2}\frac{2a-3}{a-3}r^{-1+\frac{3}{2a-3}}\right],
\end{equation}
and for a generic $r$ the rotational velocity can be written as 
\begin{equation}
v_{tg}^{2}\left( r\right) \approx \frac{B}{a-3}+\frac{C_{1}}{2}
r^{-1}+C_{2}r^{-1+\frac{3}{2a-3}}.  \label{v2out}
\end{equation}

By defining $\alpha =3/\left( 2a-3\right) =l_{2}+1$ and $\beta =B/\left(
a-3\right) $, respectively; then re-introducing the velocity of light $c$
for dimensional reasons; by replacing $C_{1}$ with a mass type constant $M+M_{0}{=C}_{1}/2Gc^{2}$, 
and also by rescaling the constant $
C_{2}$ as $C_{2}=Cc^{2}r_{b}^{1-\alpha }$, we obtain 
\begin{equation}
\left( \frac{v_{tg}\left( r\right) }{c}\right) ^{2}\approx {\frac{{G}\left( {
M}_{b}^{tot}+M_{0}^{tot}\right) }{c^{2}{r}}}+\beta +C\left( \frac{r_{b}}{r}
\right) ^{1-\alpha }~,~r>r^{\ast }.  \label{v2tg}
\end{equation}
Here we have changed the notations ($M,M_{0}$) to ($M_{b}^{tot},M_{0}^{tot}$)
for reasons to be explained in what follows. This solution is valid for
any $r$ $>r^{\ast },$ where $r^{\ast }$ represents the radius beyond which
the baryonic matter does not extend. We interpret the constant $M_{b}^{tot}$
as the total baryonic mass inside radius $r^{\ast }$, while $M_{0}^{tot}$ is
a universal constant due to the brane fluid. Thus the Weyl fluid is
characterized by three universal dimensionless constants $\alpha ,~\beta ,~C$
and a mass-type constant $M_{0}^{tot}$. Also an arbitrary scaling constant $
r_{b}$ was introduced in order to have $C$ dimensionless.

Due to the constraints established for the Weyl fluid parameters $a$ and $B$
, the newly introduced constants $\alpha ,~\beta $ obey {\it either }$
\alpha <0$ {\it or } $0<\alpha <1$ (the value $\alpha =0$ is excluded, as
it would correspond to the unphysical values $a\rightarrow \pm \infty $) 
 $\;${\it in both cases }$\beta $ {\it being a small positive number} $
0<\beta \ll 1$ (with the exception of $a \approx 3$, translating to $
\alpha \approx 1$, when $\beta $ can be an arbitrary positive number).

\section{High Surface Brightness galaxies}

\subsection{The baryonic sector: the bulge-disk decomposition}

We model the distribution of baryonic mass in High Surface Brightness
galaxies as a sum of disk and bulge components with constant, but distinct
mass-to-light ratios. We estimate the bulge parameters from a S\'{e}rsic $
r^{1/n}$ bulge model and the disk parameters from an exponential disk model,
both fitted to the optical I-band galaxy light profiles.

The surface brightness (specific intensity) of the spheroidal bulge
component of each galaxy is given by a generalized S\'{e}rsic function \citep
{Sersic}: 
\begin{equation}
I_{b}(r)=I_{0,b}\exp \left[ -\left( \frac{r}{r_{0}}\right) ^{1/n}\right] ,
\label{Ib}
\end{equation}
where $I_{0,b}$ is the central surface brightness of the bulge, $r_{0}$ is
its characteristic radius and $n$ is the shape parameter of the
magnitude-radius curve. As a rule, early-type spiral galaxy bulges have $n>1$
, while late-type spiral galaxy bulges are characterized by $n<1$. The
radius of the bulge $r_{b}$ is defined by the condition of the surface
brightness being equal to $\mu _{0}=2,64\times 10^{-4}$ mJy$\cdot $ arcsec$
^{-2}$ in the I-band images.

In a spiral galaxy, the radial surface brightness profile of the disk
exponentially decreases with the radius \citep{Freeman} 
\begin{equation}
I_{d}(r)=I_{0,d}\exp \left( -\frac{r}{h}\right) ,  \label{Id}
\end{equation}
where $I_{0,d}$ is the disk central surface brightness and $h$ is a
characteristic disk length scale. In order to measure the light and mass
distribution in the disk, the model image of the bulge is subtracted from
the original I-band image. All remaining light in these images is then
assumed to originate from the disk component.

The (bolometric) luminosity of any of the galaxy components is an integral
over surface, solid angle and frequency of $I_{b}$ and $I_{d}$,
respectively. The respective mass over luminosity is the mass-to-light 
ratio - for the Sun $\gamma _{\odot }=5133$ kg W$^{-1}$. The mass-to-light 
ratios of the bulge and disk $\sigma$ and $\tau _{b}$ will be given in units 
of $\gamma _{\odot }$ (solar units). We will also give the masses in units 
of the solar mass $M_{\odot }=1.98892\times 10^{30}$ kg. Assuming that the 
mass distribution of a spiral galaxy follows the de-projected surface 
brightness distribution (\cite{kanna07}; \cite{porti04} )(corrected with respect to the inclination, cf. \citep
{Palunas}, with constant mass-to-light ratios $\sigma $ and $\tau _{b}$ the
bulge and disk masses within the radius $r$ are%
\begin{eqnarray}
M_{b}(r) &=&\sigma \frac{\mathcal{N}(D)}{F_{\odot }}2\pi
\int\limits_{0}^{r}I_{b}(r)rdr, \\
M_{d}\left( r\right) &=&\tau _{b}\frac{\mathcal{N}(D)}{F_{\odot }}2\pi
\int\limits_{0}^{r}I_{d}(r)rdr,  \label{Md}
\end{eqnarray}
where $F_{\odot }\left( D\right) $ is the apparent flux density of the Sun
at a distance $D$ Mpc, $F_{\odot }\left( D\right) =2.635\times
10^{6-0.4f_{\odot }} \;\mathrm{mJy}{\ }$, with $f_{\odot } =4.08+5\lg \left(
D/1\;\mathrm{Mpc}\right) +25~\mathrm{mag}$, and 
\begin{equation}
\mathcal{N}(D)=4.4684\times 10^{-35}D^{-2} \;\mathrm{m}^{-2}\;\mathrm{arcsec}
^{2}.
\end{equation}

The baryonic matter at any $r$ is given by 
\begin{equation}
M_{baryonic}(r)=M_{b}(r)+M_{d}(r).  \label{bar}
\end{equation}

\begin{figure*}
\begin{center}
\begin{tabular}{ccc}
\includegraphics[height=4cm, angle=270]{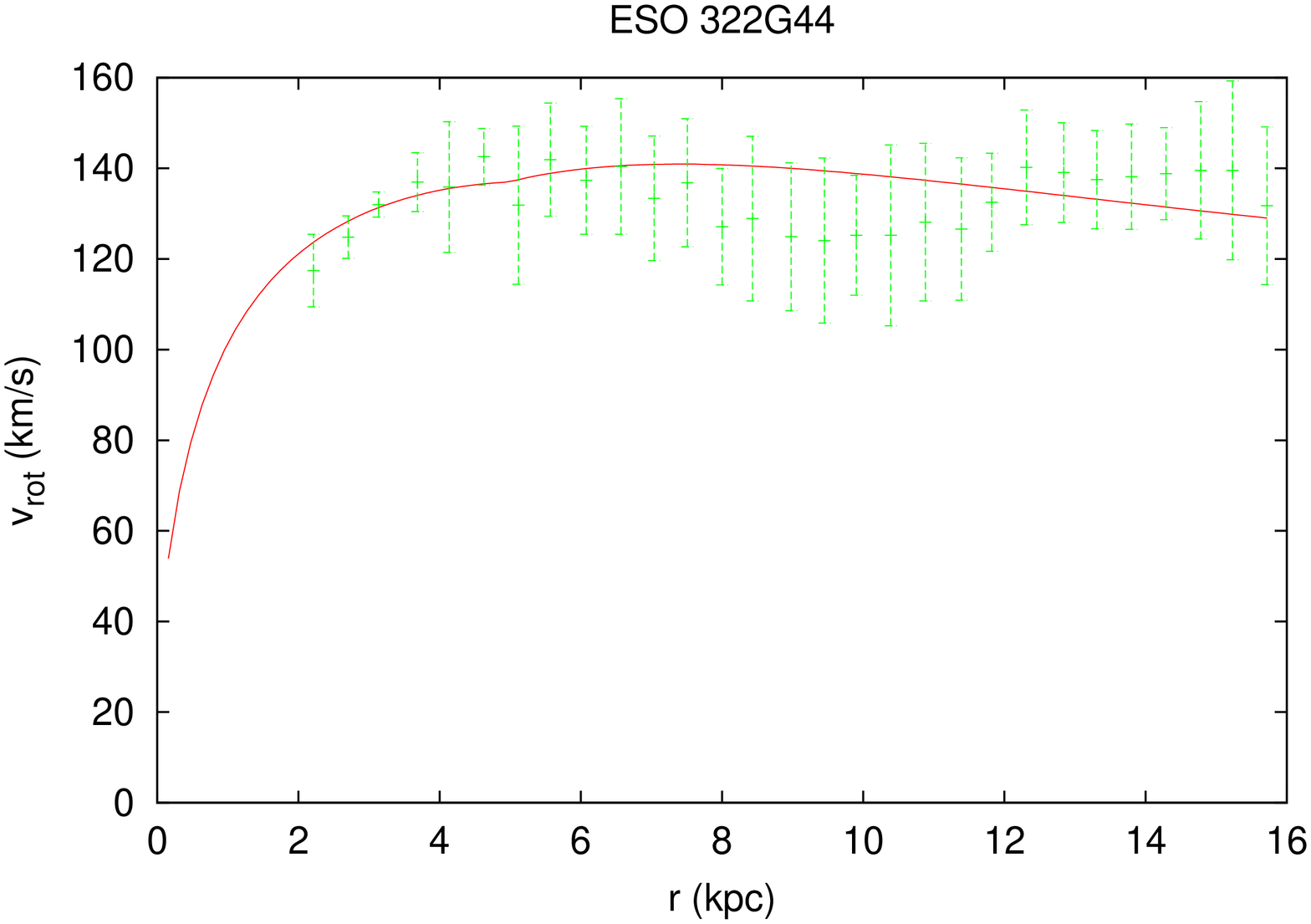} & 
\includegraphics[height=4cm, angle=270]{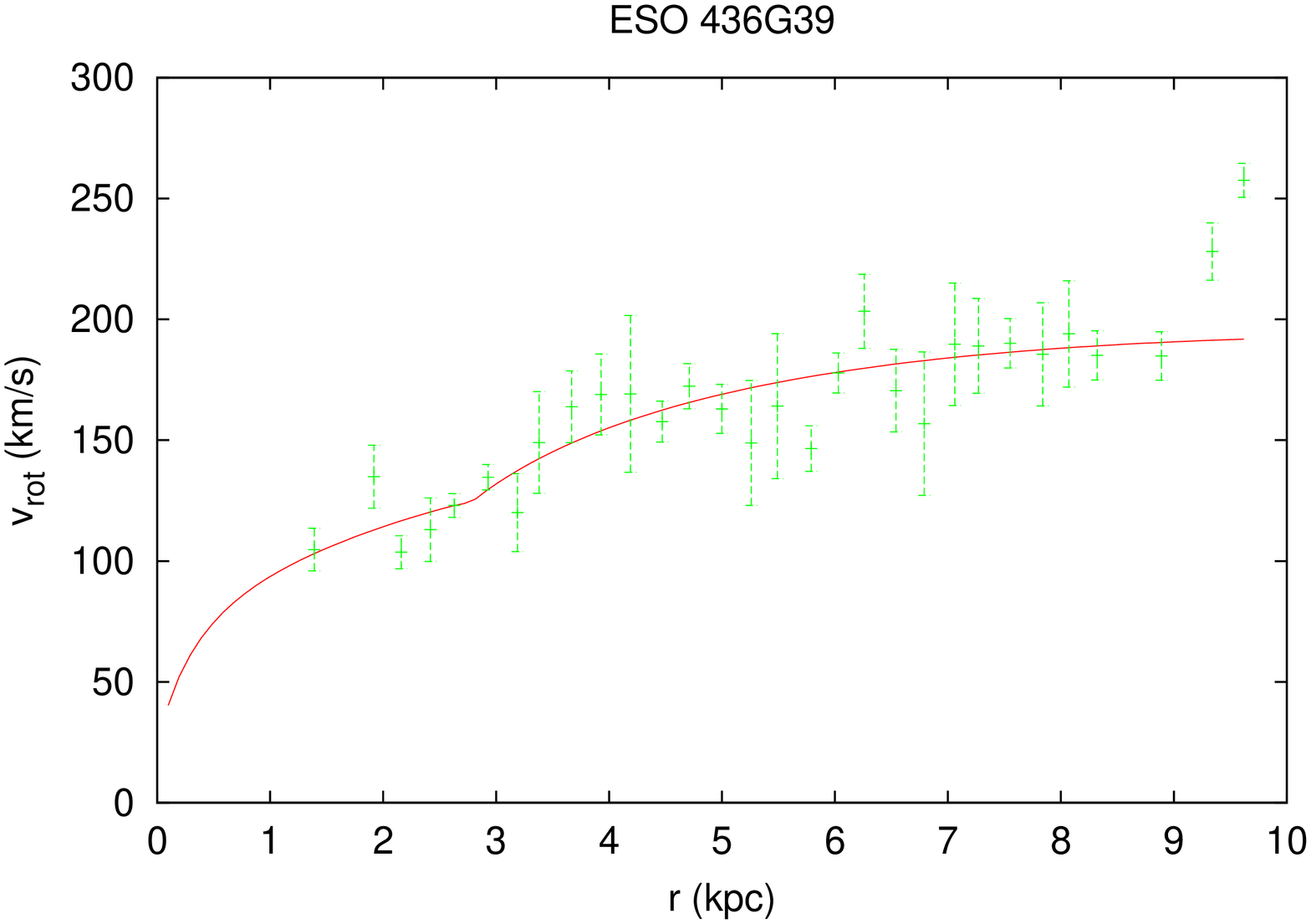} & 
\includegraphics[height=4cm, angle=270]{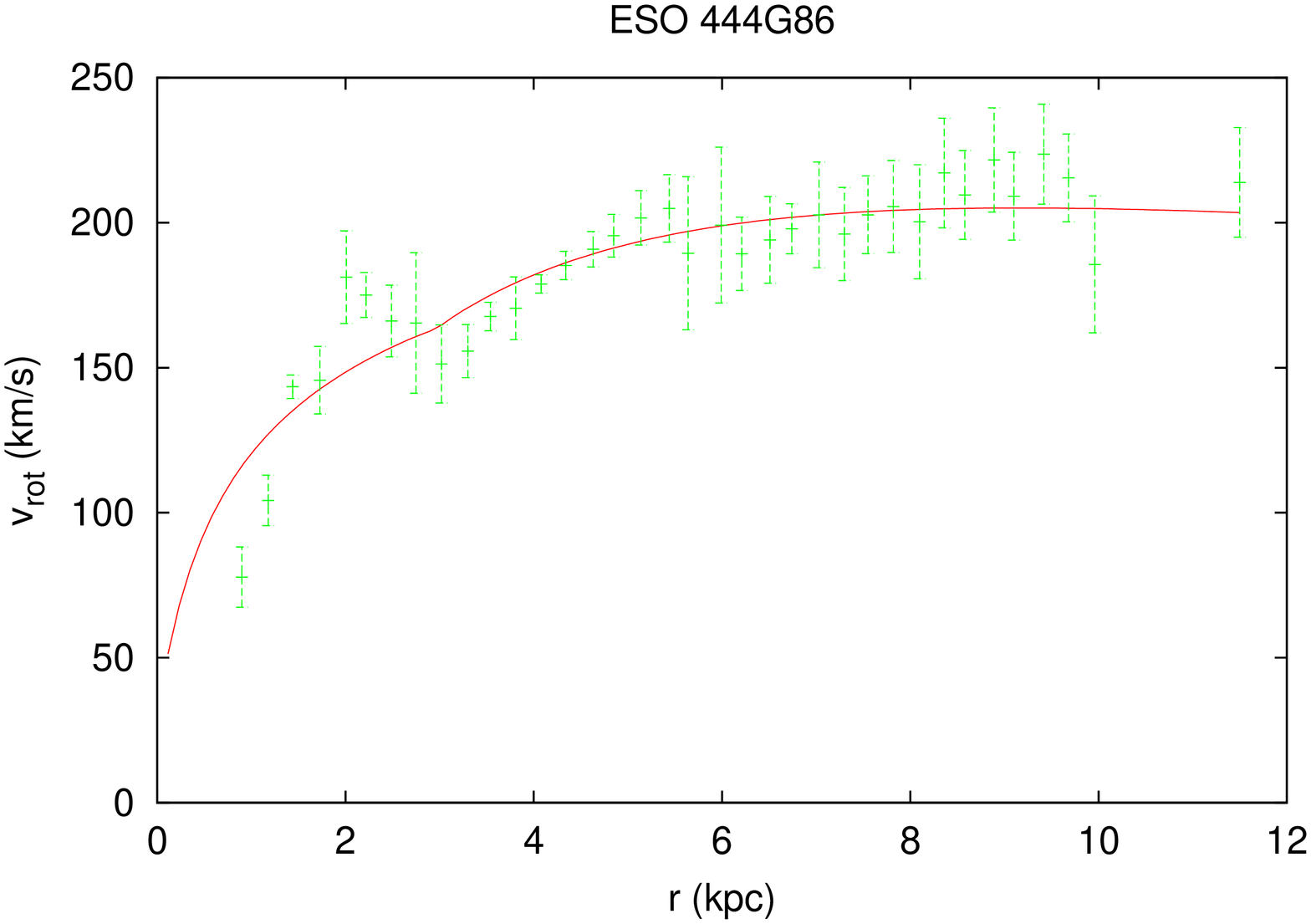}
\end{tabular}
\caption{Three examples of galaxies omitted due to the wavy pattern data
sequence. The best fits of our model are also shown, all outside the 1$
\protect\sigma$ confidence level.}
\label{fault}
\end{center}
\end{figure*}

\begin{figure*}
\begin{center}
\begin{tabular}{ccc}
\includegraphics[height=4cm, angle=270]{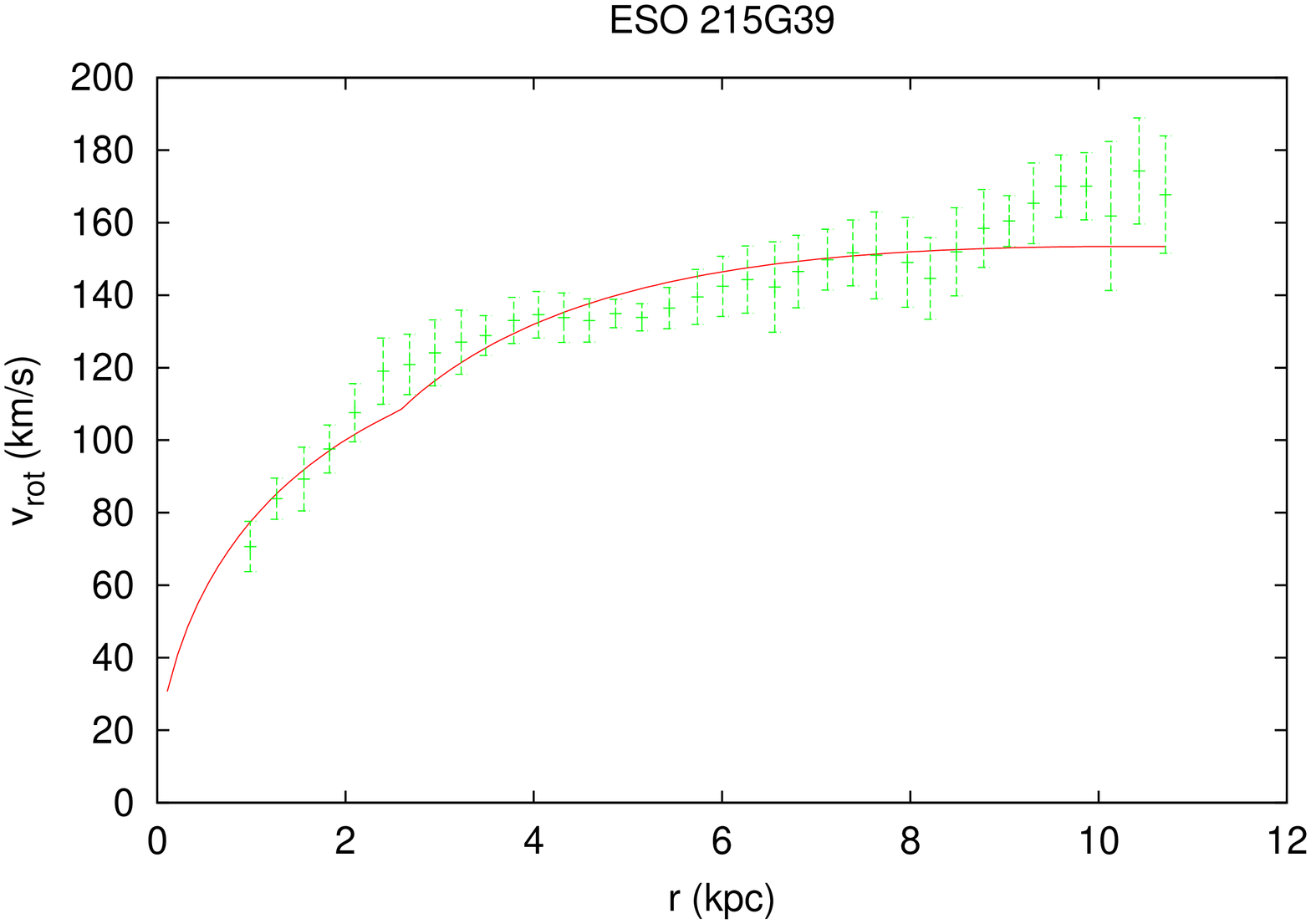} & 
\includegraphics[height=4cm, angle=270]{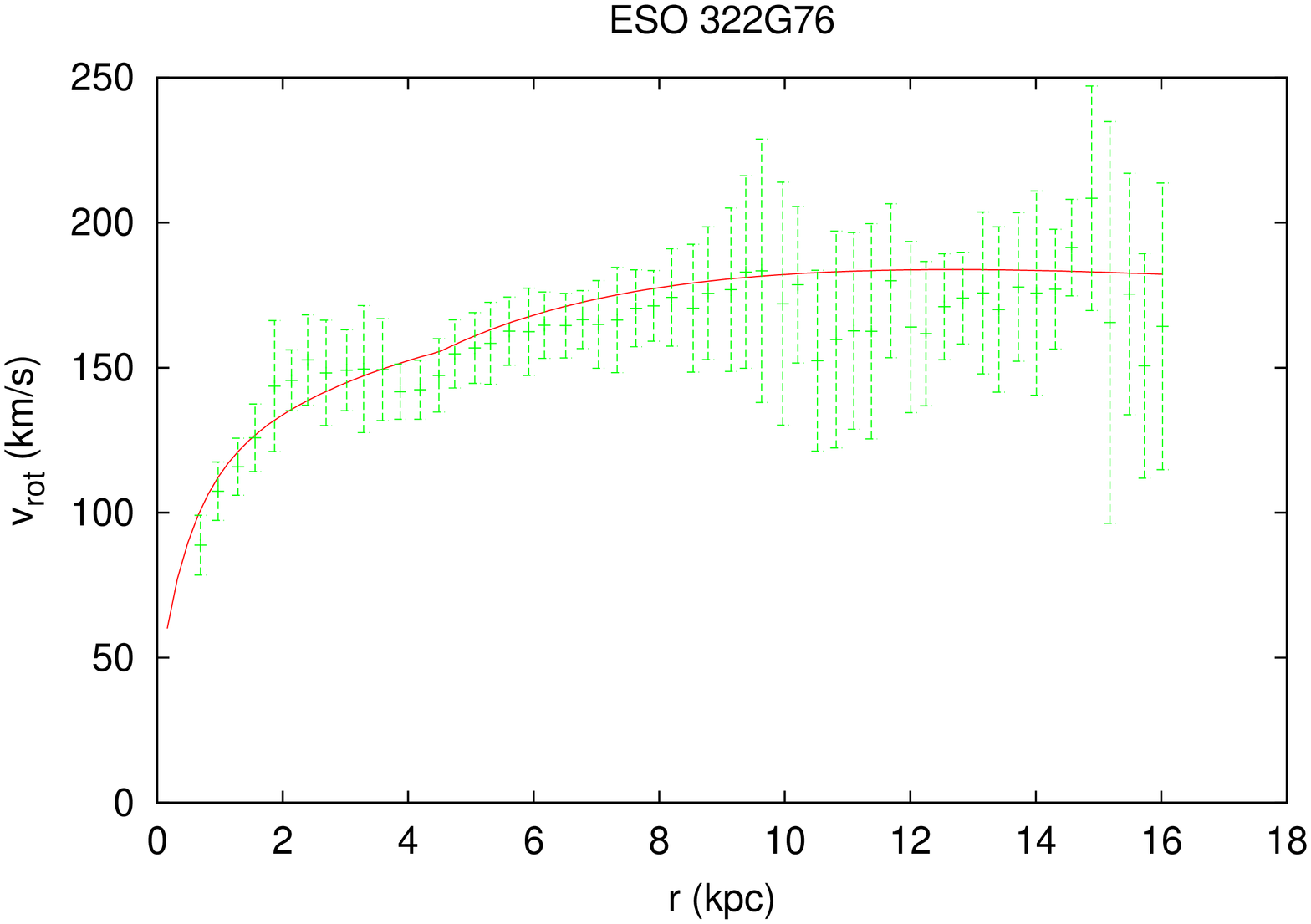} & 
\includegraphics[height=4cm, angle=270]{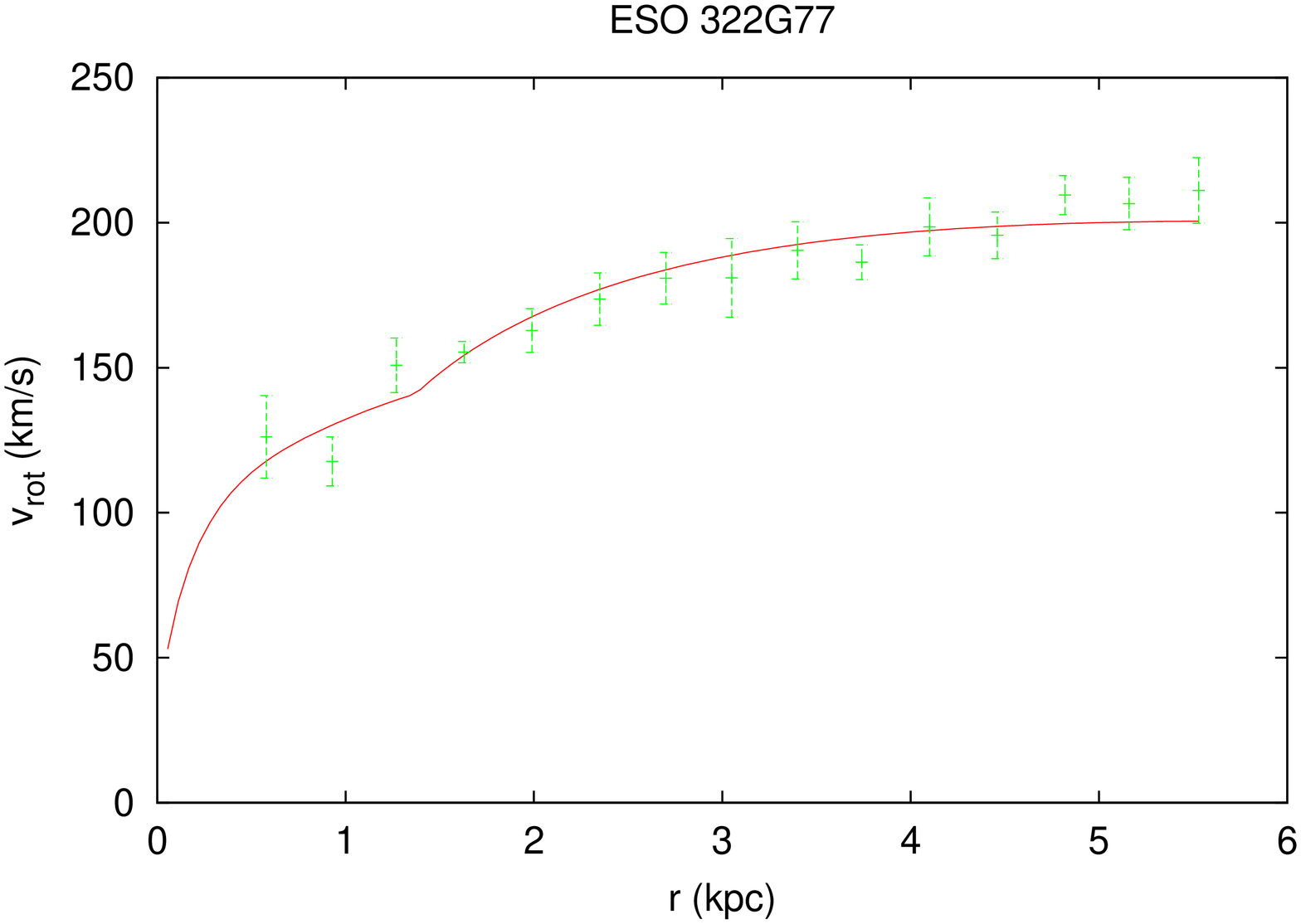} \\ 
\includegraphics[height=4cm, angle=270]{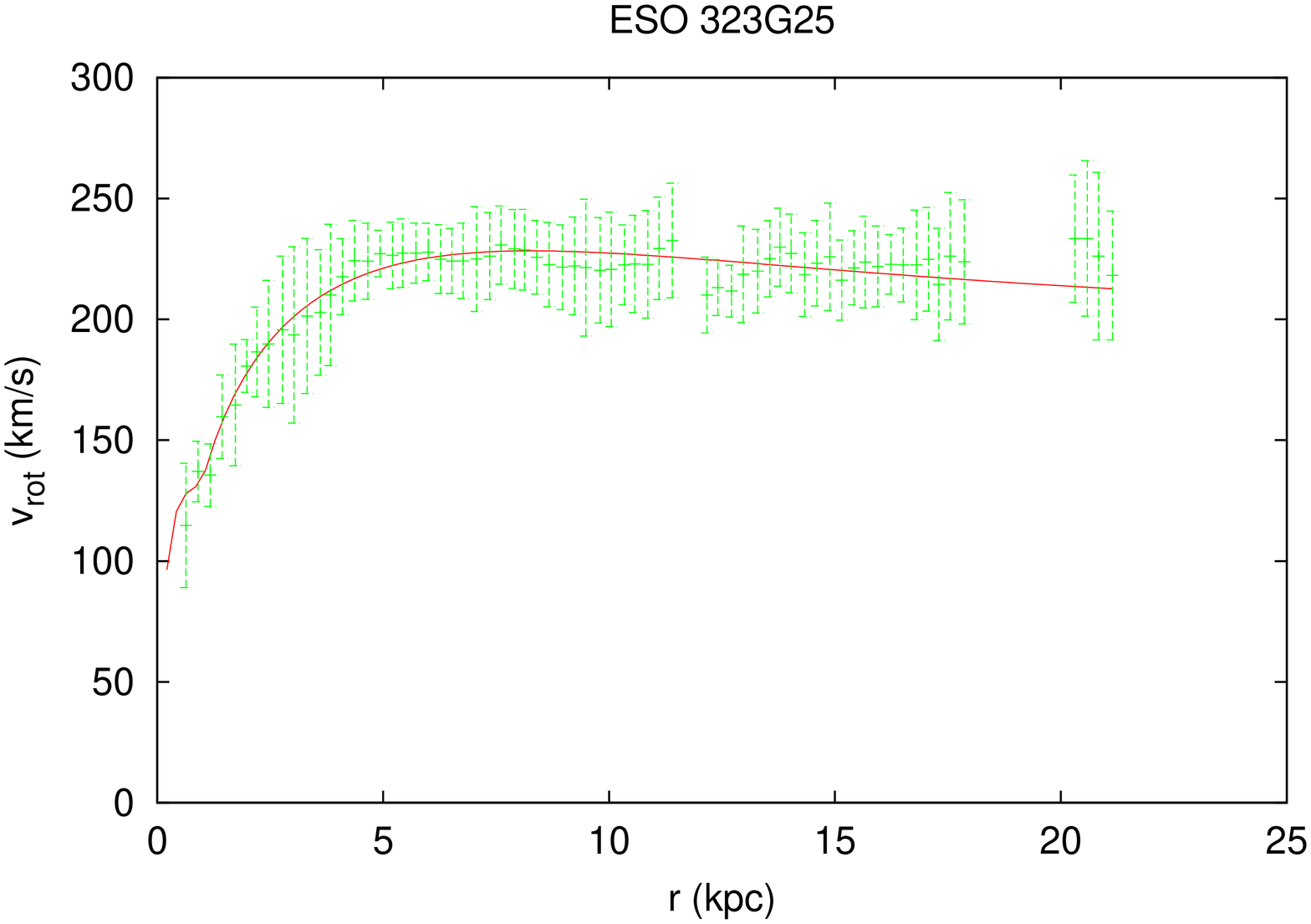} & 
\includegraphics[height=4cm, angle=270]{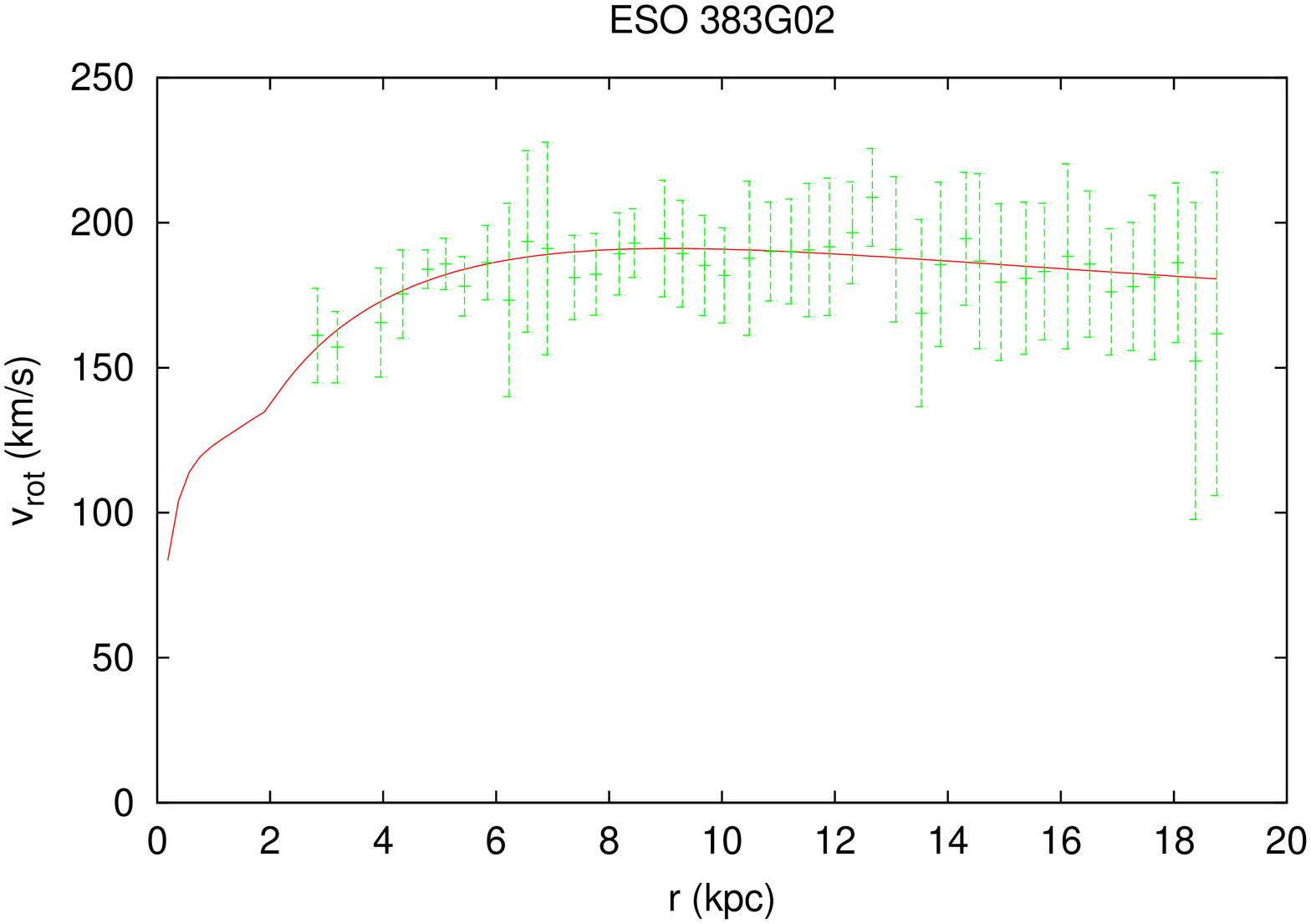} & 
\includegraphics[height=4cm, angle=270]{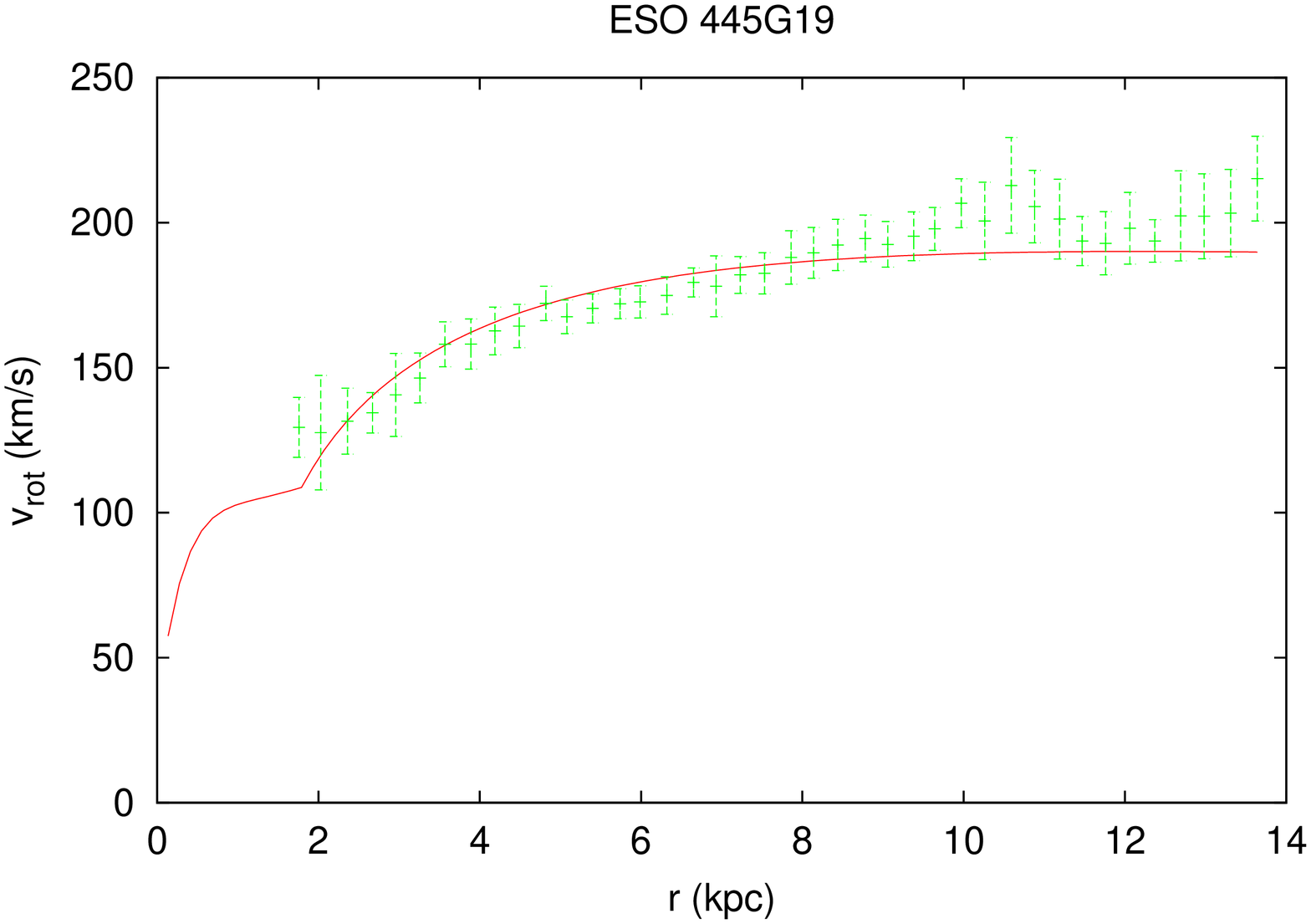} \\ 
\includegraphics[height=4cm, angle=270]{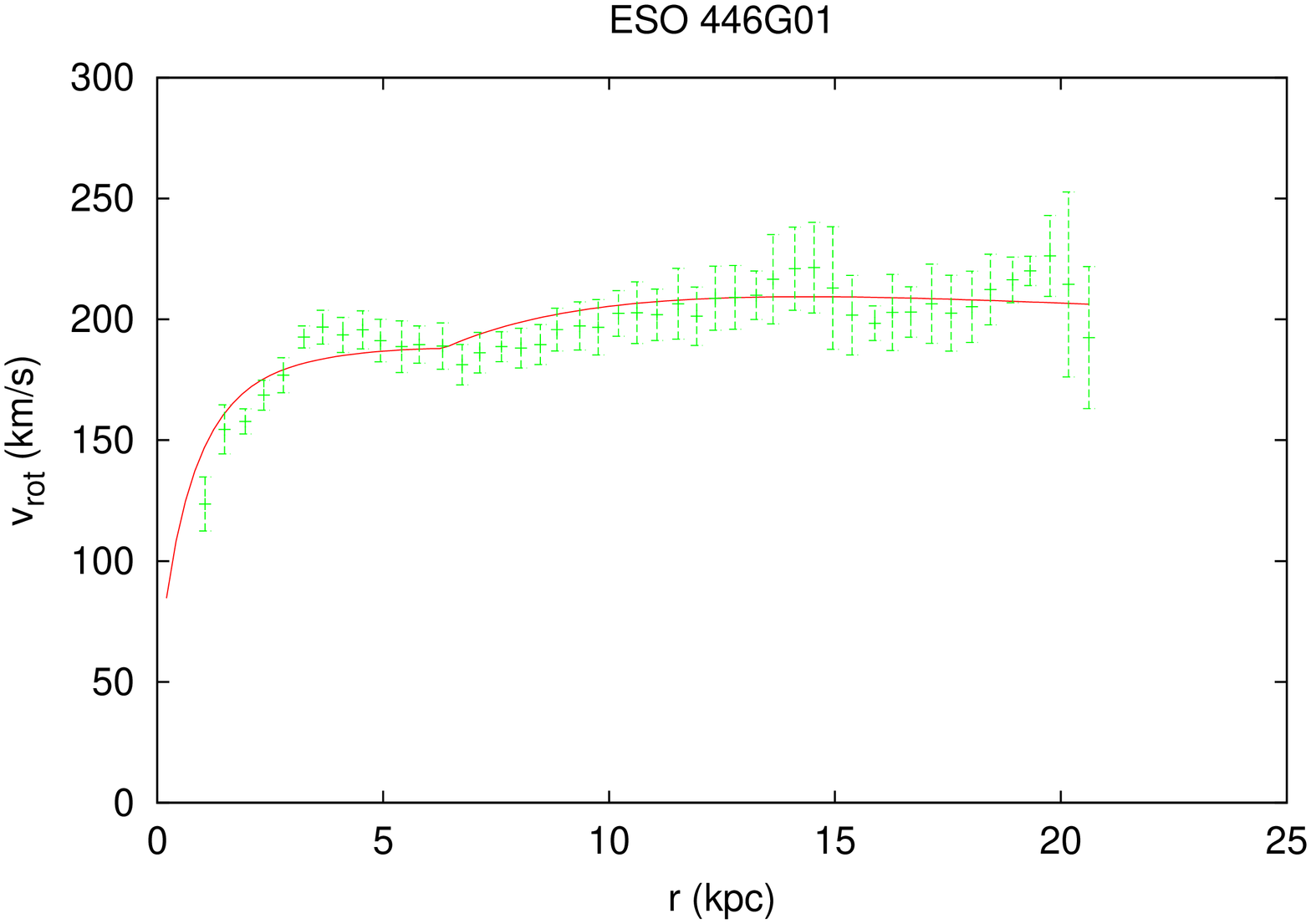} & 
\includegraphics[height=4cm, angle=270]{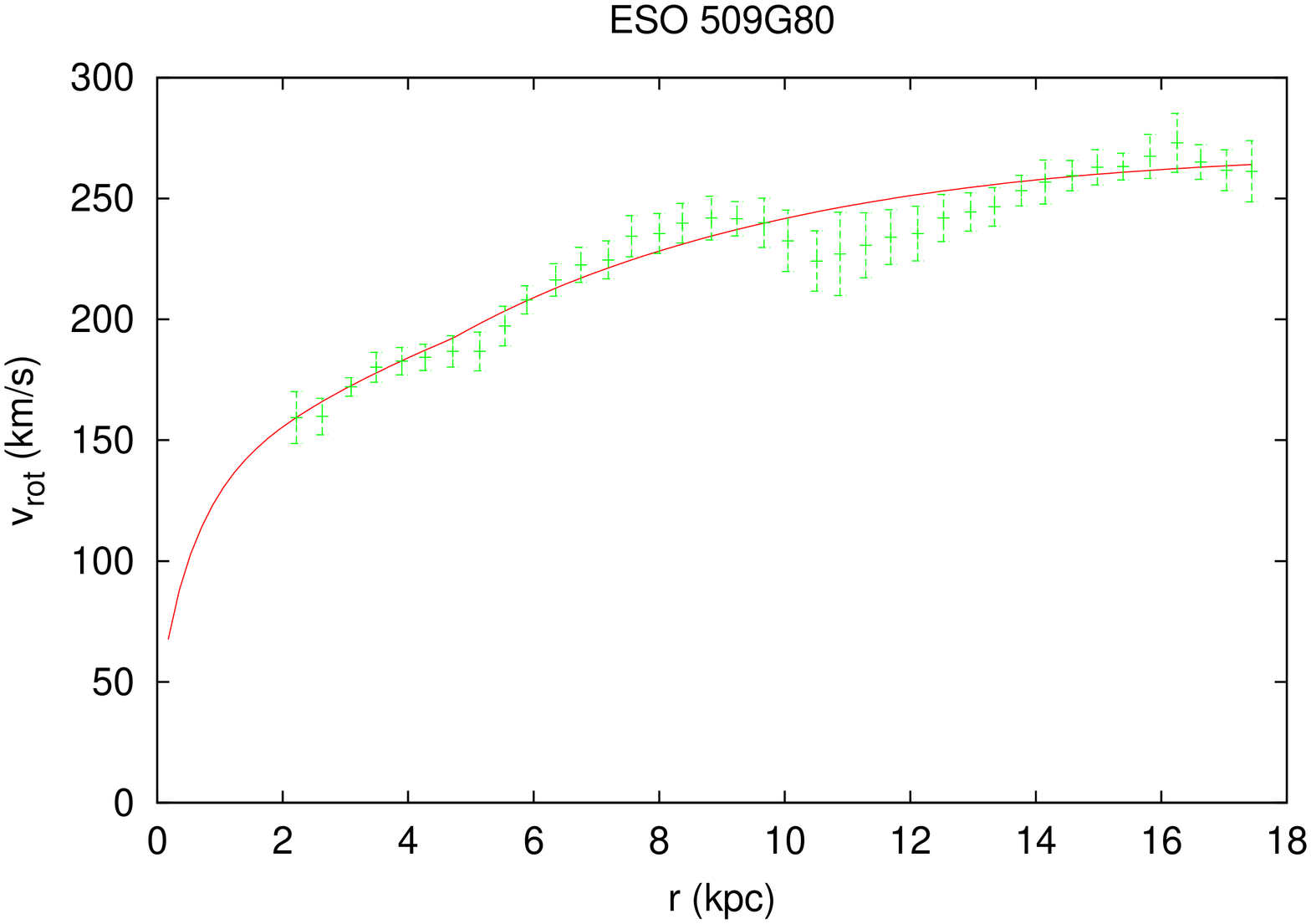} & 
\includegraphics[height=4cm, angle=270]{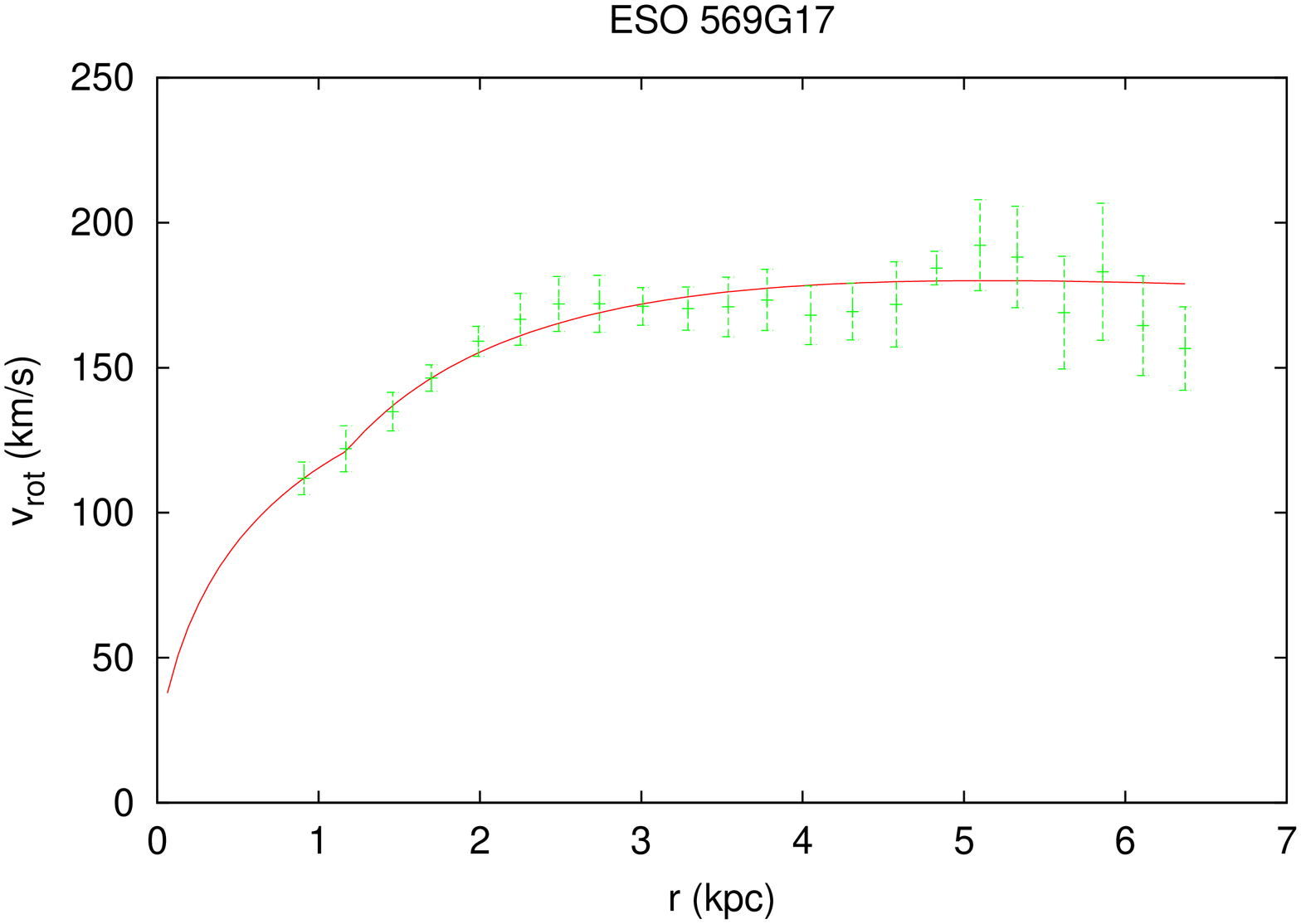} \\ 
&  & 
\end{tabular}
\caption{Best fit curves for the chosen HSB galaxies compatible with the
spherical symmetry assumption, the baryonic model and having sufficiently
accurate rotation curve and photometric data available. In the baryonic
matter dominated radial range $r<r_b$ the model curves bend as determined by
the photometric data, shown in Table~\ref{Table1}, together with the
Weyl parameters of the model. The discontinuity in the first derivative of
the model curves indicate the bulge radius $r_b$.}
\label{rotcurvHSB}
\end{center}
\end{figure*}

In \citet{Palunas} a maximum disk mass model was presented, and the
bulge-disk decompositions was performed for a sample of 74 high surface
brightness spiral galaxies, for which the I-band surface brightness profiles
and $H_{\alpha }$ velocity profiles were also given. From this set we have
left out those galaxies, which either have bars or rings, which would both
contradict the assumption of spherical symmetry. In addition we have left out the galaxies for 
which the above
baryonic model can not be applied, since they have non-vanishing bulge
surface brightness values, but vanishing bulge mass \citep{Palunas}. And
finally we have also omitted the galaxies where the observed data sequence
exhibits a wavy pattern instead of a plateau, suggesting a disk structure
strongly contradicting our assumption for rotational symmetry about the
galaxy center. Such patterns are illustrated in Fig.~\ref{fault}. As a
result of this selection process we end up with 9 galaxies. Their rotation
curves are represented on Fig.~\ref{rotcurvHSB}.

As for the bulge parameters of the 9 selected galaxies, we have derived the
best fitting values of $I_{0,b}$, $n$, $r_{0}$, $r_{b}$, $I_{0,d}$, and $h$
from the photometric data as well as the mass-to-light ratio of each
component $\sigma $ and $\tau _{b}$ by fitting Eq. (\ref{bar}) to the data
on rotation curves represented on Fig.~1 of \citet{Palunas}. These are all
collected in Table~\ref{Table1}.

\begin{table*}
\par
\begin{center}
\resizebox{16.5cm}{!} {
\begin{tabular}{|c|c|c|c|c|c|c|c|c|c|c|c|c|c|}
Galaxy & $D$ & $I_{0,b}$ & $n$ & $r_{0}$ & $r_{b}$ & $I_{0,d}$ & $h$ & $k$ &
$\sigma $ & $\tau _{b}$ & $\alpha $
& $\beta $ & $\chi _{\min }^{2}$ \\ \hline
 & Mpc & ${\rm mJy/arcsec} ^2$ & kpc & kpc  & kpc & $ {\rm mJy/arcsec} ^2$ & kpc & ${\rm kpc}^{-1}$ & $\odot $ &
$\odot $ &  &  &  \\ \hline\hline
ESO215G39 & 61.29 & 0.1171 & 0.6609 & 0.78 & 2.58 & 0.0339 & 4.11 & 26.28 &
0.04 & 2.1 & 0.67 & 2.37$\times $10$^{-7}$ & 28.29 \\
ESO322G76 & 64.28 & 0.2383 & 0.8344 & 0.91 & 4.50 & 0.0251 & 5.28 & 15.35 &
0.47 & 3.28 & 0.76 & 3.21$\times $10$^{-7}$ & 38.69 \\
ESO322G77 & 38.19 & 0.1949 & 0.7552 & 0.33 & 1.37 & 0.0744 & 2.20 & 50.32 &
1.27 & 2.7 & 0.57 & 3.92$\times $10$^{-7}$ & 10.15 \\
ESO323G25 & 59.76 & 0.1113 & 0.4626 & 0.43 & 0.99 & 0.0825 & 3.47 & 34.58 &
2.96 & 1.85 & 0.69 & 5.25$\times $10$^{-7}$ & 34.77 \\
ESO383G02 & 85.40 & 0.6479 & 0.7408 & 0.42 & 1.94 & 0.5118 & 3.82 & 17.79 &
0.47 & 0.22 & 0.70 & 3.72$\times $10$^{-7}$ & 21.55 \\
ESO445G19 & 66.05 & 0.1702 & 0.6133 & 0.57 & 1.79 & 0.0478 & 4.27 & 38.59 & 1
& 1.39 & 0.46 & 3.64$\times $10$^{-7}$ & 29.26 \\
ESO446G01 & 98.34 & 0.2093 & 0.8427 & 1.28 & 6.33 & 0.0357 & 5.25 & 10.90 &
0.93 & 2.82 & 0.54 & 4.64$\times $10$^{-7}$ & 43.35 \\
ESO509G80 & 92.86 & 0.2090 & 0.7621 & 1.10 & 4.69 & 0.0176 & 11.03 & 14.75 &
0.61 & 5.50 & 0.87 & 6.3$\times $10$^{-7}$ & 25.71 \\
ESO569G17 & 57.77 & 0.2452 & 0.4985 & 0.45 & 1.18 & 0.1348 & 2.06 & 58.74 &
0.06 & 1.4 & 0.65 & 3.18$\times $10$^{-7}$ & 7.24 \\ \hline
\end{tabular}
} 
\caption{The baryonic parameters ($D$, $I_{0,b}$, $n$, $r_{0}$, $r_{b}$, $
I_{0d}$, $h$, $k$) of the 9 HSB galaxy sample. The additional baryonic
parameters $\sigma $, $\tau _{b}$ and the Weyl parameter $
\protect\alpha $ are determined by $\protect\chi ^{2} $ fitting. The best
fit parameters ($\sigma =M_{b}/L_{b}$, $\tau _{b}=M_{d}/L_{d}$, $\alpha $) 
and the minimum value of the $\chi ^{2}$ statistic $\chi 
_{\min }^{2}$ are also given. All $\chi _{\min }^{2}$ are within 1$
\sigma $.}
\label{Table1}
\end{center}
\par
\end{table*}
\begin{table*}

\par
\begin{center}
\begin{tabular}{|c|c|c|c|c|c|c|}
Galaxy & $k$ & $M_{0}$ & $r_{c}$ & $\alpha $ & $\beta $ & $\chi _{\min }^{2}$
\\ \hline
& ${\rm kpc}^{-1}$ & $\odot $ & kpc &  &  &  \\ \hline\hline
DDO 189 & 57.5 & 4.05$\times $10$^{8}$ & 1.25 & 0.3 & 6.43$\times $10$^{-8}$
& 0.742 \\ 
NGC 2366 & 46.0 & 1.05$\times $10$^{9}$ & 1.47 & 0.8 & 1.12$\times $10$%
^{-7} $ & 2.538 \\ 
NGC 3274 & 138.1 & 4.38$\times $10$^{8}$ & 0.69 & -0.4 & 6.73$\times $10$%
^{-8}$ & 18.099 \\ 
NGC 4395 & 30.0 & 2.37$\times $10$^{8}$ & 0.71 & 0.9 & 3.43$\times $10$^{-7}$
& 27.98 \\ 
NGC 4455 & 99.7 & 2.26$\times $10$^{8}$ & 1.03 & 0.9 & 2.72$\times $10$^{-7}$
& 7.129 \\ 
NGC 5023 & 86.3 & 2.69$\times $10$^{8}$ & 0.74 & 0.9 & 4.53$\times $10$^{-7}$
& 10.614 \\ 
UGC 10310 & 36.4 & 1.28$\times $10$^{9}$ & 2.6 & 0.4 & 1.12$\times $10$%
^{-7} $ & 0.729 \\ 
UGC 1230 & 15.3 & 3.87$\times $10$^{9}$ & 3.22 & -1.7 & 1.12$\times $10$%
^{-7} $ & 0.539 \\ 
UGC 3137 & 34.5 & 5.32$\times $10$^{9}$ & 3.87 & -0.5 & 1.23$\times $10$%
^{-7} $ & 4.877 \\ \hline
\end{tabular}
\caption{The best fit parameters of the 9 LSB galaxy sample ($M_{0}$, $r_{c}$
, $\alpha $, $\beta $). The minimum values of the $
\chi ^{2}$ statistic $\chi _{\min }^{2}$ are also given. All $
\chi _{\min }^{2}$ are within 1$\sigma $.}
\label{Table2}
\end{center}
\par
\end{table*}

\subsection{Combined baryonic and Weyl model}

We assume that within the bulge radius $r_{b}$ the contribution of the
Weyl-fluid can be neglected and the part of the observed rotation curves
lying below $r_{b}$ could be explained with baryonic matter alone. Outside
this radius we switch on the Weyl-fluid and we take into account the
exponential disk as a perturbation that does not affect the geometry. We
add, however, its contribution to the rotation velocity, 
\begin{eqnarray}
v_{tg}^{2}(r)&=&\frac{G\left[ M_{b}(r)+M_{d}(r)\right] }{r}+\nonumber\\
&&c^{2}\left[ \beta 
+C\left(\frac{r_{b}}{r}\right) ^{1-\alpha }\right] H_{k}\left( 
r_{b}\right) ,  \label{v2tgnew1}
\end{eqnarray}
where 
$H_{k}\left( r_{b}\right) $ is some function smoothly approaching the
Heaviside step function: 
\begin{equation}
H\left( r_{b}\right) =\lim_{k\longrightarrow \infty }H_{k}\left(
r_{b}\right) =\left\{ 
\begin{array}{c}
0~,\quad r<r_{b} \\ 
1~,\quad r\geq r_{b}
\end{array}
\right..  \label{Heaviside}
\end{equation}
The parameter $M_{0}^{tot}$ of the Weyl fluid in Eq.~(\ref{v2tg}) represents
the contribution of the equivalent Weyl mass from the spheres within a given
radius. As such a contribution is taken care by the first terms of Eq.~(\ref
{v2tgnew1}), we have chosen $M_{0}^{tot}=0.$

For $H_{k}\left( r_{b}\right) $ we choose the logistic function:%
\begin{equation}
H_{k}\left( r_{b}\right) =\frac{1}{1+\exp \left( -2k\left( r-r_{b}\right)
\right) }~.  \label{H}
\end{equation}
The value of $k$ gives the sharpness of the transition. We determine the
parameter $k$ by imposing the conditions $H_{k}\left(0.95r_{b}\right)
=0.001 $, which also implies $H_{k}\left(1.05r_{b}\right) =0.999$ and $
H_{k}\left(r_{b}\right) =0.5$ and its values are given in Table \ref{Table1}
for each of the 9 selected galaxies \citep{Palunas}.

The baryonic parameters $\tau _{b}$, $\sigma $ and the Weyl parameter $
\alpha $ have to be determined by a $\chi ^{2}$-minimization fitting of Eqs.
(\ref{Ib}) - (\ref{H}), with the rotation curve data for each individual
galaxy. The parameter $\beta $ of Weyl sector gives the asymptotic rotation
velocity. We identify $\beta $ with the average of the points of the
rotation curves for $r>r_{b}$ (these are on the plateau).

At the end of this subsection we prove that the "truncation" of the
Weyl fluid at $r_{b}$ does not induce any distributional source layer at
\thinspace $r_{b}$, thus it is consistent with the junction conditions
across the sphere with radius \thinspace $r_{b}$. For this we first remark
that at a formal level the truncation can be imposed in all equations by the
replacements $\beta \rightarrow \beta H_{k}\left( r_{b}\right) $. By also
keeping in mind that \thinspace $C=C_{2}c^{-2}r_{b}^{\alpha -1}=-\beta $ was
chosen, Eq. (\ref{U}) shows that $U$~smooths to zero with $H_{k}\left(
r_{b}\right) $ across $r_{b}$. The equation of state (\ref{eqstate}) and $
B=\left( 3\beta /2\right) \left( 1/\alpha -1\right) $ then guarantees that $
P $ has the same property. As both inside and outside $r_{b}$ we have the
same spherically symmetric metric and the energy-momentum tensor changes
smoothly across $r_{b}$, there is no discontinuity in either the metric or
its first derivative, therefore the even stronger Lichnerowicz continuity
conditions are obeyed. The Israel-Lanczos-Darmois junction conditions of the
continuity of induced metric and extrinsic curvature thus follow. Our
approach is different from that used by \citet{Wise}, in which a given
density profile is considered, together with the condition of the isotropy
on the brane. Therefore, while in \citet{Wise} the matching conditions
impose these requirements as boundary conditions, in our approach the
matching conditions are automatically satisfied.

\subsection{HSB galaxy rotation curves}

Despite of the differences in the surface brightness profiles and rotation
curves for the chosen HSB galaxies, the combined baryonic+Weyl model Eq. (\ref{v2tgnew1})
fits well the sample.

A restriction on the parameter space emerges as $\beta $ $=-C$ from the
condition that the Weyl contribution to $v_{tg}^{2}\left(r\right) $
vanishes at $r_{b}$ (by switching on the Weyl fluid at $r=r_{b}$). Therefore
Eq.~(\ref{v2tgnew1}) simplifies to 
\begin{eqnarray}
v_{tg}^{2}(r)&=&\frac{G\left[ M_{b}(r)+M_{d}(r)\right]}{r}+\nonumber\\
&&c^{2}\beta \left[ 
1-\left( \frac{r_{b}}{r}\right) ^{1-\alpha }\right] H_{k}\left( r_{b}\right) 
.  \label{rot1}
\end{eqnarray}
The rotation curves themselves together with the data are shown on Fig.~\ref
{rotcurvHSB}.

\begin{table*}
\par
\begin{center}
\begin{tabular}{|c|c|c|}
Galaxy & $M_{Weyl}$ & $M_{simulation}$ \\ \hline
& $10^{12}M_{\odot }$ & $10^{12}M_{\odot }$ \\ \hline\hline
ESO215G39 & 1.27 & 1.89 \\ 
ESO322G76 & 2.43 & 8.53 \\ 
ESO322G77 & 2.19 & 1.487 \\ 
ESO323G25 & 3.75 & 3.76 \\ 
ESO383G02 & 2.35 & 3.12 \\ 
ESO445G19 & 2.33 & 1.97 \\ 
ESO446G01 & 4.69 & 9.7 \\ 
ESO509G80 & 10.7 & 205 \\ 
ESO569G17 & 1.55 & 1.17 \\ \hline
\end{tabular}
\caption{The calculated halo masses for the investigated sample of HSB galaxies.}
\label{Table5}
\end{center}
\par
\end{table*}

\section{Low Surface Brightness galaxies}

\subsection{LSB galaxy rotation curves}

A typical LSB galaxy resembles a normal late-type spiral, usually with some
ill-defined spiral arms. They usually have HI masses of a few times $10^{9}$ 
$M_{\odot }$. \citet{Thuan} and \citet{Bothun} have found that, although LSB
galaxies follow the spatial distribution of HSB galaxies, they tend to be
more isolated from their nearest neighbors than HSB galaxies. LSB galaxies
can be distinguished from the galaxies defining the Hubble sequence by their
low surface brightness, rather than small size (LSB galaxies are not
necessarily dwarfs). The central surface brightness of LSB galaxies is much
lower than $\mu _{B}(0)=21.65\pm 0.3$ mag$\cdot $arcsec$^{2} $ - the typical
B-band value for HSB galaxies, established by the Freeman law (\cite{Freeman}; \cite{hulst}).

The mass density distribution of LSB galaxies at small radii is dominated by
a nearly constant density core with a total mass $M_{0}$, and a radius $
r_{c} $ of only a few kpc, as established by \citet{de blok et al}. Hence we
ignore the mass contributions of the stellar and gas components. At low
radii $r<r_{c}$, the constant density implies $v^{2}(r)=G\rho V/r$, where $V$
is the volume of a sphere with radius $r$, and $\rho =3M_{0}/4r_{c}^{3}\pi $,
i.e., $v\sim ~r$. The Weyl fluid can reproduce this behavior if we choose
the equation of state $p(\mu)=0$ (i.e. $a=2$ and $B=0$ in Eq.~(\ref{eqstate}). Then the dark anisotropic stress/pressure also vanishes ($P=0$), and the
remaining part of the Weyl fluid energy momentum tensor (as $Q_{\mu }=0$)
then represents radiation. Using Eqs.~(\ref{m}) - (\ref{n}), (\ref{v0}) - (\ref{l12}) and (\ref{vtgin}) - (\ref{eta}), for $C_{2}=0$ the tangential
velocity, given by Eq.~(\ref{v2core}), yields to $v_{tg}^{2}~\approx \gamma
r^{2}$, the desired behavior. For $r>r_{c}$ we use Eq.~(\ref{v2tg}), without
baryonic matter ($M_{b}^{tot}=0$), and apply a new notation: $
M_{0}^{tot}=M_{0}$. Requiring the continuity of the rotation velocity
through $r=r_{c}$ we have 
\begin{eqnarray}
v_{tg}^{2}(r) &=&\frac{GM_{0}}{r}\left(\frac{r}{r_{c}}\right) ^{3}\left[
1-H_{k}(r_{c})\right]  \nonumber \\
&&+\left\{ \frac{GM_{0}}{r}+c^{2}\beta \left[1-\left( \frac{r_{c}}{r}
\right) ^{1-\alpha }\right] \right\} H_{k}\left(r_{c}\right) ,  \label{rot}
\end{eqnarray}
where $H_{k}\left( r_{c}\right) $ is the logistic function given by Eq.~(\ref
{H}).

\begin{figure*}
\begin{center}
\begin{tabular}{ccc}
\includegraphics[height=3.8cm, angle=270]{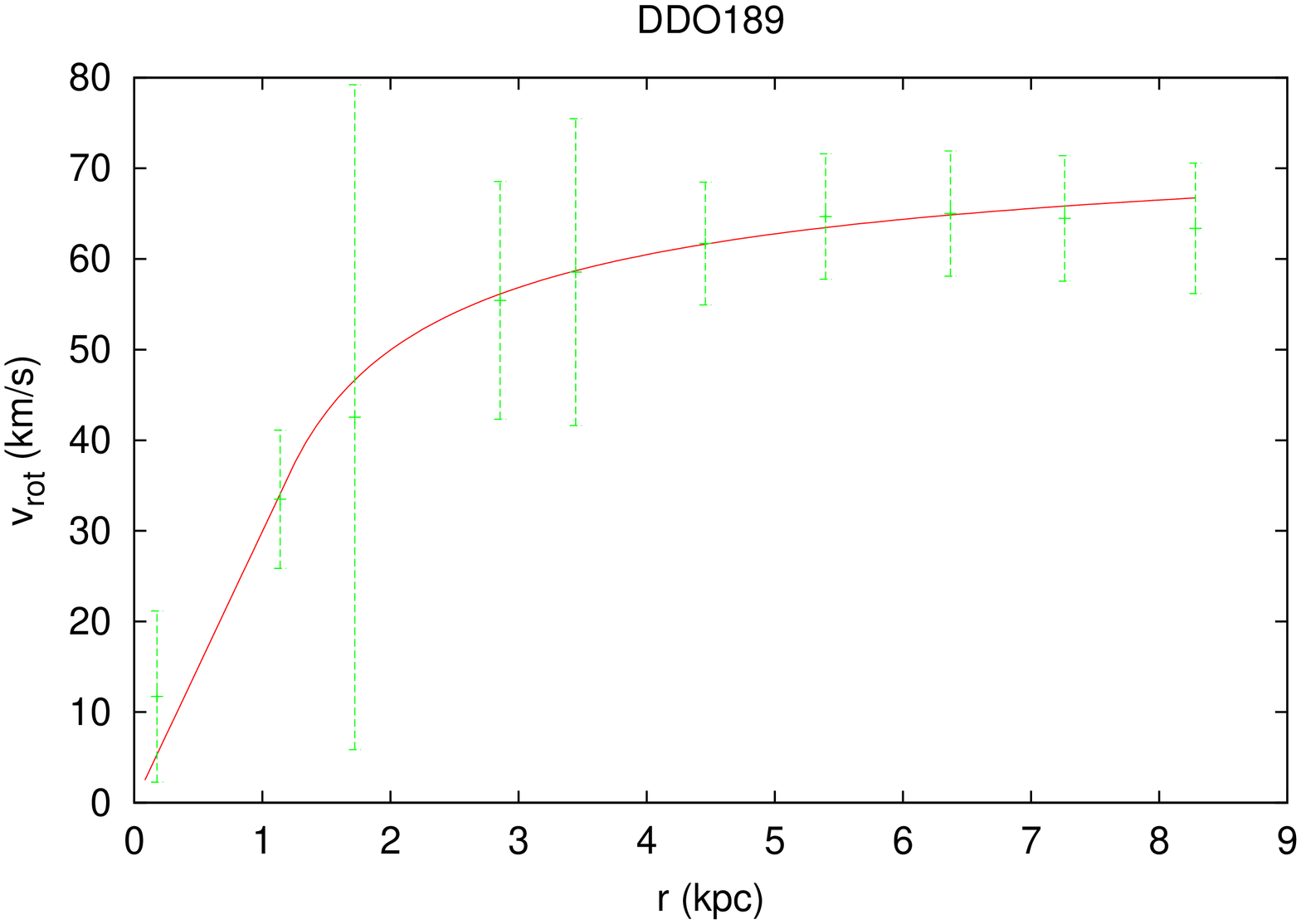} & 
\includegraphics[height=3.8cm, angle=270]{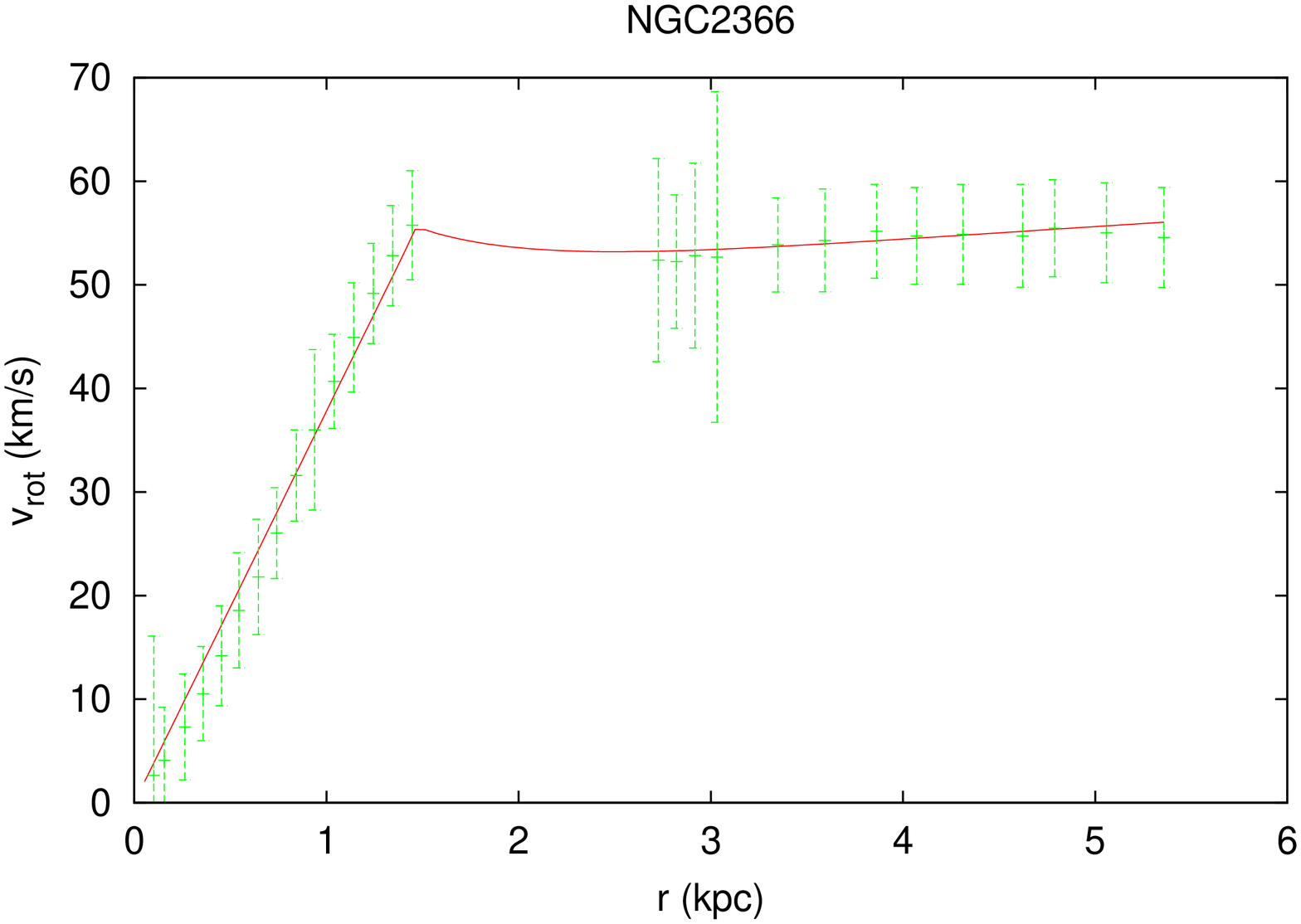} & 
\includegraphics[height=3.8cm, angle=270]{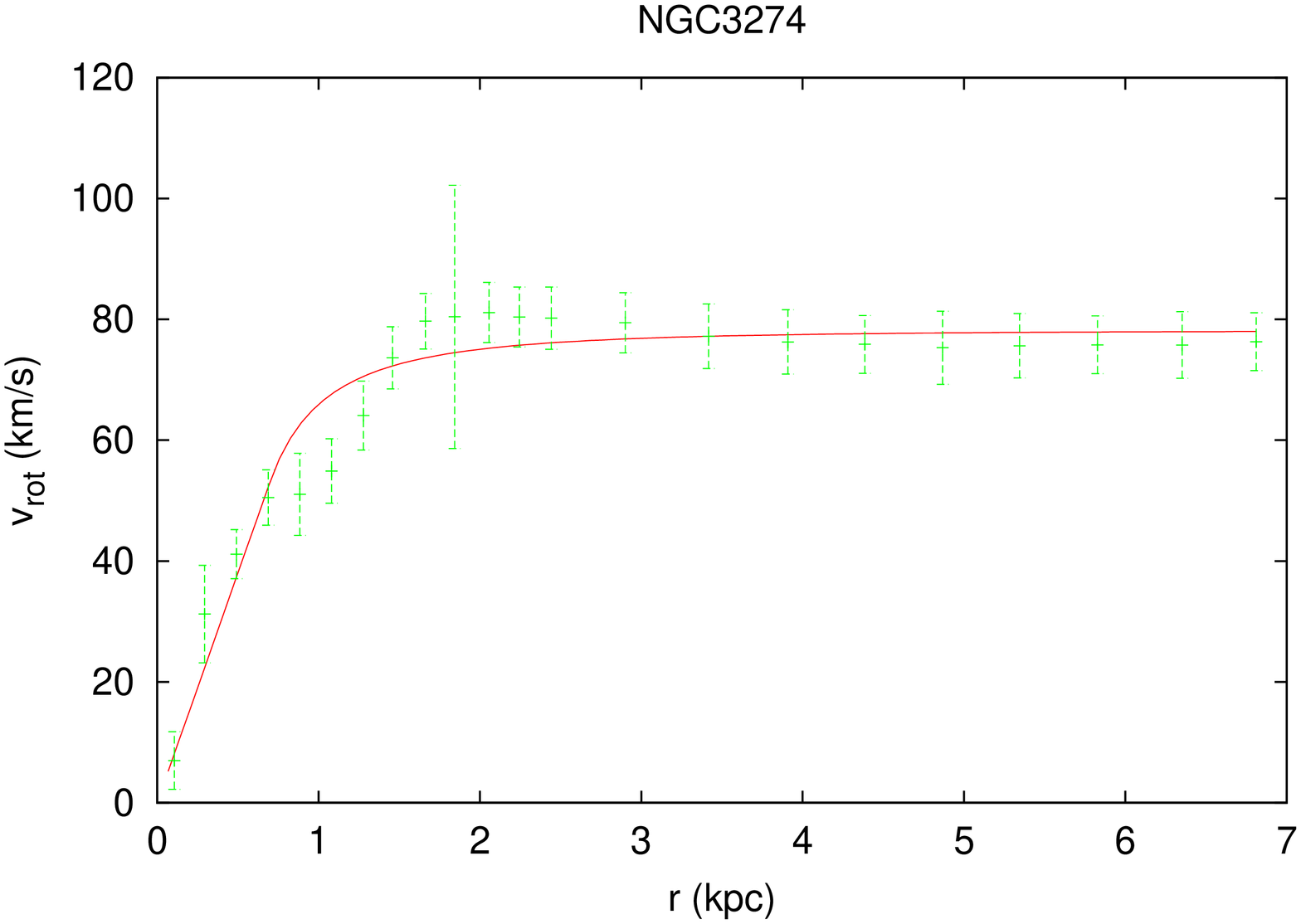} \\ 
\includegraphics[height=3.8cm, angle=270]{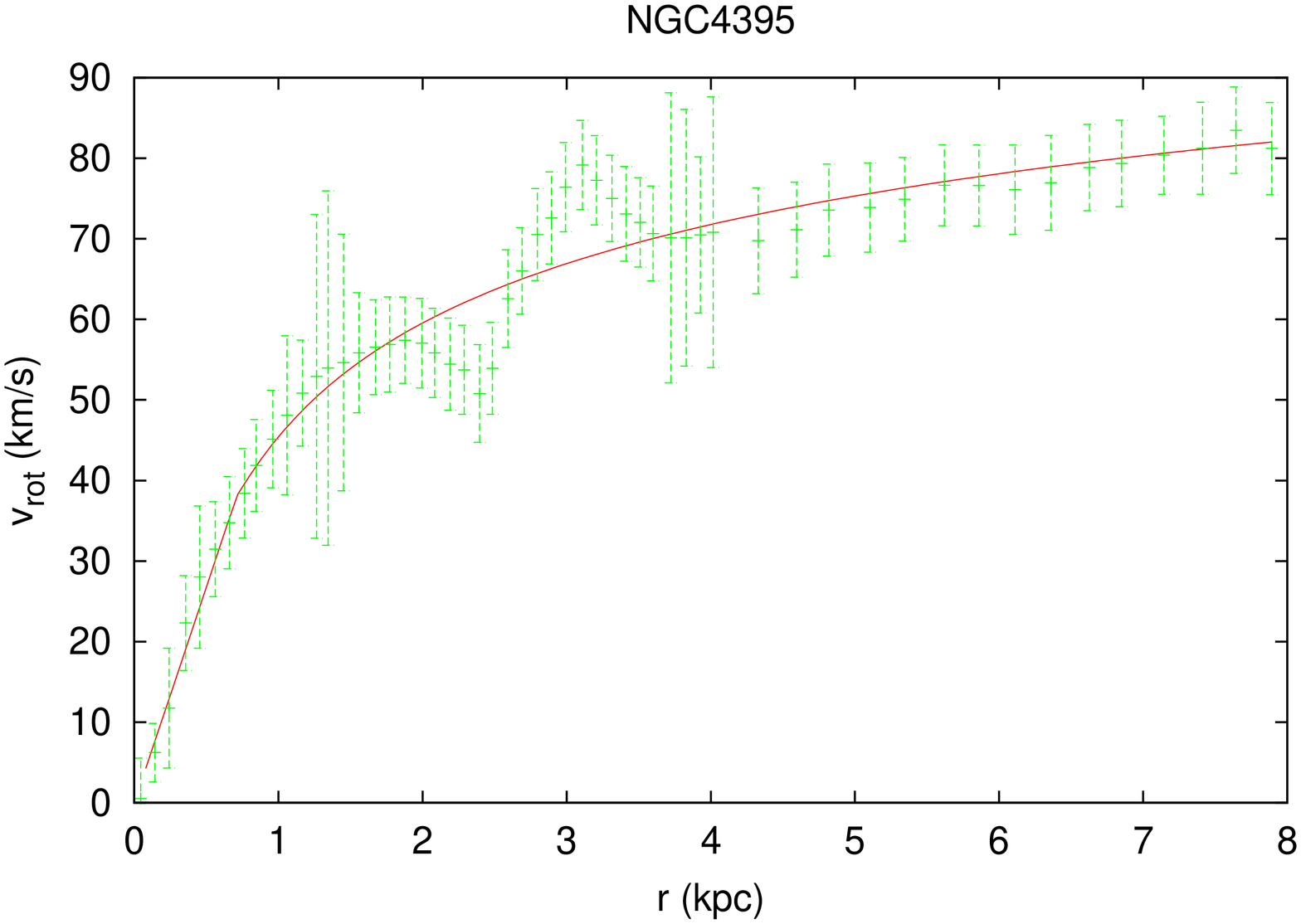} & 
\includegraphics[height=3.8cm, angle=270]{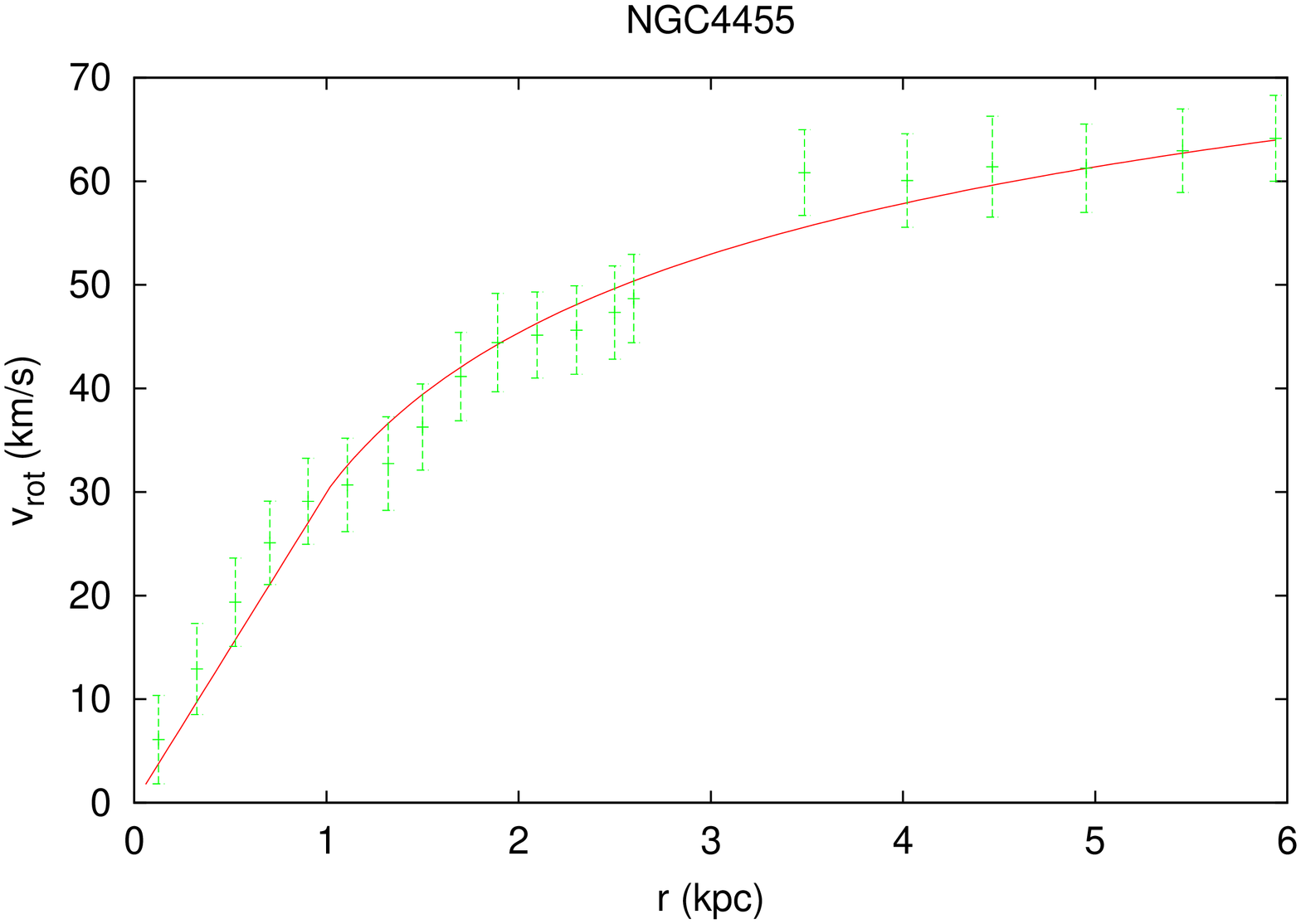} & 
\includegraphics[height=3.8cm, angle=270]{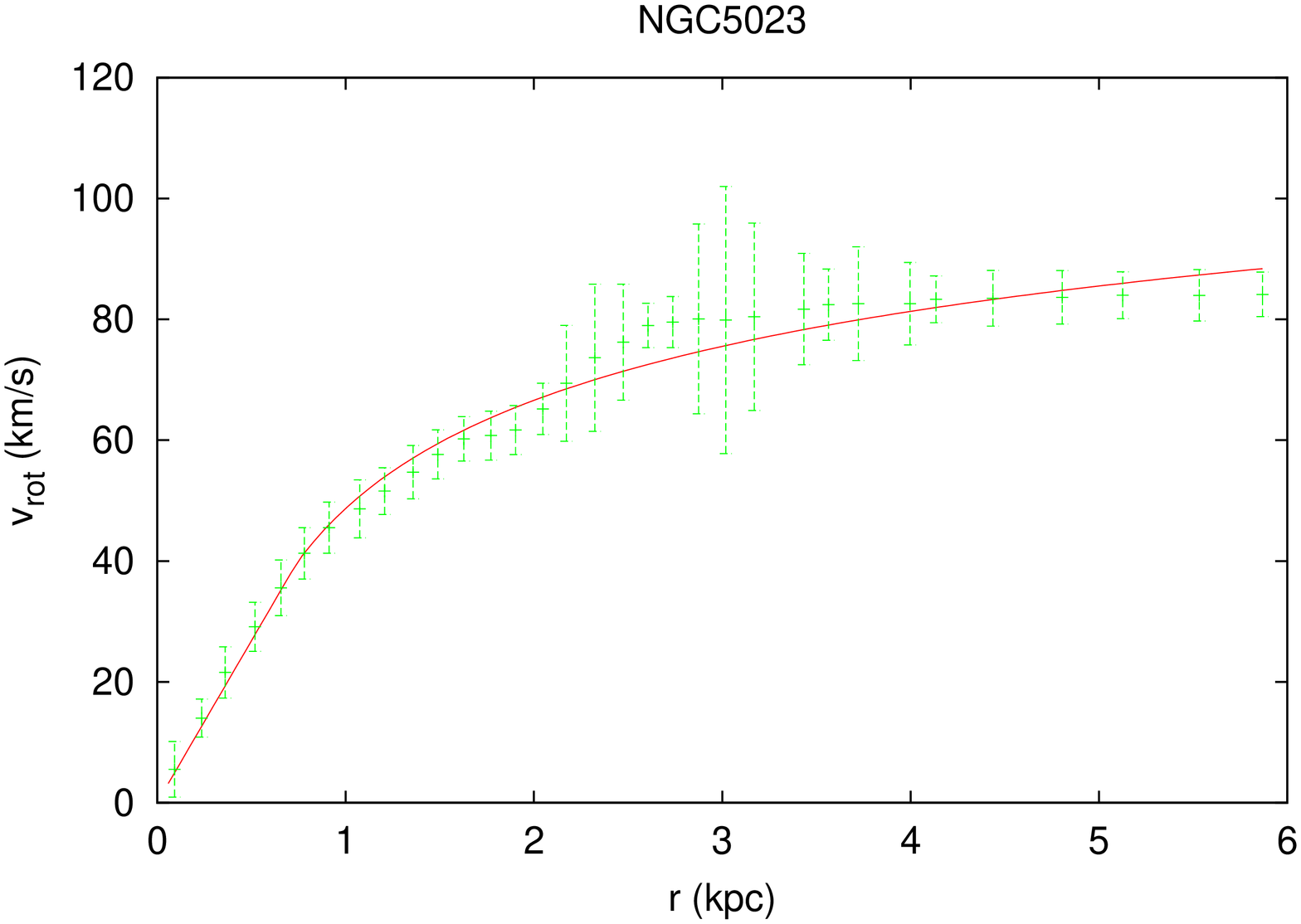} \\ 
\includegraphics[height=3.8cm, angle=270]{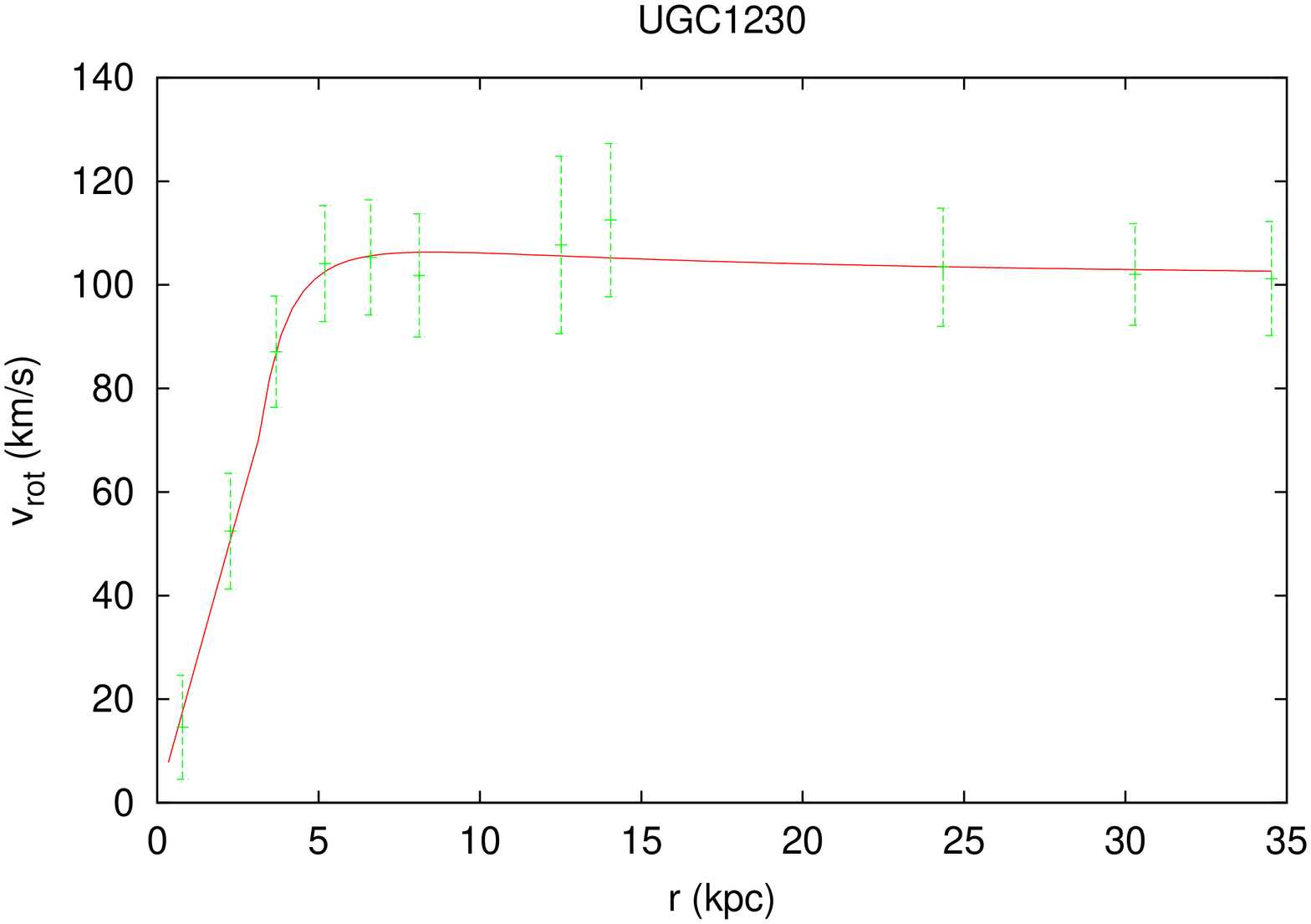} & 
\includegraphics[height=3.8cm, angle=270]{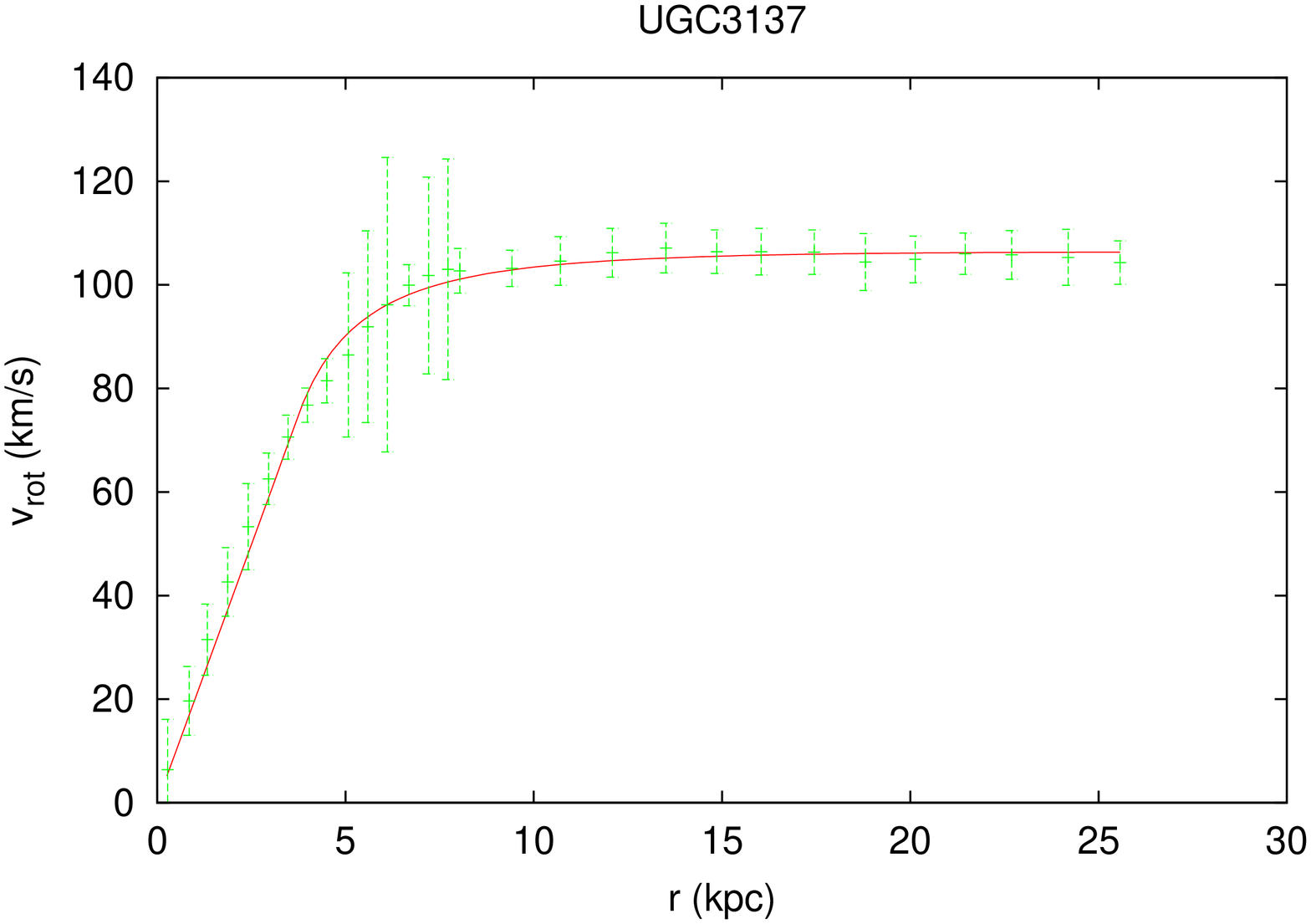} & 
\includegraphics[height=3.8cm, angle=270]{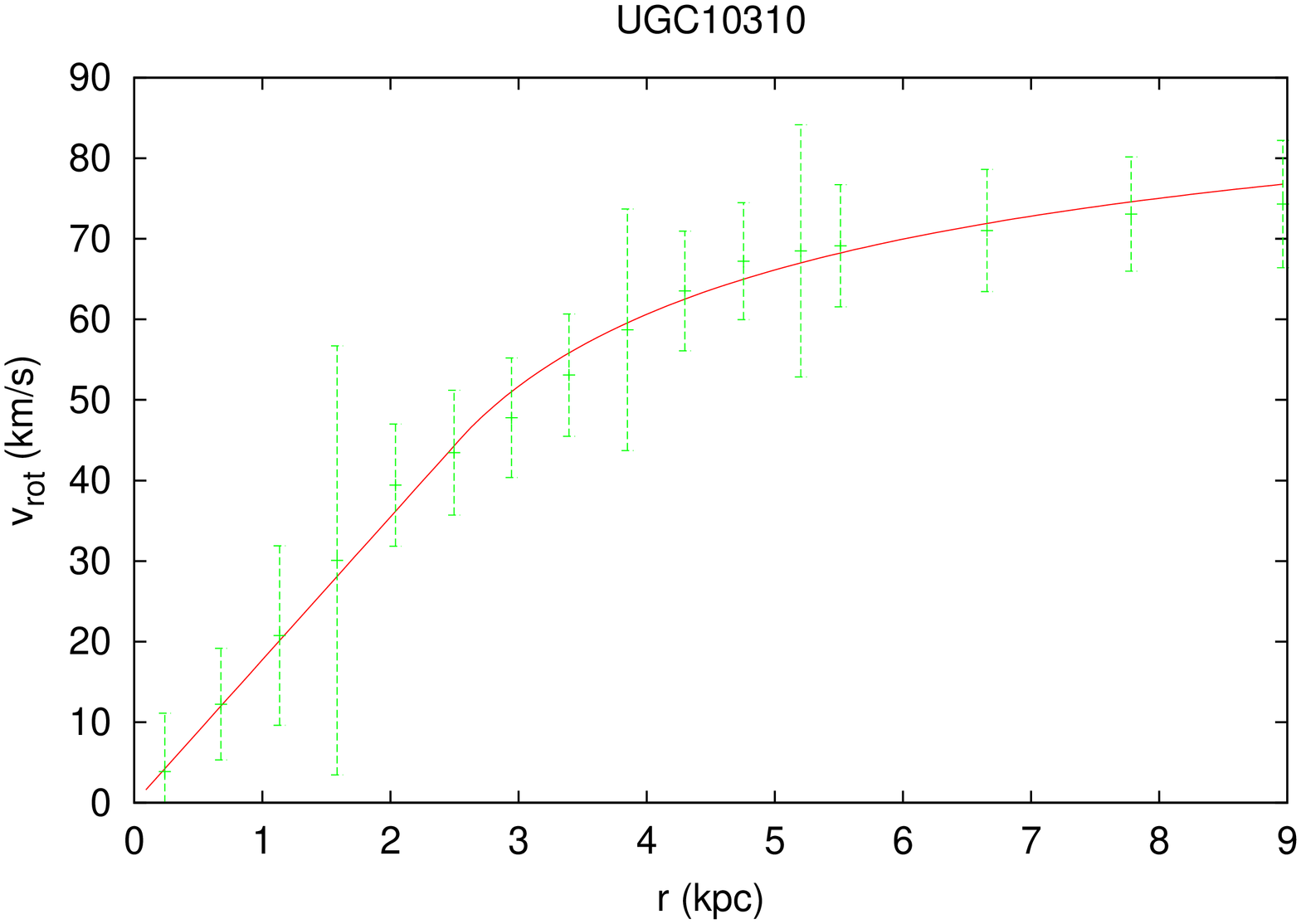} \\ 
&  & 
\end{tabular}
\caption{Best fit curves for the LSB galaxy sample. The parameters of the
model velocity curves are given in Table~\ref{Table2}.}
\label{rotcurvLSB}
\end{center}
\end{figure*}

We test the model with a sample of $9$ LSB galaxies, extracted from a larger
sample from \citet{deblok2002} as typical galaxies exhibiting the plateau
region. We fit Eq.~(\ref{rot}) with the rotation curve data taken from
combined $HI$ and $H\alpha $ measurements. The fitted curves are represented
on Fig.~\ref{rotcurvLSB}. In all cases we find remarkably good agreement
between the model and observations. Outside the core radius the parameters
of the equation of state $a$ and $B$ are determined by fitting through $
\alpha $ and $\beta $ (see Section 5).

\begin{figure*}[tbp]
\begin{center}
\begin{tabular}{ccc}
\includegraphics[height=6cm, angle=270]{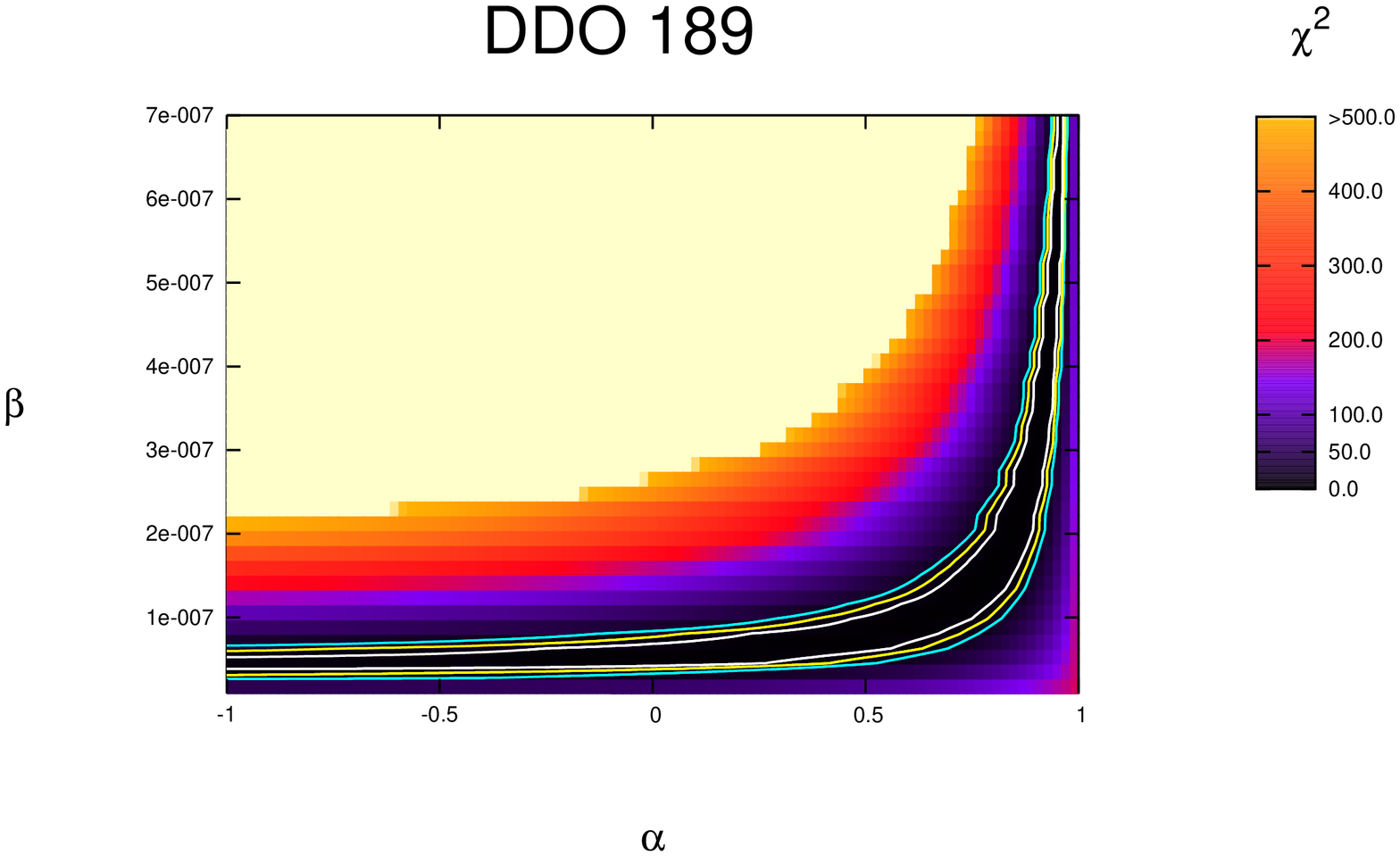} & %
\includegraphics[height=6cm, angle=270]{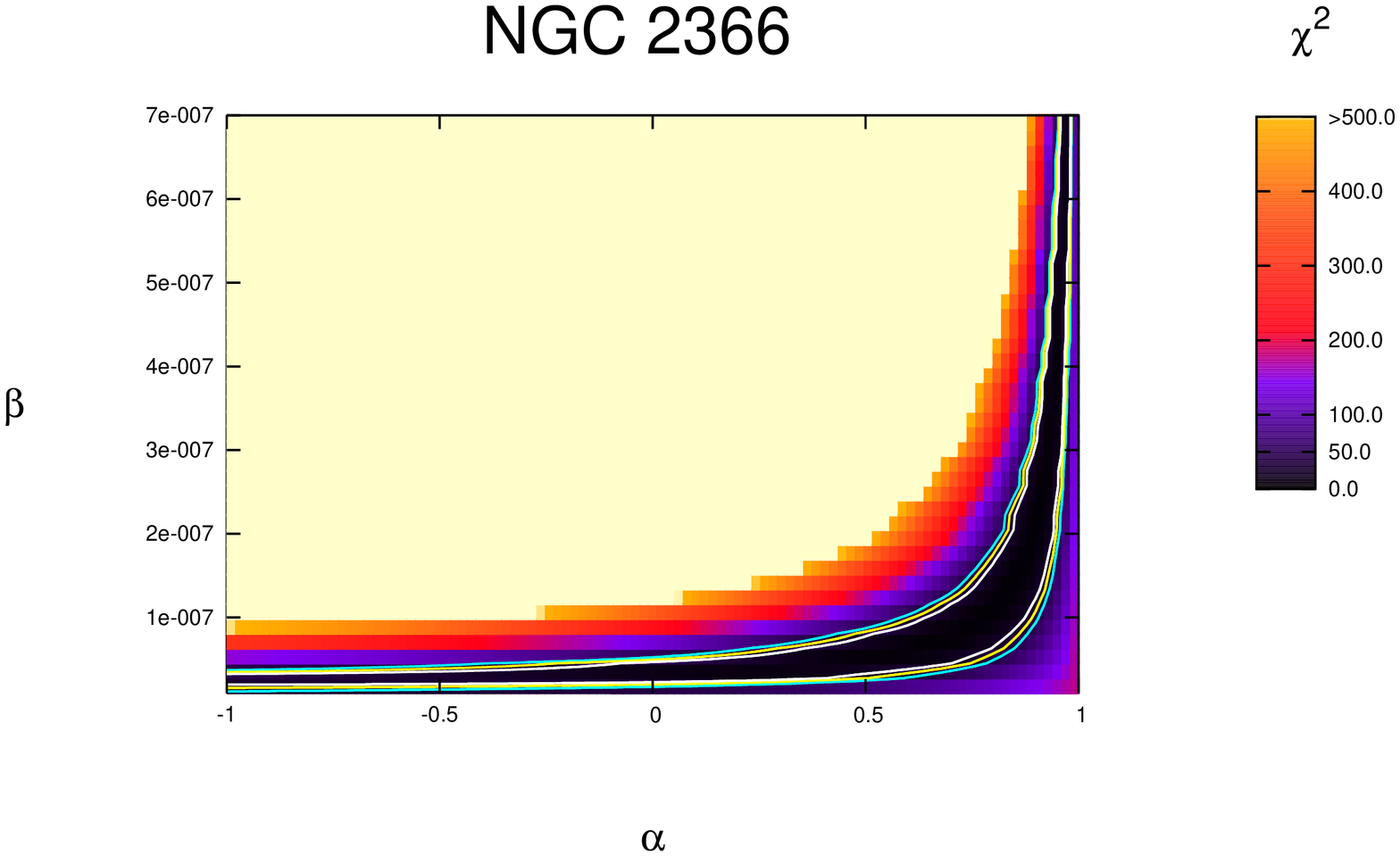} & %
\includegraphics[height=6cm, angle=270]{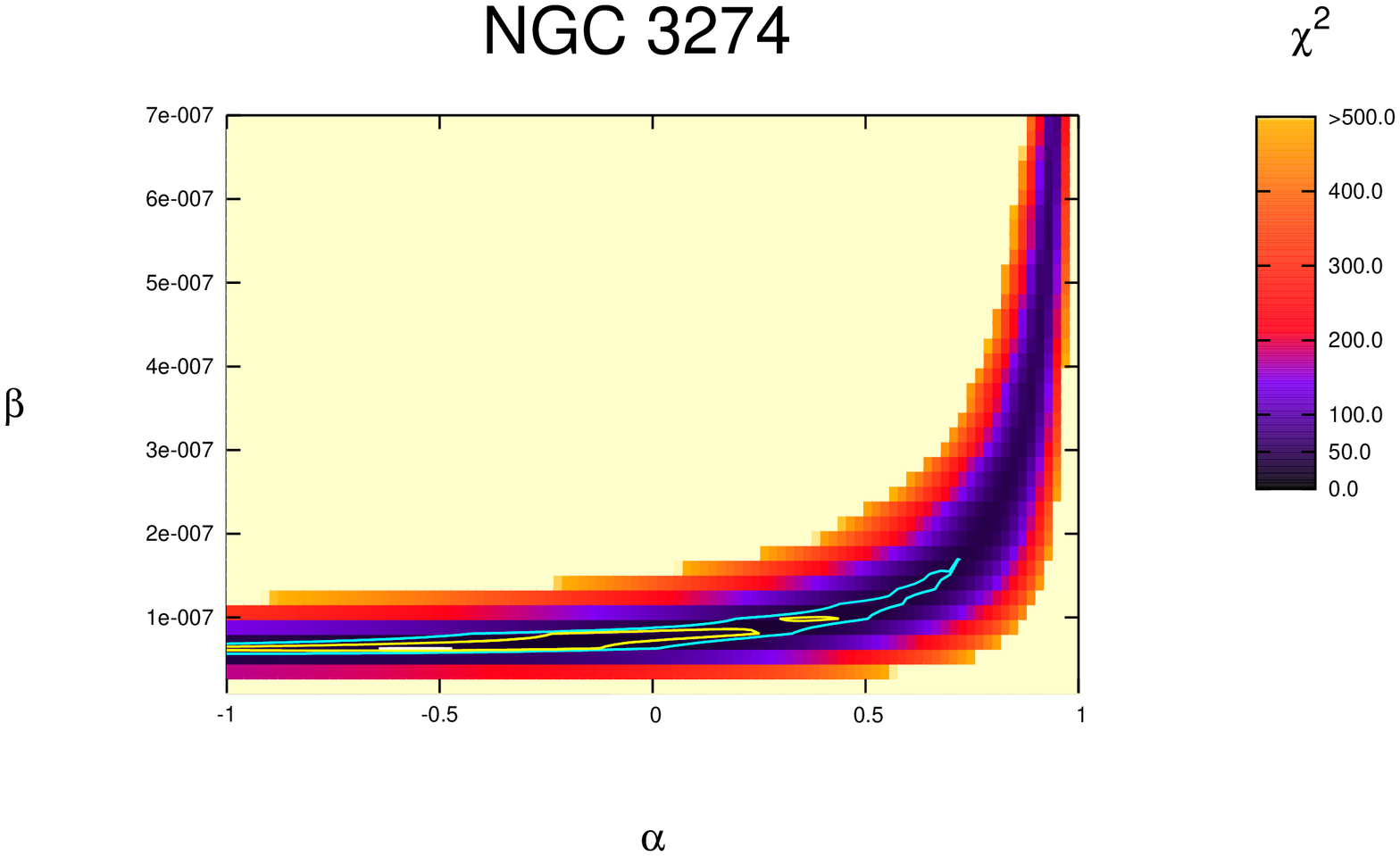} \\ 
\includegraphics[height=6cm, angle=270]{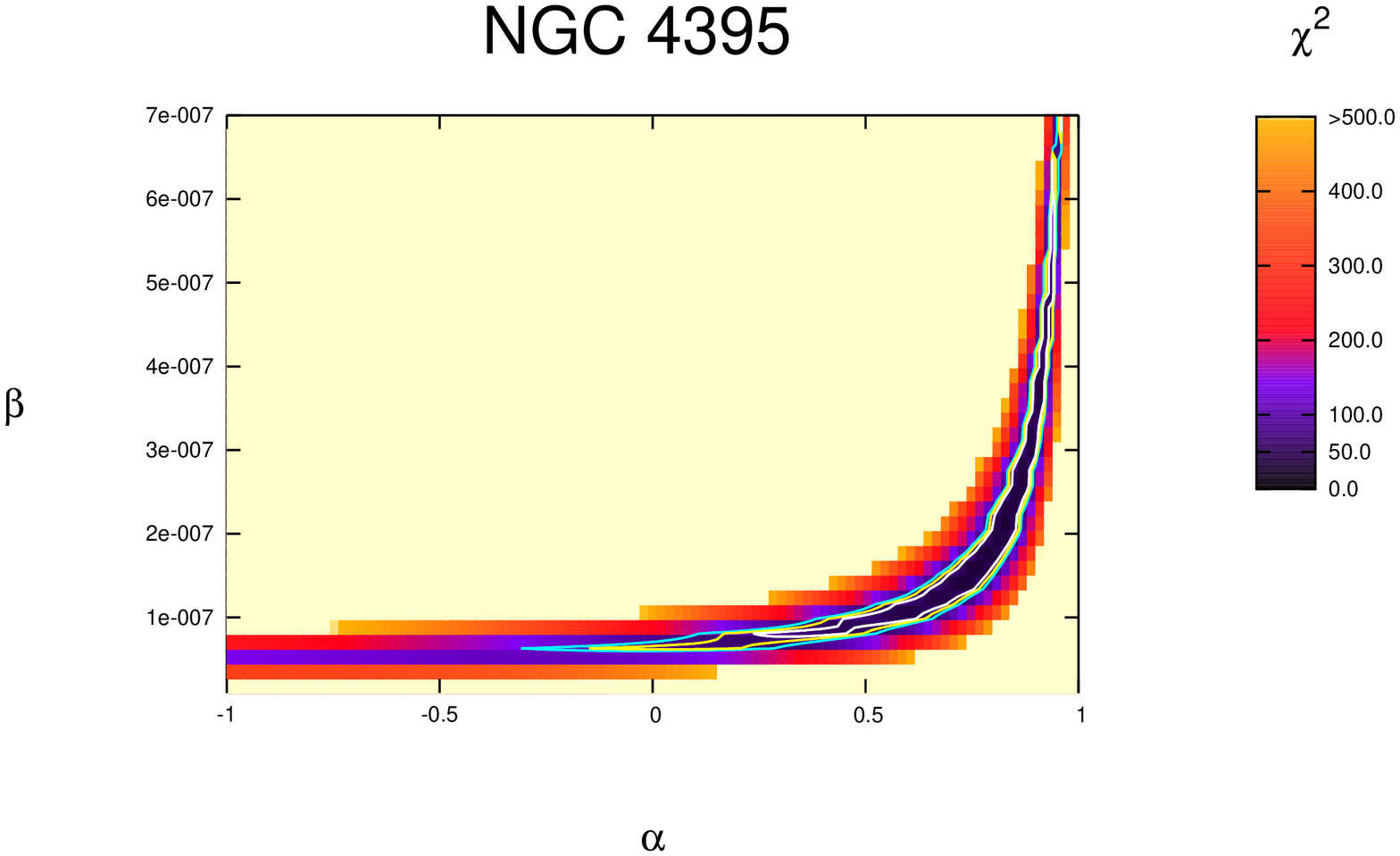} & %
\includegraphics[height=6cm, angle=270]{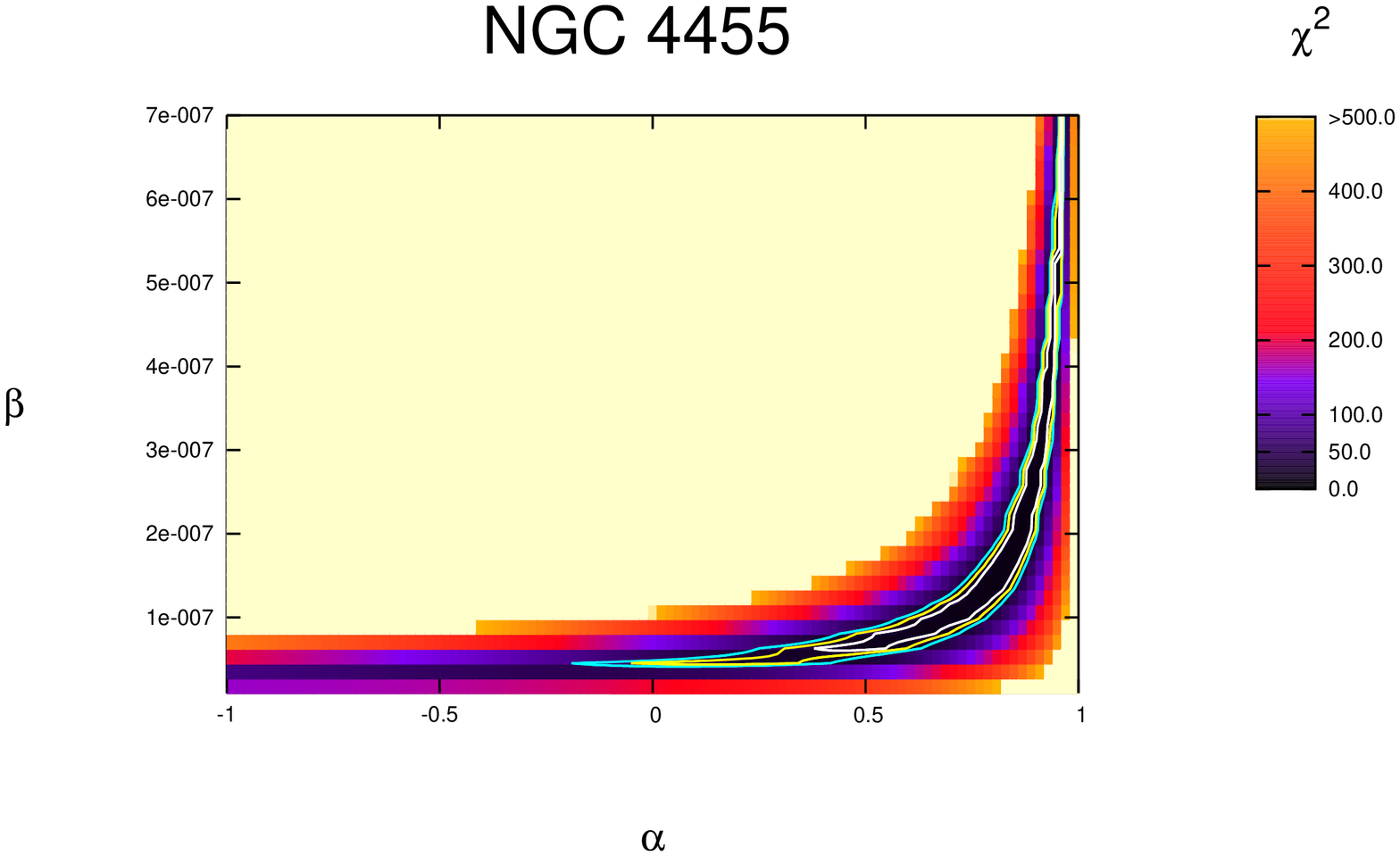} & %
\includegraphics[height=6cm, angle=270]{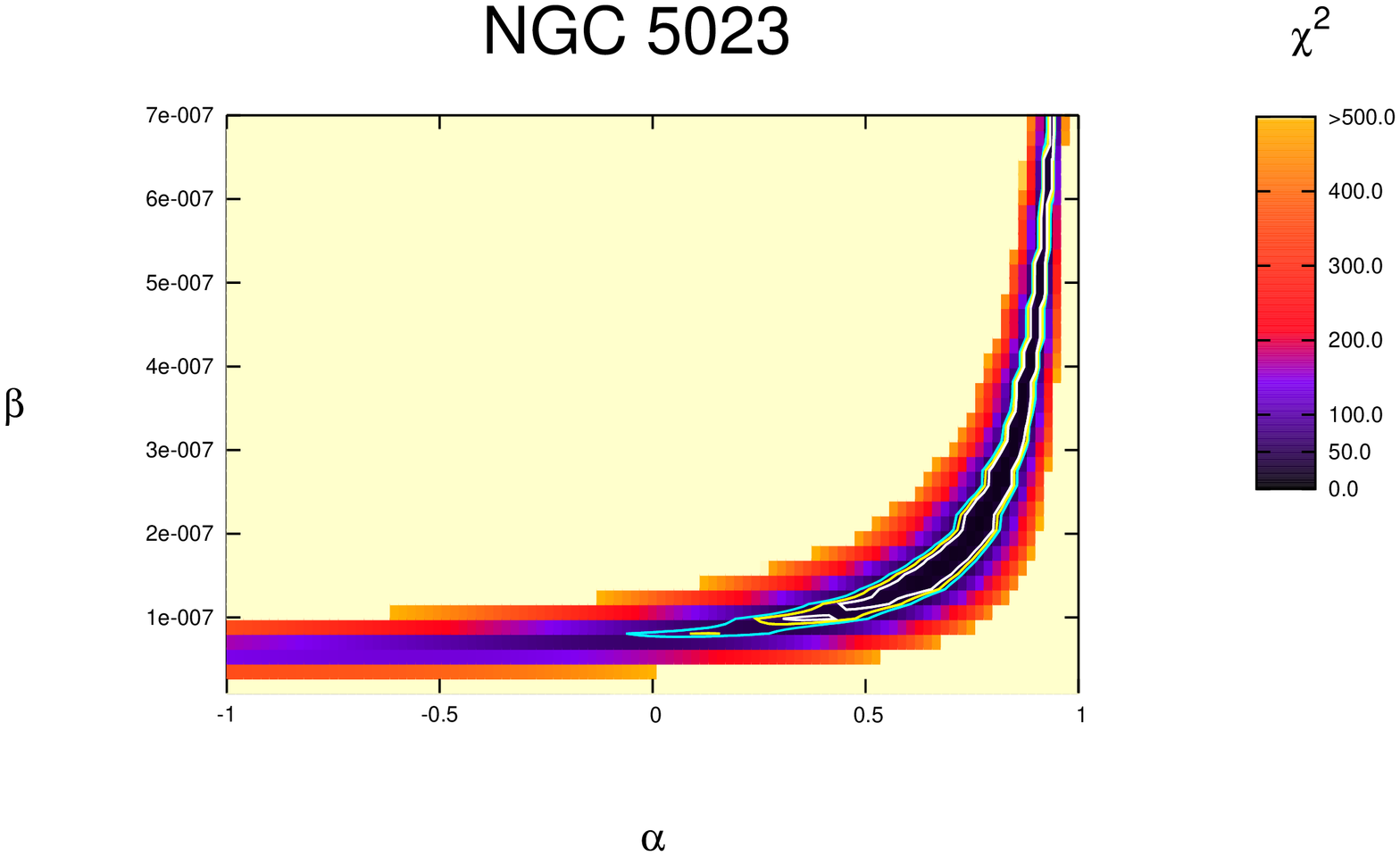} \\ 
\includegraphics[height=6cm, angle=270]{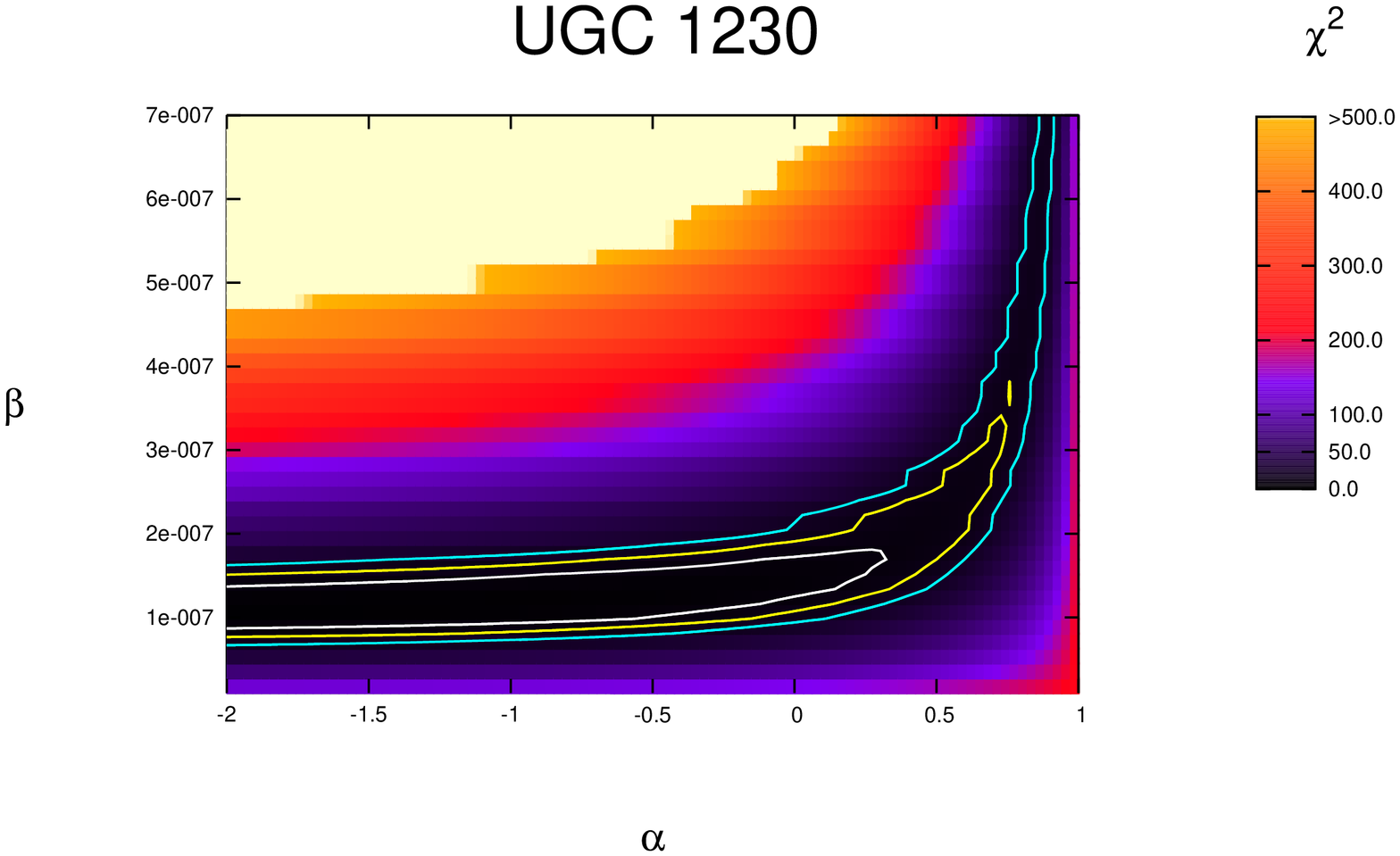} & %
\includegraphics[height=6cm, angle=270]{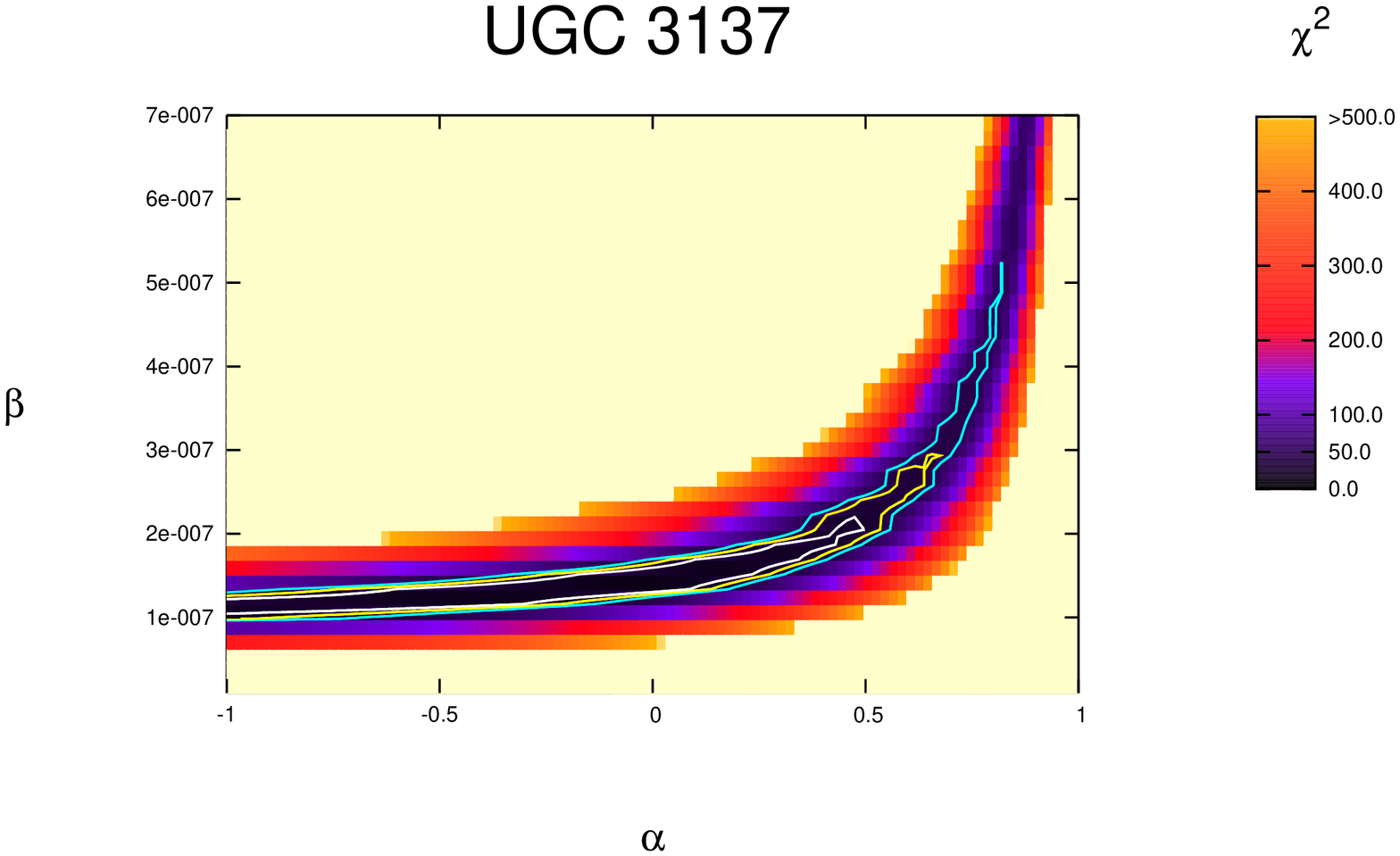} & %
\includegraphics[height=6cm, angle=270]{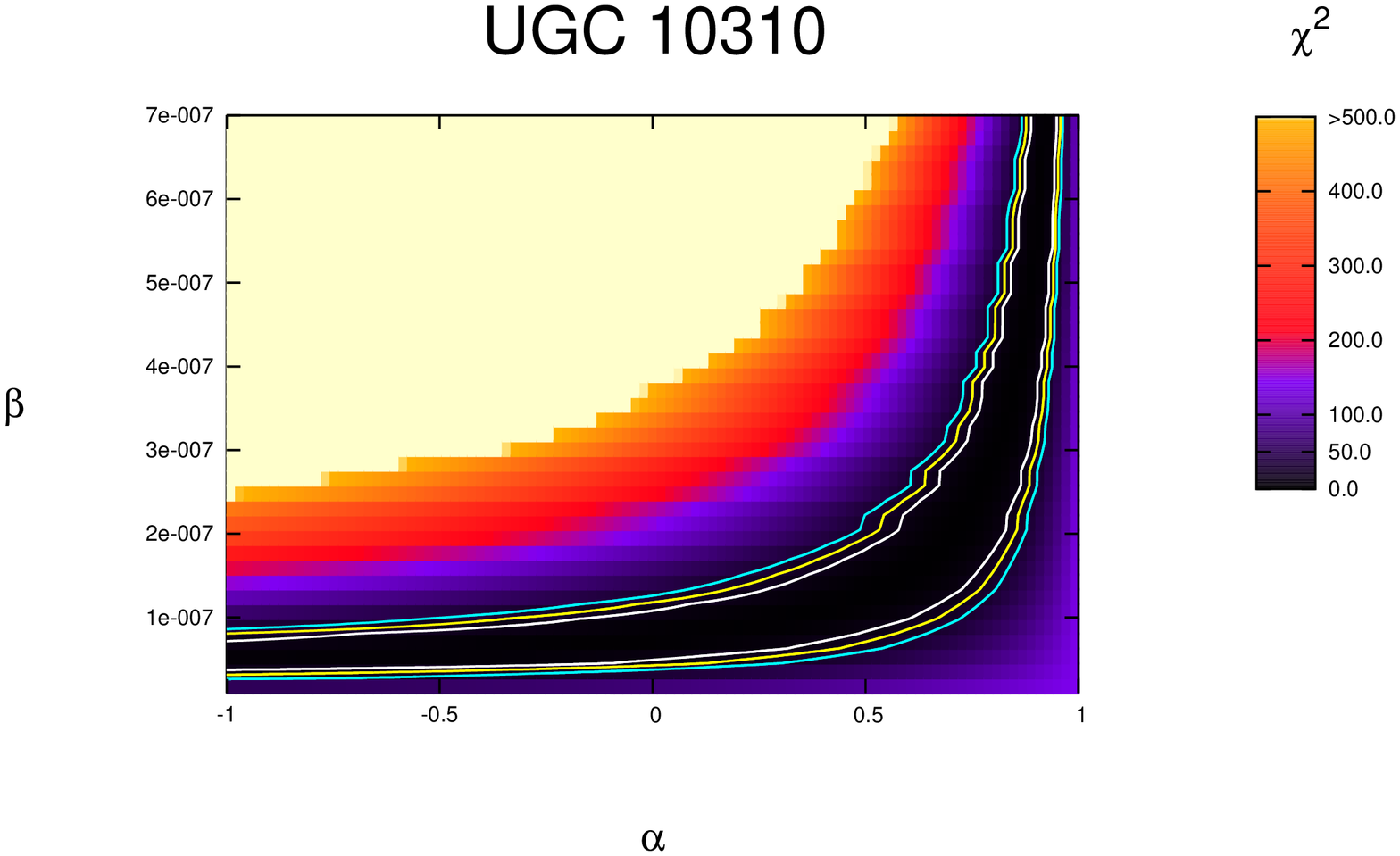} \\ 
&  & 
\end{tabular}%
\end{center}
\caption{The fit of the rotation curves of LSB galaxies. The contours refer
to the $68.3\%$ ($1\protect\sigma $), $95.4\%$ ($2\protect\sigma $) and $%
99.7\%$ ($3\protect\sigma $) confidence levels. The color code for $\protect%
\chi ^{2}$ is indicated on the vertical stripes. The light grey vertical
line denotes the forbidden region ($\protect\alpha =0$).}
\label{chi2LSB}
\end{figure*}

From a $\chi ^{2}$-test we determine $M_{0}$ and$\ r_{c}$, and also the Weyl
parameters $\alpha $ and $\beta $, shown on Table~\ref{Table2} and on Fig. 
\ref{chi2LSB}.

\section{Discussions and final remarks}

The shape of the observed rotation curves strongly indicate the need for
dark matter or equivalent modifications of gravity on galactic scale and
above. However there are no universally accepted candidates explaining the
whole amount of dark matter needed for agrement with observations. Because
dark matter does not interact with ordinary matter, except gravitationally,
the question comes whether dark matter is rather an effect of modified
gravity. We have investigated whether extra dimensions could provide this
type of modification in the simplest, codimension one brane-world scenario. 
We have derived an analytical expression for the rotational velocity
of a test particle on a stable circular orbit in the exterior region to a
galaxy, with Weyl fluid contributions (the dark pressure and dark radiation)
included. For this we have assumed a linear equation of state for the Weyl
fluid and we have argued for the rightness of this as follows.

In subsection 4.1 we have derived a system of two first order
differential equations for three brane variables with straightforward
geometrical meaning. These variables are: (1) the acceleration of the normal
congruence to the t=const hypersurfaces; (2) the expansion of the normal
congruences to the r=const spheres in the t=const hypersurfaces; and (3) the
scalar characterizing the electric part of the Weyl curvature of the brane
(due to its construction, this is defined on the t=const hypersurfaces). The
dark pressure and dark radiation do not enter these equations.

We also derived additional algebraic relations for these three
quantities valid in Schwarzschild space-time. As the brane-world spherically
symmetric space-time is a modification of the Schwarzschild solution, we
have slightly modified one of these relations (by introducing two continuous
deformation parameters, which take the value 1 for Schwarzschild), closing
in this way the system of equations for these variables. The Cauchy-Peano
theorem then proves the existence of a solution for this system.

We have shown then, how the dark pressure and dark radiation are
determined algebraically in terms of the above mentioned two kinematical and
one brane Weyl quantities. We find that the above mentioned choice induces a
linear relation between them, the desired equation of state.

There are two types of evolutions, which in principle could destroy
this solution. No problem is posed by the temporal evolution, due to the 
static character of the space-time. 
On the other hand, the above determined solution of the brane equations stands as a set of "initial" conditions for the sytem of first order off-brane evolution (propagation) equations (for example along some suitably defined parameter associated to the brane normal). This in turn again allows for solutions due to the Cauchy-Peano theorem. 
Thus we only have to check then, whether such initial conditions are
allowed, in other words whether there is any constraint involving these initial conditions,
which has to be obeyed. Such a constraint indeed exists: it is the
4-dimensional (twice contracted) Bianchi identity on the brane. This
identity however trivially follows from the equations solved before,
therefore the linear equation of state is compatible with both types of evolution.

The model has several free parameters, which in principle could be
fixed in such a way to explain the observed galactic rotation curve behavior.

In order to much closely test this assumption, we have considered a sample
of $9$ HSB and $9$ LSB galaxies with well measured combined $HI$ and $%
H\alpha $ rotation curves. Since LSB galaxies are known to be dark matter
dominated (\cite{MRB}), reproducing their rotation curves without the need of any dark
matter component would be a major achievement. Fitting the model to rotation
curve data and photometric measurements allowed us to constrain the Weyl
parameters $\alpha $ and $\beta =-C$; also determine the mass-to-light
ratios $M/L$ of the baryonic components in HSB galaxies, and the total core
mass $M_{0}$ and radius $r_{c}$ of LSB galaxies. The fit was in all
cases within 1$\sigma $ confidence level, which supports the
choice of the equation of state (\ref{EOS}) from a physical point of view.
In all cases we found $\beta \ll 1$ and in most cases $\alpha \in
\left(0.4,0.9\right) $, which for the constants $\mathcal{A}
=\left(1+\alpha \right) /2$, $\mathcal{B}=\mathcal{A}\left(1-\beta \right) +\alpha \beta $ introduced in Section 4.1 give $
\mathcal{A}\approx \mathcal{B}\in \left( 0.7,0.95\right) $. These
express the difference of our model from a Schwarzschild solution, for which 
$\mathcal{A}=\mathcal{B}=1$.

With the parameters determined from the fit the theoretical rotation curves
will have an almost flat (slightly increasing) asymptotic behavior at larger
radii then the available observations of $HI$ and $H\alpha $ velocities for
these galaxies. This tendency is somewhat contradictory with the shape of
the universal rotation curve, scaling only with the virial mass \citep{URC},
at least for the well-fitting model parameters. The asymptotic shape of the
universal rotation curve shows a decreasing tendency, as has been found from
N-body simulations, which assume the Navarro-Frenk-White CDM model %
\citep{NFW}. We note though that parameters of the Weyl fluid reproducing
the asymptotics of the universal rotation curve can be found, but for these
the fit with the $HI$ and $H\alpha $ velocity data falls outside 3$\sigma $
confidence level.

Observationally, the galactic rotation curves remain flat to the farthest
distances that can be observed. On the other hand there is a simple way to
estimate an upper bound for the cutoff of the constancy of the tangential
velocities. The idea is to consider the point at which the decaying density
profile of the dark radiation associated to the galaxy becomes smaller than
the average energy density of the Universe. Let the value of the coordinate
radius at the point where the two densities are equal be $R_{U}^{\max }$.
Then at this point $3\alpha _bU\left( R_{U}^{\max }\right) =\left(8\pi
G/c^2\right)\rho _{univ}$, where $\rho _{univ}c^2$ is the mean energy
density of the Universe. In the limit of large $r$, with the use of Eq.~(\ref%
{U}), we can approximate the dark radiation term as $3\alpha _b U\approx
v_0/r^2$. Hence for $R_U$ we obtain 
\begin{equation}
R_{U}^{\max }=c\sqrt{\frac{v_0}{8\pi G\rho _{univ}}}.
\end{equation}

The mean density of the Universe is given by $\rho _{univ}=\rho
_{crit}=3H_{0}^{2}/8\pi G$, where $H_{0}$ is the Hubble constant, given by $%
H_{0}=100h$ km/sec Mpc, $1/2\leq h\leq 1$. Therefore 
\begin{equation}  \label{ru}
R_{U}^{\max }\approx cH_{0}^{-1}\sqrt{\frac{v_0}{3}}\approx 3\times
10^3\times\sqrt{\frac{v_0}{3}}h^{-1}\;\mathrm{Mpc}.
\end{equation}

A numerical evaluation of $R_{U}^{\max }$ requires the knowledge of the
constant $v_0$ in the flat region, and of the basic fundamental cosmological
parameters. In the case of the HSB galaxies $v_0\in \left(0.237\times
10^{-6}, 0.63\times 10^{-6}\right)$, while in the case of the LSB galaxies, $%
v_0\in \left(0.316\times 10^{-7}, 0.316\times 10^{-6}\right)$. This gives
for $R_U^{\max}$ a range of $R_U^{\max}\in \left(0.84h^{-1}\;\mathrm{Mpc}
,1.37h^{-1}\;\mathrm{Mpc}\right)$ for the HSB galaxies, and $R_U^{\max}\in
\left(0.307h^{-1}\;\mathrm{Mpc},0.97h^{-1}\;\mathrm{Mpc}\right)$ for the LSB
galaxies, respectively. The measured flat regions are about $R\approx
2\times R_{opt}$, where $R_{opt}$ is the radius encompassing $83\%$ of the
total integrated light of the galaxy \citep{Bi87,Pe96,Bo01}. If we take as a
typical value $R\approx 30$ kpc, then it follows that $R<<R_{U}^{\max }$.
However, according to our model, the flat rotation curves region should
extend far beyond the present measured range.

An alternative estimation of $R_{U}^{\max }$ can be obtained from the
observational requirement that at the cosmological level the energy density
of the dark matter represents a fraction $\Omega _{m}\approx 0.3$ of the
total energy density of the Universe $\Omega =1$. Therefore the dark matter
contribution inside a radius $R_{U}^{\max }$ is given by $4\pi \Omega
_{m}\left(R_{U}^{\max }\right) ^{3}\rho _{crit}/3$, which gives 
\begin{equation}  \label{ru1}
R_{U}^{\max }\approx \sqrt{\frac{1}{2\Omega_{m}}}\frac{c}{H_{0}}\left(\frac{
v_{tg}}{c}\right).
\end{equation}

Therefore, by assuming that the dark radiation contribution to the total
energy density of the Universe is of the order of $\Omega _{m}\approx 0.3$
we have $R_{U}^{\max }\in \left(388h^{-1},3881h^{-1}\right) $ kpc for $%
v_{tg}\in \left(10^{-4},10^{-3}\right) $.

The limiting radius at which the effects of the extra-dimensions extend, far
away from the baryonic matter distribution, is given in the present model by
Eqs.~(\ref{ru}) or (\ref{ru1}). In the standard dark matter models this
radius is called the truncation parameter $s$, and it describes the extent
of the dark matter halos. Values of the truncation parameter by weak lensing
have been obtained for several fiducial galaxies by \citet{Ho104}. In the
following we compare our results with the observational values of $s$
obtained by fitting the observed values with the truncated isothermal sphere
model, as discussed in some detail in \citet{Ho104}. The truncation
parameter $s$ is related to $R_U^{max}$ by the relation $s=R_U^{max}/2\pi $
(see Eq.~(4) in \citet{Ho104}). Therefore, generally $s$ can be obtained
from the relation 
\begin{equation}
s\approx \frac{\sigma }{\sqrt{6}\pi }\sqrt{\frac{1}{2\Omega _{m}}}H_{0}^{-1},
\label{s}
\end{equation}
where $\sigma $ is the velocity dispersion, expressed in km/s. Hence the
truncation parameter is a simple function of the velocity dispersion and of
the cosmological parameters only. For a velocity dispersion of $\sigma =146$
km/s and with $\Omega _{m}=0.3$, Eq. (\ref{s}) gives $s\approx 245h^{-1}$
kpc, while the truncation size obtained observationally in \citet{Ho104} is $%
s=213h^{-1}$ kpc. For $\sigma =110$ km/s we obtain $s\approx 184h^{-1}$ kpc,
while $\sigma =136$ km/s gives $s\approx 228h^{-1}$ kpc. All these values
are consistent with the observational results reported in \citet{Ho104}, the
error between prediction and observation being of the order of $20\%$. We
have also to mention that the observational values of the truncation
parameter depend on the scaling relation between the velocity dispersion and
the fiducial luminosity of the galaxy. Two cases have been considered in %
\citet{Ho104}, the case in which the luminosity $L_B$ does not evolve with
the redshift $z$ and the case in which $L_B$ scales with $z$ as $L_B\propto
(1+z)$. Depending on the scaling relation slightly different values of the
velocity dispersion and truncation parameter are obtained.

One of the most straightforward evidence for dark matter comes from the
radial Tully-Fisher relation \citep
{Yegorova}: at a given galactocentric distance (measured in unit of the
optical radius) there is a relation between rotation velocities and the
absolute magnitudes of the galaxies. From this relation we can extract
information about the mass distribution of spiral galaxies. We have studied
whether the Weyl model satisfies this relation. The plot on Fig.~\ref%
{Tully-Fisher} is in good agreement with the result of \citet{Yegorova}.

Statistical studies of N-body simulations reveal that there is a link
between halo masses and galaxy properties (e.g. luminosity, baryonic mass) %
\citep{Shankar, Moster}. We have compared the equivalent halo mass predicted
by the Weyl model $M_{Weyl}$ with the halo mass arising from the
simulations, $M_{simulation}$ \citep{Moster} (\ref{Table5}). From the
baryonic mass $M_{b}(\infty )+M_{d}(\infty )$ we have derived the predicted
halo mass $M_{simulation}$ with Eq.~(2) in \citet{Moster}. With this mass we
have calculated the virial radius of the galaxy $R_{vir}$ (see Eq.~(4) in %
\citet{URC}) and the Weyl halo mass, defined as 
\begin{equation}
M_{Weyl}=c^{2}\beta \left[ 1-\left( \frac{r_{b}}{r}\right) ^{1-\alpha } %
\right] \frac{R_{vir}}{G}\ .
\end{equation}
With one exception the emerging masses are of same order of magnitude, in
one case being also equal. These derived data signals some tension between
the Weyl fluid model and the numerical simulations.

\begin{figure}[tbp]
\begin{center}
\begin{tabular}{c}
\includegraphics[height=8.3cm, angle=270]{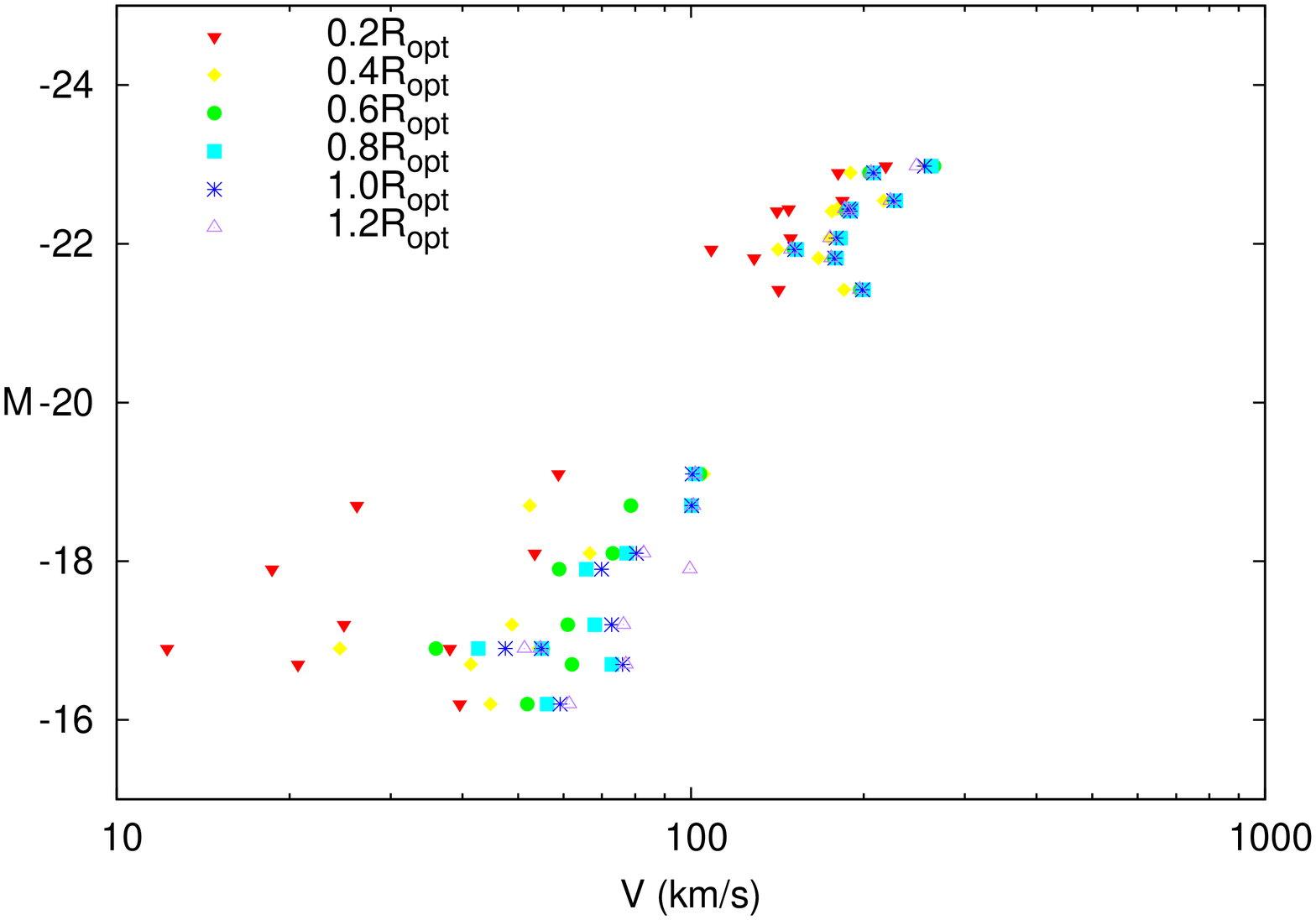}%
\end{tabular}%
\end{center}
\caption{The radial Tully-Fisher relation of our HSB and LSB Galaxy samples:
absolute magnitude of the galaxies as function of the rotational velocities
predicted by the Weyl model at $0.2$ (red), $0.4$ (yellow), $0.6$ (green), $%
0.8$ (cyan), $1$ (navy), $1.2$ (lilac) optical radii. }
\label{Tully-Fisher}
\end{figure}

\section{Acknowledgements}

We would like to thank to the anonymous referee for comments and suggestions
that helped us to significantly improve the manuscript. We acknowledge
discussions with J\'{o}zsef Vink\'{o} at an earlier stage of this work. L%
\'{A}G is grateful to Tiberiu Harko for kind hospitality during his visit at
the University of Hong Kong. L\'{A}G and ZK\ were partially supported by
COST Action MP0905 "Black Holes in a Violent Universe". TH was supported by
the GRF grant No.~701808P of the government of the Hong Kong SAR.

\section*{Appendix I: 3+1+1 decomposition of the 5D Weyl curvature }

\label{one}

With the brane normal $n^{A}$ and the temporal normal $
u^{B}$ to the spatial hypersurfaces singled out, the
five-dimensional Weyl tensor $\widetilde{C}_{ABCD}$ admits a 3+1+1
decomposition \citep{3+1+1}, generalizing the corresponding decomposition of
the four-dimensional Weyl tensor $C_{abcd}$.

The projections 
\begin{eqnarray}
-k_{4}^{4}U &=&\widetilde{C}_{ABCD}n^{A}u^{B}n^{C}u^{D}, \\
k_{4}^{4}Q_{K} &=&\widetilde{C}_{ABCD}h_{K}^{\,\,\,\,A}n^{B}u^{C}n^{D}\ , \\
-k_{4}^{4}P_{AB} &=&\widetilde{C}_{ABCD}h_{\langle
K}^{\,\,\,\,\,\,A}n^{B}h_{L\rangle }^{\,\,\,\,C}n^{D}\ .
\end{eqnarray}
of the 5-dimensional Weyl tensor combine to an\ effective fluid\ on
the brane, the Weyl fluid. (Angular brackets $\langle ~\rangle $ on indices
denote tensors which are projected in all indices with the 3-metric 
$h_{IJ}=\ ^{(5)}g_{IJ}-n_{I}n_{J}+u_{I}u_{J}$, symmetrized and
trace-free.)

The projections 
\begin{equation}
\mathcal{E}_{KL}=\widetilde{C}_{ABCD}h_{\langle
K}^{\,\,\,\,\,A}u^{B}h_{L\rangle }^{\,\,\,\,C}u^{D}
\end{equation}
and
\begin{equation}
\mathcal{H}_{KL}=\frac{1}{2}\varepsilon _{(K}^{\ \ \ \
AB}h_{L)}^{\,\,\,\,C}\ \widetilde{C}_{ABCD}u^{D}
\end{equation}
are related to the electric and magnetic parts of the
four-dimensional Weyl tensor on the brane, $\widetilde{E}
_{ac}=C_{abcd}u^{b}u^{d}$ and $\widetilde{H}_{kl}=\frac{1}{2}
\varepsilon _{(k}^{\ \ \ \ ab}h_{l)}^{\,\,\,\,c}C_{abcd}u^{d}$, as 
\begin{eqnarray}
\mathcal{E}_{ab} &=&E_{ab}-\frac{k_{4}^{4}}{2}P_{ab}-\frac{1}{2}\left( 
\widehat{K}+\frac{\widehat{\Theta }}{3}\right) \widehat{\sigma }_{ab} 
\nonumber \\
&&+\frac{1}{2}\widehat{\sigma }_{c\langle a}\widehat{\sigma }_{b\rangle
}^{\,\,\,\,c}+\frac{1}{2}\widehat{K}_{\langle a}\widehat{K}_{b\rangle }\ ,
\label{Eab} \\
\mathcal{H}_{ab} &=&H_{ab}-\varepsilon _{\langle a}^{\ \ \ \ cd}\widehat{
\sigma }_{b\rangle c}\widehat{K}_{d}\ ,  \label{Hab}
\end{eqnarray}%
Here $\widehat{K}=u^{B}u^{C}\widetilde{\nabla }_{C}n_{B}$, $
\widehat{K}_{A}=h_{A}^{\,\,\,\,K}u^{C}\widetilde{\nabla }_{C}n_{K}$, $
\widehat{\Theta }=h^{AB}\widetilde{\nabla }_{A}n_{B}$, and $
\widehat{\sigma }_{AB}=h_{\langle A}^{\,\,\,\,\,K}h_{B\rangle }^{\,\,\,\,L} 
\widetilde{\nabla }_{K}n_{L}$ are kinematic quantities emerging as
various projections of the extrinsic curvature of the brane ($\widehat{
\Theta }$ and $\widehat{\sigma }_{AB}$ being the expansion
and shear of the brane normal), while $\widetilde{\nabla }_{A}$ is
the covariant derivative in the 5-dimensional space-time.

The rest of the components 
\begin{eqnarray}
\mathcal{E}_{K} &=&\widetilde{C}_{ABCD}h_{K}^{\,\,\,\,A}u^{B}n^{C}u^{D}, \\
\mathcal{H}_{K} &=&\frac{1}{2}\varepsilon _{K}^{\ \ AB}\widetilde{C}
_{ABCD}u^{C}n^{D}\ ,~ \\
\mathcal{F}_{KL} &=&\widetilde{C}_{ABCD}h_{(K}^{\,\,\,\,\,\,A}u^{B}h_{L)}^{
\,\,\,\,C}n^{D}, \\
\widehat{\mathcal{H}}_{KL} &=&\frac{1}{2}\varepsilon _{(K}^{\ \ \
AB}h_{L)}^{\,\,\,\,C}\,\,\widetilde{C}_{ABCD}n^{D}\ 
\end{eqnarray}
do not enter the equations on the brane.

\section*{Appendix II: Dynamics on the brane}

\label{two}

In the 2+1+1 formalism on a brane with spherical symmetry the
covariant derivative of any scalar field $f\left(r\right)$, taken
along the integral curves of the radial vector field $r^{a}$ is
defined \citep{3+1+1} as $f^{\star}=r^{a}D_{a}f\ $. The nontrivial
dynamical equations are four ordinary differential equations: 
\begin{equation}
\widetilde{\Theta }^{\star }+\frac{\widetilde{\Theta }^{2}}{2}+\frac{2 
\widetilde{E}}{3}+\frac{k_{4}^{4}}{3}\left( 2U+P\right) =0\,,  \label{beq1}
\end{equation}
\begin{equation}
\!\!\left( U+2P\right) ^{\star }+\!4AU+\left( 2A+3\widetilde{\Theta }
\!\right) P=0\ ,\,  \label{beq2}
\end{equation}
\begin{equation}
A^{\star}+A\left( \widetilde{\Theta}+A\right) -k_{4}^{4}U=0\ ,
\label{beq3}
\end{equation}
\begin{equation}
\!A^{\star }+A\left( A-\frac{\widetilde{\Theta }}{2}\right) -\widetilde{\!E}%
+ \frac{k_{4}^{4}}{2}P=0\ ,  \label{beq4}
\end{equation}
for the five variables $\widetilde{\Theta },~A,~\widetilde{E},~U,~P$
. Eq. (\ref{beq2}) is equivalent with the constraint Eq. (\ref{DU}).
The difference of Eqs. (\ref{beq3}) and (\ref{beq4}) gives a simple
algebraic relation between the dark radiation $U$, dark pressure $
P$\, acceleration $A$ of temporal normals, expansion of
radial geodesics $\widetilde{\Theta }$ and electric part of the
4-dimensional Weyl tensor $\widetilde{E}$, respectively:
\begin{equation}
k_{4}^{4}\left(2U+P\right) =3A\widetilde{\Theta }+2\widetilde{E}\,.
\label{beq5}
\end{equation}
This enables us to eliminate the Weyl constributions from Eq. (\ref%
{beq3}), obtaining 
\begin{equation}
\widetilde{\Theta }^{\star }+\frac{\widetilde{\Theta }^{2}}{2}+\frac{4 
\widetilde{E}}{3}+A\widetilde{\Theta }=0\,.  \label{beq1b}
\end{equation}

Another algebraic relation emerges as follows. From (a) the relation
between the Ricci scalar of the sphere and the Riemann tensor of the
3-space: 
\begin{eqnarray}
^{2}\mathcal{R} &=&\left( h^{ac}-r^{a}r^{c}\right) \left(
h^{bd}-r^{b}r^{d}\right) \Biggl[\,^{3}\mathcal{R}_{abcd}  \nonumber \\
&&+\left( D_{a}r_{c}\right) \left( D_{b}r_{d}\right) -\left(
D_{a}r_{d}\right) \left( D_{b}r_{c}\right) \Biggr]  \nonumber \\
&=&\frac{2}{3}\left[ k_{4}^{4}\left( U-P\right) -2\widetilde{E}\right] + 
\frac{\widetilde{\Theta }^{2}}{2}\,,  \label{R2scalar}
\end{eqnarray}
where we have used $D_{a}r_{b}=\widetilde{\Theta }\left(
h_{ab}-r_{a}r_{b}\right) /2$, (b) the 3-dimensional Riemann tensor
given in \citet{3+1+1}\footnote{
We note that there is a missing factor 2 in front of $\mathcal{E}$ in Eq.
(23) of \citet{3+1+1}. The conversion of
notations to the notations of the present paper is: $\left(\mathcal{E},~
\widehat{\mathcal{E}}_{a},~ \widehat{\mathcal{E}}_{ab},~E_{ab}\right)
~\rightarrow ~\left(-k_{4}^{4}U,~k_{4}^{4}Q_{a},~-k_{4}^{4}P_{ab},~
\widetilde{E}_{ab}\right) $.}, which in the spherically symmetric case
simplifies to 
\begin{eqnarray}
^{3}\mathcal{R}_{abcd} &=&\frac{2k_{4}^{4}}{3}U\,h_{c[a}h_{b]d}-2\left( 
\widetilde{E}_{d[a}h_{b]c}-\widetilde{E}_{c[a}h_{b]d}\right)  \nonumber \\
&&-k_{4}^{4}\left( P_{d[a}h_{b]c}-P_{c[a}h_{b]d}\right) \,,
\end{eqnarray}
and (c) the Gaussian curvature of the two-dimensional spacelike
group orbits orthogonal to $r^{a}$ and $u^{a}$ being  $
^{2}\mathcal{R}=2/r^{2}$, we obtain 
\begin{equation}
\widetilde{E}=\frac{k_{4}^{4}}{2}\left( U-P\right) +\frac{3\widetilde{\Theta 
}^{2}}{8}-\frac{3}{2r^{2}}\,\,.  \label{rcoord}
\end{equation}
We need to relate the newly introduced curvature coordinate to the $
\star$-derivative. For this we take the $\star$
-derivative of Eq. (\ref{rcoord}), employ Eqs. (\ref{beq1b}), (\ref{beq2}),
(\ref{beq3}); also the $\star$-derivative of Eq. (\ref{beq5}) for
eliminating $\widetilde{E}^{\star}$, finding
\begin{equation}
\left( \ln r^{2}\right) ^{\star }=\widetilde{\Theta }\,.  \label{rstar}
\end{equation}
This relation allows to replace the $\star$-derivative in
all equations by $r$-derivatives, denoted by a prime. Thus Eqs. (\ref{beq1b}), (\ref
{beq2}), (\ref{beq3}) can be rewritten as
\begin{equation}
\frac{r\widetilde{\Theta }}{2}\widetilde{\Theta }^{\prime }+\frac{\widetilde{
\Theta }^{2}}{2}+\frac{4\widetilde{E}}{3}+A\widetilde{\Theta }=0\,,
\label{beqr1}
\end{equation}
\begin{equation}
\frac{r\widetilde{\Theta }}{2}\!\!\left( U+2P\right) ^{\prime }+\!4AU+\left(
2A+3\widetilde{\Theta }\!\right) P=0\ ,\,  \label{beqr2}
\end{equation}
\begin{equation}
\frac{r\widetilde{\Theta }}{2}A^{\prime }+A\left(\widetilde{\Theta }
+A\right) -k_{4}^{4}U=0\ ,  \label{beqr3}
\end{equation}
where the prime denotes the derivative with respect to $r$.

In summary, the variables $U$, $P$, $A$,
$\widetilde{E}$ and $\widetilde{\Theta }$ are constrained
by two independent first order ordinary differential equations (\ref{beqr1}),
(\ref{beqr3}) and two algebraic equations (\ref{beq5}), (\ref{rcoord}).
Eq. (\ref{beqr2}) follows from these.

\end{document}